\documentclass[a4paper,11pt]{article}
\pdfoutput=1 

\usepackage{jinstpub} 

\title{\boldmath Simulation study of energy resolution, position resolution and $\pi^0$-$\gamma$ separation of a sampling electromagnetic calorimeter at high energies}

\author[a]{Ashim Roy,}
\author[b,1]{Shilpi Jain,\note{Corresponding author.}}
\author[c]{Sunanda Banerjee,}
\author[a]{Satyaki Bhattacharya,}
\author[d]{Gobinda Majumder}

\affiliation[a]{Saha Institute of Nuclear Physics, HBNI,\\1/AF Bidhannagar, Kolkata, India}
\affiliation[b]{National Central University,\\Taiwan}
\affiliation[c]{Fermi National Accelerator Lab,\\United States}
\affiliation[d]{Tata Institute of Fundamental Research,\\India}

\emailAdd{shilpi.jain@cern.ch}

\abstract{A simulation study of energy resolution, position resolution, and $\pi^0$-$\gamma$ separation using multivariate methods 
of a sampling calorimeter is presented. 
As a realistic example, the geometry of the calorimeter is taken from the design geometry of the Shashlik calorimeter 
which was considered as a candidate for CMS endcap for the phase II of LHC running. 
The methods proposed in this paper can be easily adapted to various geometrical
layouts of a sampling calorimeter.
Energy resolution is studied for different layouts and different absorber-scintillator combinations of the Shashlik detector.
It is shown that a boosted decision tree using fine grained information of the calorimeter can perform three times better than 
a cut-based method for separation of  $\pi^0$ from $\gamma$ over a large energy range of 20 GeV-200 GeV.}


\arxivnumber{1703.05246} 

\begin{document}
\maketitle
\flushbottom

\section{Introduction} \label{sec:Shashlikwork}

Sampling calorimeters serve as a cost effective, yet highly performing, energy 
measurement device in many major high energy experiments
and have been studied in the recent past for calorimetry in the future 
experiments. One such example is the CMS \cite{ref:cms} experiment where the 
Shashlik detector was considered as a replacement of the existing PbWO$_4$ 
crystal calorimeter. 
In this study we use the material and geometry used in the CMS prototype for 
endcap calorimetry \cite{ref:tdr}. In this design LYSO is used as the sensitive 
detector.
LYSO (cerium doped lutetium yttrium silicate) is a
radiation hard, high light yield (about 4 times of BGO), high 
stopping power ($\rho$ = 7.4 g/cm$^3$ , X$_0$ = 1.14 cm and R$_{Moliere}$ = 2.07 
cm) and fast response ($\tau$ = $~$ 40 ns) inorganic scintillator~\cite{ref:lysopaper1,ref:lysopaper2}.

For absorber, lead and tungsten are the two possible choices. 
For this study, the baseline option uses 4~mm thick lead layers
interleaved with 2~mm thick LYSO. The alternative scenario considered
uses 2.5~mm thick tungsten with 1.5~mm thick LYSO \cite{bib:rhenstalk}.
The scintillation light is read out using four wavelength shifting fibers going 
all the way through a Shashlik tower.

The following properties of the Shashlik calorimeter are studied:
\begin{enumerate}
\item the sampling resolution 
as a function of number of absorber/scintillator layers
as well as total energy resolution; 
\item position resolution of the impact point of photons on the calorimeter;
\item $\pi^{0}$/$\gamma$ separation using the information from the four fibers. 
\end{enumerate}
These are discussed in Sections \ref{sec:ShashlikResoStudy}, \ref{sec:position} 
and \ref{sec:Shashlikpi0gammaStudy} respectively.
\section{Simulation} \label{sec:simulation}
A stand-alone detector setup consisting of alternative layers of absorbers
(either 4 mm thick lead or 2.5 mm thick tungsten) and scintillators (2.0~mm or
1.5~mm thick LYSO) is defined in the framework of \textsc{Geant4}~\cite{ref:g4}.
\textsc{Geant4} version 9.6.p02 is used with the physics list QGSP\_FTFP\_BERT.
Light saturation effect is introduced through the use of Birk's law
~\cite{ref:birk1}:
\begin{eqnarray*}
w & = & \frac{k_{0}}{\left(1 + k_{1}\cdot\left(\frac{dE}{dX}\right) + k_{2}\cdot\left(\frac{dE}{dX}\right)^{-1}\right)}
\end{eqnarray*}
with $k_0$ = $0.883$, $k_1$ = $6.50\times 10^{-3}$ MeV$^{-1}\cdot$g$\cdot$cm$^{-2}$ 
and $k_2$ = $-0.241$ MeV$\cdot$g$^{-1}\cdot$cm$^{2}$ as measured in \cite{ref:birk2}. 
The weight factor $w$ is restricted in the range 0.1:1.0.

For studies of energy resolution, a single Shashlik tower, with no lateral segmentation 
and with each layer of  transverse size  $100\times100$ cm$^{2}$ is used. 
The gun particle is directed along the central axis of the tower.
The transverse size of the tower is sufficient to avoid any lateral leakage of the shower.
Monochromatic electrons of energy 50, 150 and 200 GeV and monochromatic 
photons of energy 50, 100, 150, 200, 300, 400 and 500 GeV are produced for 
sampling resolution study as a function of number of layers. To study total
energy resolution, electrons of energies 10, 20, 30, 50, 70, 100, 150, 200, 
250, 300,  500, 700 and 1000 GeV are generated with 28 layers of absorber
and 29 layers of scintillators. 
Ten thousand events are generated for each sample. 
Also, LYSO is considered to be undamaged by 
radiation ({\it i.e.} its attenuation length is taken to be 100 cm in this
scenario). 

For $\pi^0$-$\gamma$ separation studies, a detector setup with 28 layers of absorber and 29 layers
of scintillators in a 11$\times$11 matrix is defined. Transverse size of each 
tower in each layer is chosen to be $14 \times 14$ mm$^{2}$. Five fiber paths 
are defined of
which the central fiber is for calibration and the other four fibers
are at positions ($\pm$3.5~mm, $\pm$3.5~mm) with respect to the central axis.
They are read out individually or to a combined output. The fibers
are of diameter 1.6~mm and are inserted in holes of diameter 1.6 mm. 
The hit position is uniformly distributed in X and Y directions between 
$-$7~mm and $+$7~mm with respect to the centre of the central Shashlik tower. 
All the samples are generated using single particle gun with momentum along the Z direction. 
The gun is placed at a distance of 3.2~m from the calorimeter. 
For this study, photons and $\pi^{0}$'s of energy 
10, 20, 30, 50, 70, 100, 150 and 300 GeV are shot at the calorimeter. 

Energy deposited at a given point in the scintillator plate is shared 
unequally by the four fibers.
The closer a fiber is to the point of impact of the photon, the higher its 
probability of collecting light is.
Also the probability distribution of scintillator light 
among the fibers depends on the transmission coefficient of the scintillator 
and hence on integrated luminosity. This is estimated in a separate 
study \cite{ref:shasha} using SLitrani \cite{bib:slitrani}. 

\begin{figure}[htbp]
  \begin{center}
    \begin{tabular}{cc}
      \includegraphics[width=0.5\textwidth]{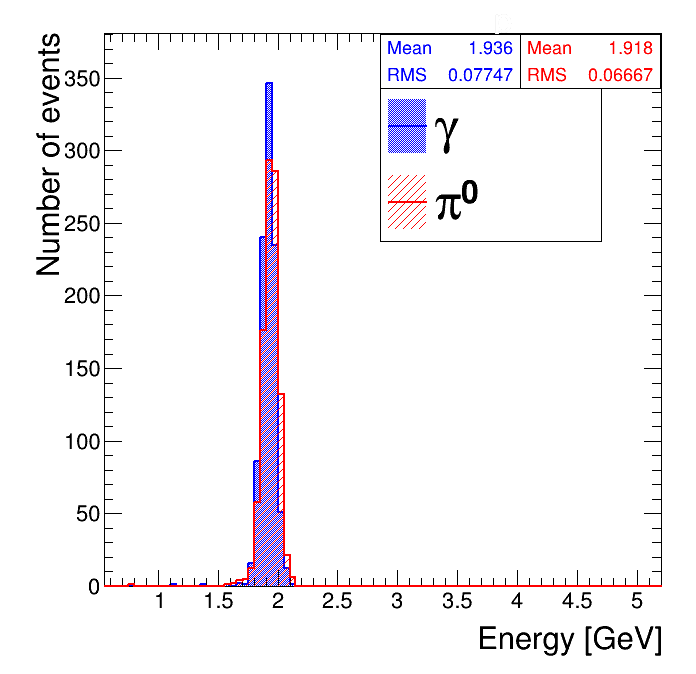}
      \includegraphics[width=0.5\textwidth]{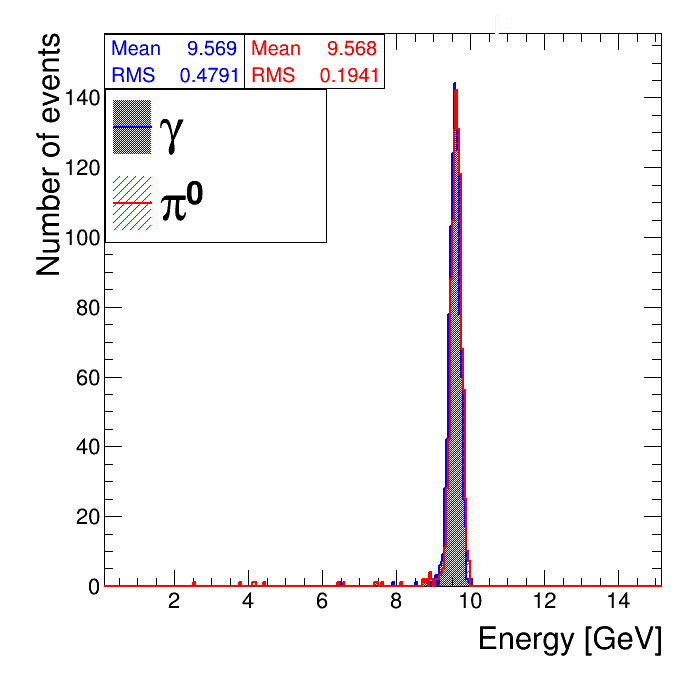}
    \end{tabular}
  \end{center}
  \caption{
    Total energy deposit in the scintillator of the 11 $\times$ 11 matrix for 
    10 GeV photons and $\pi^{0}$'s on the left and 50 GeV photons and $\pi^{0}$'s on the right. 
    Blue shaded histogram is for photons and red hatched histogram is for $\pi^{0}$'s.}
  \label{fig:totEphopi0_10_50}
\end{figure}

To validate the geometry, total energy deposit in the scintillator of the 11 $\times$ 11 
matrix is looked at. As shown in Figure \ref{fig:totEphopi0_10_50}, the 
total energy deposit in the scintillator for $\pi^{0}$ and for $\gamma$ are very 
similar as expected. 

%
\section{Energy resolution} \label{sec:ShashlikResoStudy}
The energy resolution in the Shashlik detector depends on the following factors:
\begin{enumerate}
 \item energy leakage;
 \item sampling fluctuation;
 \item photo statistics;
 \item electronic noise;
 \item other sources which include contributions from pile-up and 
       inter-calibration.
\end{enumerate}
The second term is specific for a sampling calorimeter while the other terms 
contribute also to any homogenous calorimeter like the one used in the CMS experiment.
Sampling resolution depends on the number of layers and the relative 
thickness between absorber and scintillator. A study is performed to
obtain the optimum number of layers to achieve good sampling resolution. 
\subsection{Sampling resolution} \label{sec:samplReso}
\begin{figure}[htbp]
  \begin{center}\begin{tabular}{cc}
\includegraphics[width=0.5\textwidth]{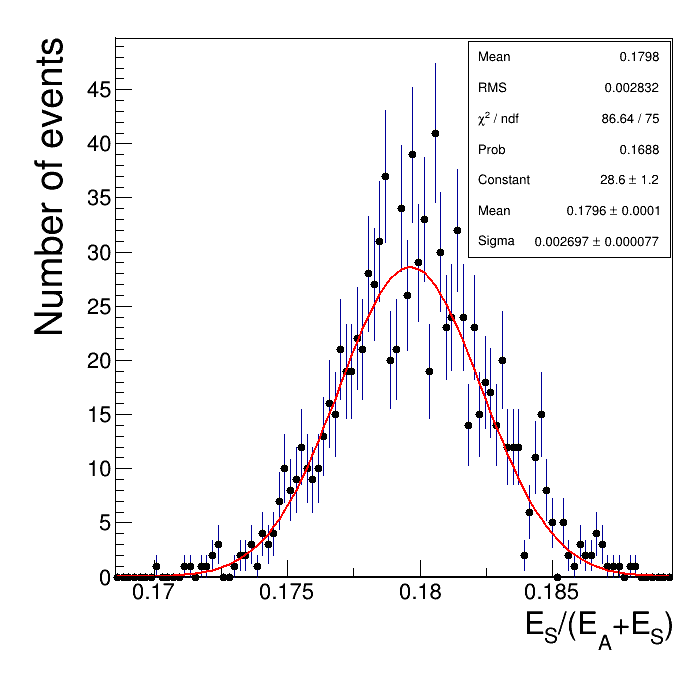}&
\includegraphics[width=0.5\textwidth]{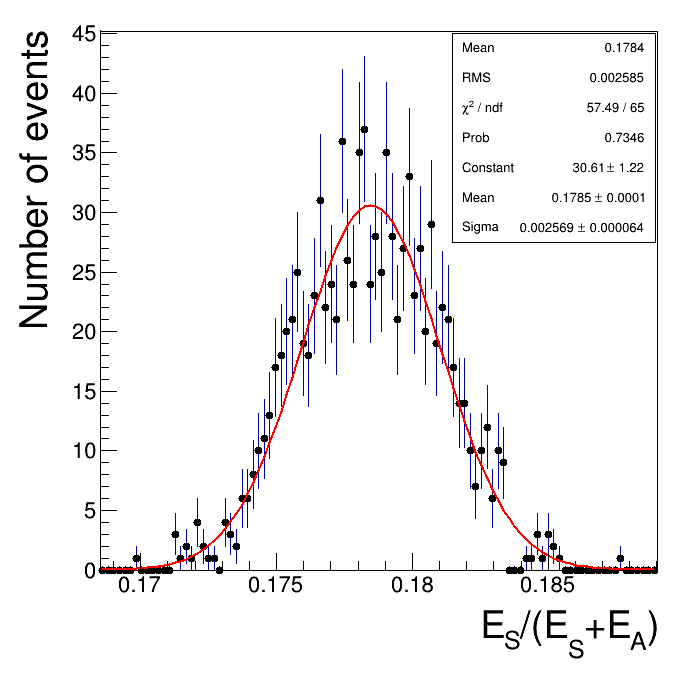}\\
\includegraphics[width=0.5\textwidth]{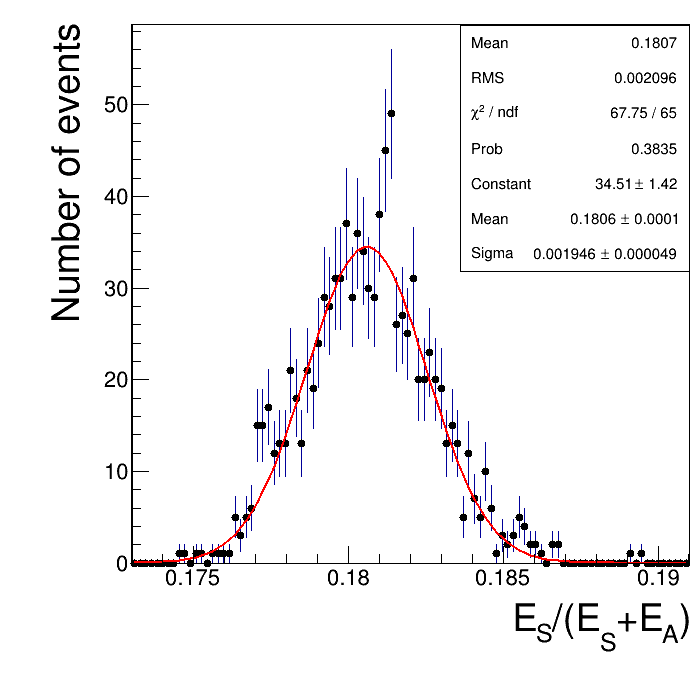}&
\includegraphics[width=0.5\textwidth]{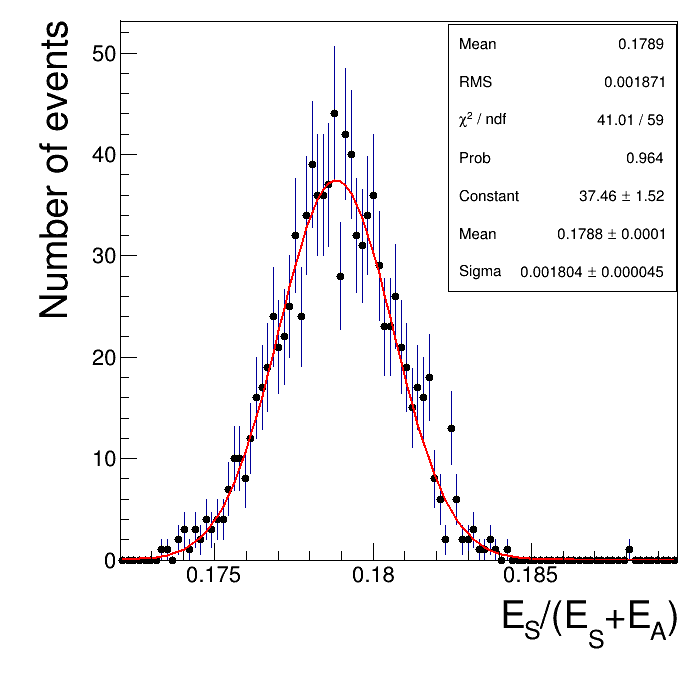}
  \end{tabular}\end{center}
\caption{The top-left and bottom-left plots show the distributions of 
         sampling fraction (F$_{S}$) for 18 layers and the top-right and 
         bottom-right plots are for 30 layers. Top plots are for 50 GeV 
         electrons and bottom plots are for 100 GeV electrons. These 
         distributions are fitted to Gaussian distribution functions.}
\label{fig:samplfrac_ele50150gev}
\end{figure}

\begin{figure}[htbp]
  \centering
  \begin{tabular}{cc}
\includegraphics[width=0.45\textwidth]{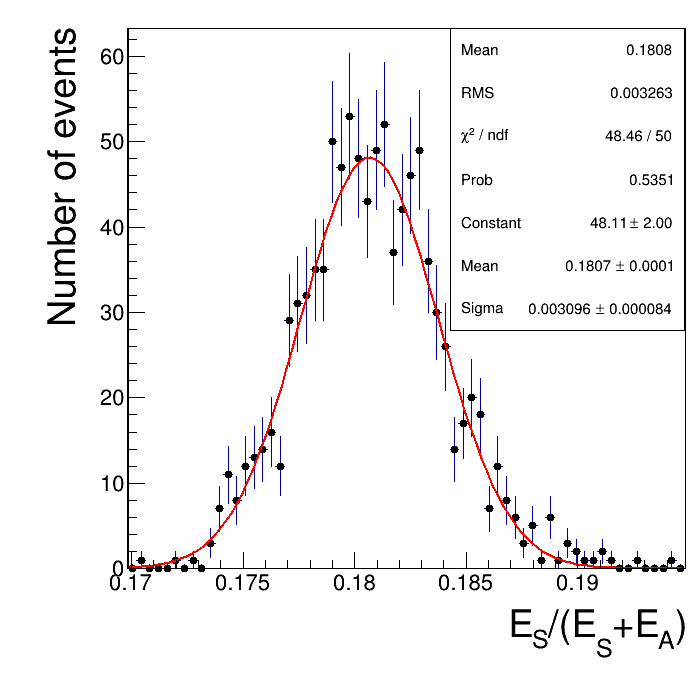}&
\includegraphics[width=0.45\textwidth]{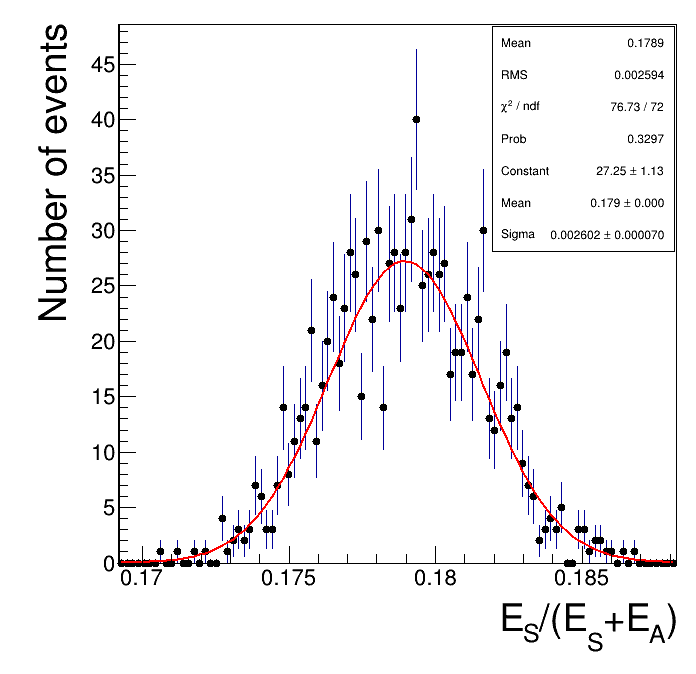}\\
\includegraphics[width=0.45\textwidth]{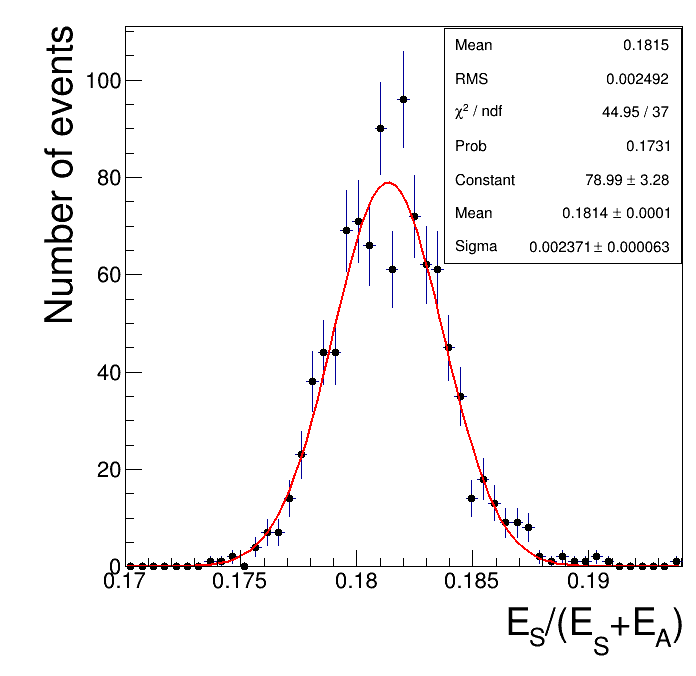}&
\includegraphics[width=0.45\textwidth]{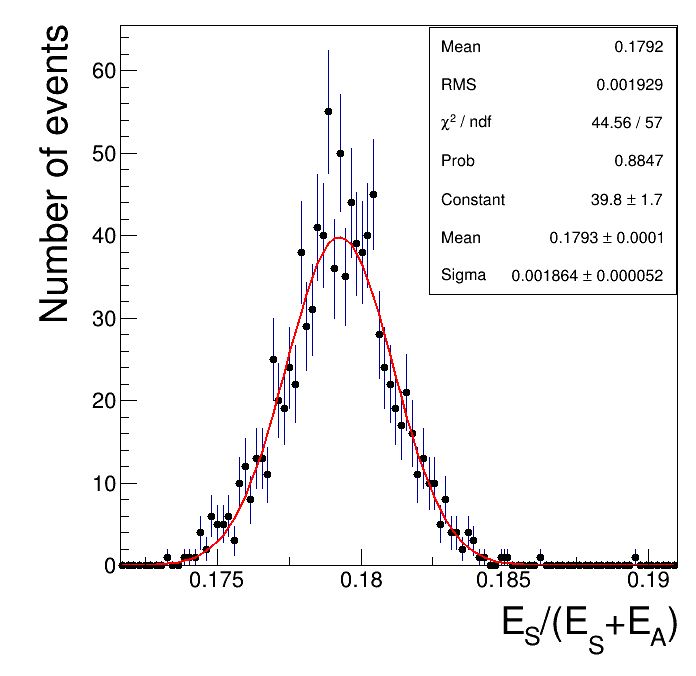}\\
  \end{tabular}
\caption{The top-left and bottom-left plots show the distributions of  
         sampling fraction (F$_{S}$) for 18 layers and the top-right and 
         bottom-right plots are for 30 layers. Top plots are for 50 GeV photons 
         and bottom plots are for 100 GeV photons. These distributions are 
         fitted to Gaussian distribution functions.}
\label{fig:samplfrac_pho50150gev}
\end{figure}

For a given thickness of the absorber and sensitive layer and a given energy of electron/photon gun
the sampling fraction is defined as

\begin{equation}
\label{eqn:samplingfrac}
F_{s} = \frac{E_{S}}{E_{S}+E_{A}}
\end{equation}

where E$_{S}$ and E$_{A}$ are the energies deposited in the scintillator layer 
and the absorber layer. The deposited energies in the scintillator as well as 
in the absorber layers follow Gaussian distribution. Figure 
\ref{fig:samplfrac_ele50150gev} shows distributions of sampling fractions for 
50 GeV and 100 GeV electrons in a configuration with 18 and 30 layers.  
The absorber in both the configurations is lead. Figure \ref{fig:samplfrac_pho50150gev} shows 
similar distributions for 50 GeV and 100 GeV photons in a detector
configuration with 18 and 30 layers.
This configuration is with tungsten as absorber. 
Gaussian function provides good description of these distributions and the fits 
are used to estimate the sampling resolutions.
One advantage of estimating the sampling resolution from $F_{S}$ and 
not estimating from  the distribution of E$_{S}$ is that E$_{S}$ contains also 
the contribution to resolution due to leakage. In case of sampling fraction,
the term due to leakage appears in the numerator and the denominator and gets
canceled.

\begin{figure}[htbp]
\centering
\begin{tabular}{cc}
\includegraphics[width=0.45\textwidth]{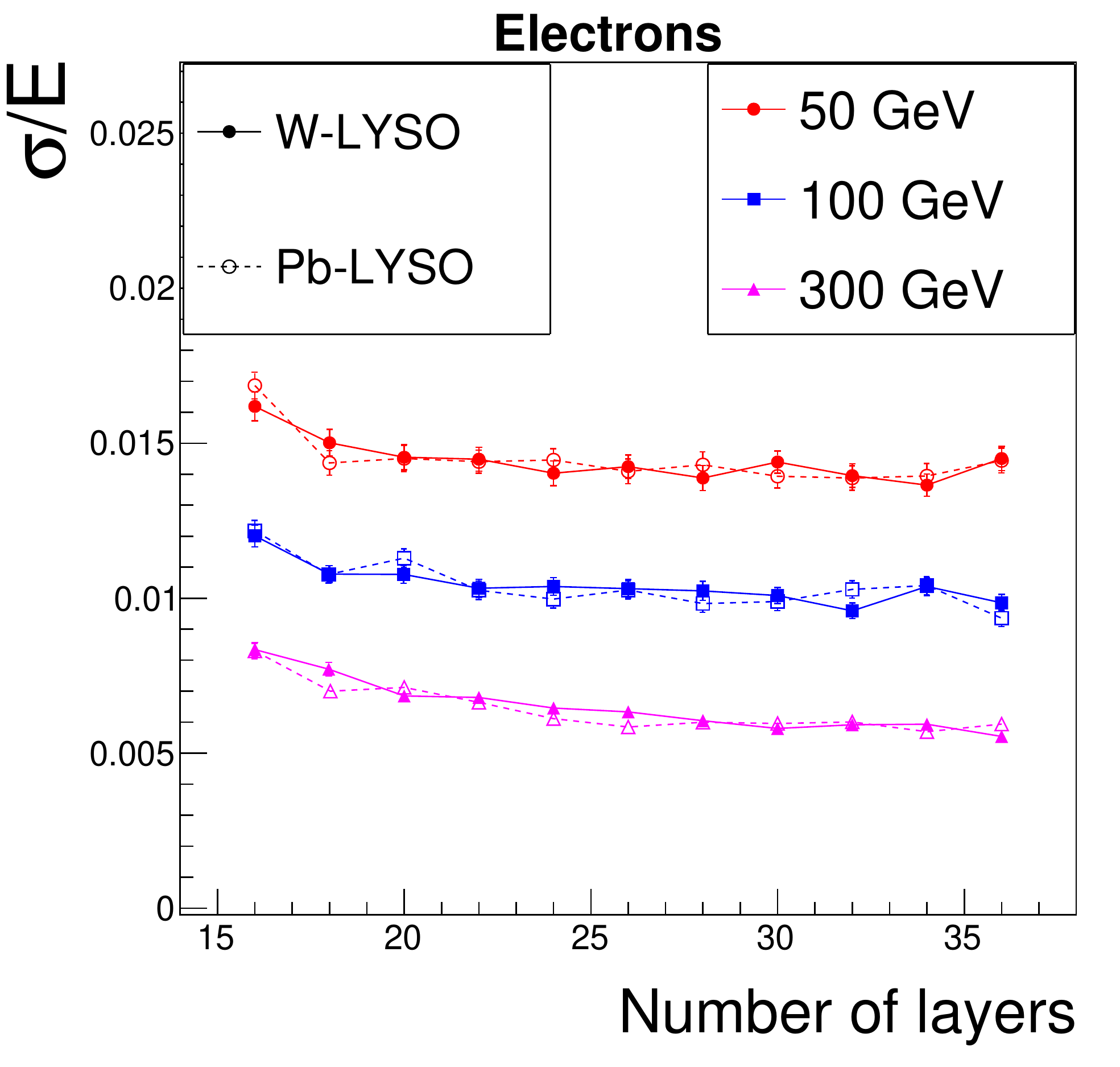}&
\includegraphics[width=0.45\textwidth]{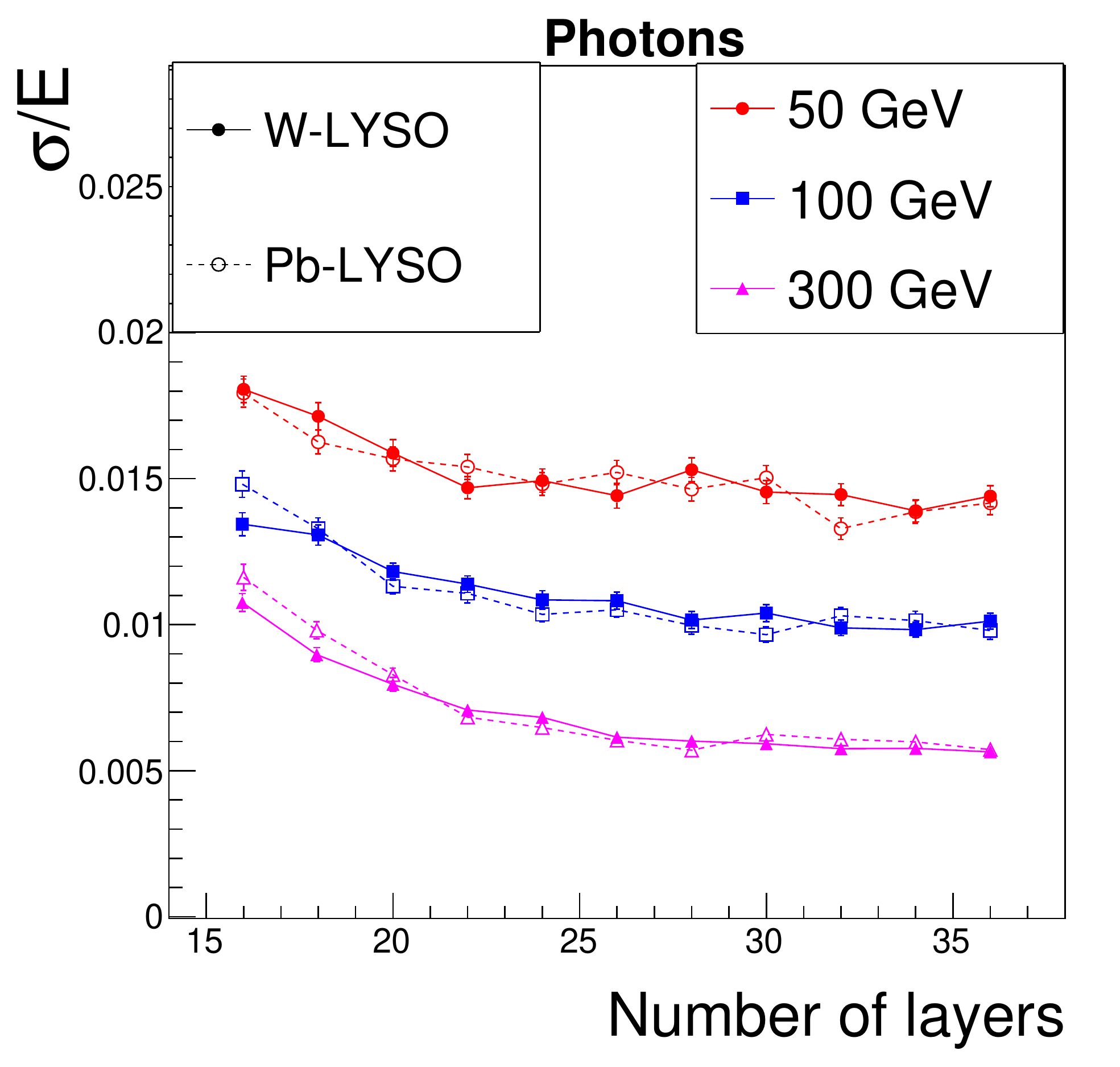} \\
\end{tabular}
\caption{Plot on the left shows the sampling resolution of electrons of various 
         energies for different number of layers of Pb-LYSO (dashed line with 
         hollow triangle) and W-LYSO (solid line with solid circle). The plot 
         on the right shows the same for photons.}
\label{fig:samplreso_elepho}
\end{figure}

Fits are performed for each energy point and for each configuration. 
The fitted mean ($\bar{E}$) and the fitted width (${\sigma} $) of the Gaussian are used to 
estimate the energy resolution:
\begin{equation}
\label{eqn:resosampl}
\textit{sampling-resolution} = {\sigma} / \bar{E},
\end{equation}
Figure \ref{fig:samplreso_elepho} shows the sampling resolution for electrons and photons as a function of number of layers of 
absorber and scintillator. The following points are to be noted:
\begin{enumerate}
\item For same number of layers, sampling resolution becomes better as the 
energy increases. 
\item As the number of layers increases, sampling resolution becomes 
better. 
\item For number of layers larger than 28, the resolution becomes nearly flat. 
This corresponds to around 25 radiation lengths (X$_{0}$) in both the 
configurations (Pb-LYSO and W-LYSO), when longitudinal shower leakage is very 
small. Both configurations seem to give similar results within 
the statistics.
\item 25X$_{0}$ corresponds to $\sim$16.8~cm for Pb-LYSO and 11.8~cm in case of W-LYSO. If detector length is not 
an issue, then one can go with the less expensive configuration.
\end{enumerate}

\subsection{Total energy resolution} \label{sec:totEnReso}
Following are the terms contributing to the total energy resolution of the Shashlik detector:
\begin{description}
\item[Energy leakage:] To estimate this term, distributions of 
E$_{A}$+E$_{S}$ are plotted in Figure \ref{fig:eTot_el10501003005001000gev} for different electron beam energies. The $\sigma/\bar{E}$ of this distribution gives the 
energy resolution due to leakage, where $\sigma$ is the measure of spread of 
the shower distribution and $\bar{E}$ is the mean of the shower distribution. 
It has low energy tail due to  energy leakage because of limited length of the detector. Since the 
transverse size is 100 cm which is much larger than the Moliere radius ( 2.07 cm ) of the LYSO, transverse energy leakage is negligible. 
The distributions of E$_{A}$+E$_{S}$ are fitted with the Crystal Ball (CB) 
function (as given in equation \ref{eqn:cb}) because of the presence of low 
energy tail arising from shower leakage. 
\begin{eqnarray}
f(x;\alpha, n, \bar{E}, \sigma) &=& N \times exp\left(\frac{-(x-\bar{E})^{2}}{2\sigma^{2}}\right),
~for~\frac{x-\bar{E}}{\sigma}>-\alpha\nonumber\\
&=& N \times A \times \left(B-\frac{x-\bar{E}}{\sigma}\right)^{-n},~for~\frac{x-\bar{E}}{\sigma}<-\alpha 
\label{eqn:cb}
\end{eqnarray}
where $N$ is the normalization factor and $\alpha$, $n$, $\bar{E}$, $\sigma$
are parameters of the fit. $A$ is given as:
\begin{equation}
A = \left(\frac{n}{|\alpha|}\right)^{n} \times exp\left(-\frac{|\alpha|^{2}}{2}\right)
\end{equation}
and $B$ is given as:
\begin{equation}
B = \frac{n}{|\alpha|} - |\alpha|
\end{equation}

Figure \ref{fig:eTot_el10501003005001000gev} shows the CB fit to the 
distribution of E$_{A}$+E$_{S}$. The CB parametrization fits the distribution well. The fitted mean of the CB function is 
taken as  $\bar{E}$. To estimate the $\sigma_{68}$, 68$\%$ interval around $\bar{E}$,
is constructed using the parameters of the CB fit. The interval is formed 
in such a way that for each side of the mean, the area covered is 68$\%$  
of the area of that side. $\sigma_{68}$ is the half width of the interval as
obtained by the above construction. Table \ref{tab:sigma68_Ebar} shows the values of 
$\sigma_{68}$ and $\bar{E}$ for different energies.
\begin{table}[htbp]
\begin{center}
\begin{tabular}{|c|c|c|}
\hline
Energy  & $\sigma_{68}$ & $\bar{E}$           \\ 
(GeV) & (GeV) & (GeV) \\ \hline
10   & $ 0.23  \pm  0.03 $  & $  9.52   \pm  0.01 $     \\ \hline
20   & $ 0.29  \pm  0.05 $  & $  19.28  \pm  0.01 $     \\ \hline
30   & $ 0.29  \pm  0.02 $  & $  29.14  \pm  0.02 $     \\ \hline
50   & $ 0.35  \pm  0.02 $  & $  48.81  \pm  0.03 $     \\ \hline
70   & $ 0.40  \pm  0.03 $  & $  68.47  \pm  0.03 $     \\ \hline
100  & $ 0.49  \pm  0.03 $  & $  98.04  \pm  0.04 $     \\ \hline
200  & $ 0.66  \pm  0.06 $  & $  196.22 \pm  0.05 $     \\ \hline
300  & $ 1.00  \pm  0.09 $  & $  294.44 \pm  0.10 $     \\ \hline
500  & $ 2.17  \pm  0.32 $  & $  490.95 \pm  0.12 $     \\ \hline
1000 & $ 5.53  \pm  0.81 $  & $  980.97 \pm  0.32 $     \\ \hline
\end{tabular}
\end{center}
\caption{The values of $\sigma_{68}$ and $\bar{E}$ of the distribution of E$_{A}$+E$_{S}$ for electrons with different energies.} 
\label{tab:sigma68_Ebar}
\end{table}
The values of $\sigma_{68}/\bar{E}$ are then plotted as a function of 
energy. Fluctuations due to energy leakage are parametrized using 
Grindhammer-Peter's parametrization \cite{bib:fastsim:grindh}: 
\begin{equation}
\sigma_{leakage}(\ln (E/E_{c})) = (s1 + s2 \times \ln (E/E_{c}))^{-1}, 
\label{eqn:resoleakagegp}
\end{equation}
where $E_{c}$ is the critical energy and is dependent on the 
material of the detector. 
The above equation is expanded up to the third power in $\ln (E/E_{c})$ 
and the resolution due to leakage is fitted with a function of type $p_{0} 
+ p_{1} \times \ln E + p_{2} \times (\ln E)^{2} + p_{3} \times (\ln E)^{3}$. 
Fitted values of $p_{0}$, $p_{1}$, $p_{2}$ and $p_{3}$ are shown in 
Table \ref{tab:fitleak}.
\begin{table}[htbp]
\begin{center}
\begin{tabular}{|c|c|}
\hline
parameter & Fitted value           \\ \hline
$p_{0}$ & $0.118 \pm 0.010$      \\ \hline
$p_{1}$ & $-0.058 \pm 0.007$   \\ \hline
$p_{2}$ & $(9.8 \pm 1.6)\times10^{-3}$      \\ \hline
$p_{3}$ & $(-5.3 \pm 1.2)\times10^{-4}$   \\ \hline
\end{tabular}
\end{center}
\caption{Fitted values of 
          the parameters
          when $\sigma_{68}/\bar{E}$
         due to leakage is fitted with function $p_{0} + p_{1} \times \ln E + 
         p_{2} \times (\ln E)^{2} + p_{3} \times (\ln E)^{3}$. }
\label{tab:fitleak}
\end{table}
\begin{figure}[htbp]
  \begin{center} \begin{tabular}{cc}
\includegraphics[width=0.45\textwidth]{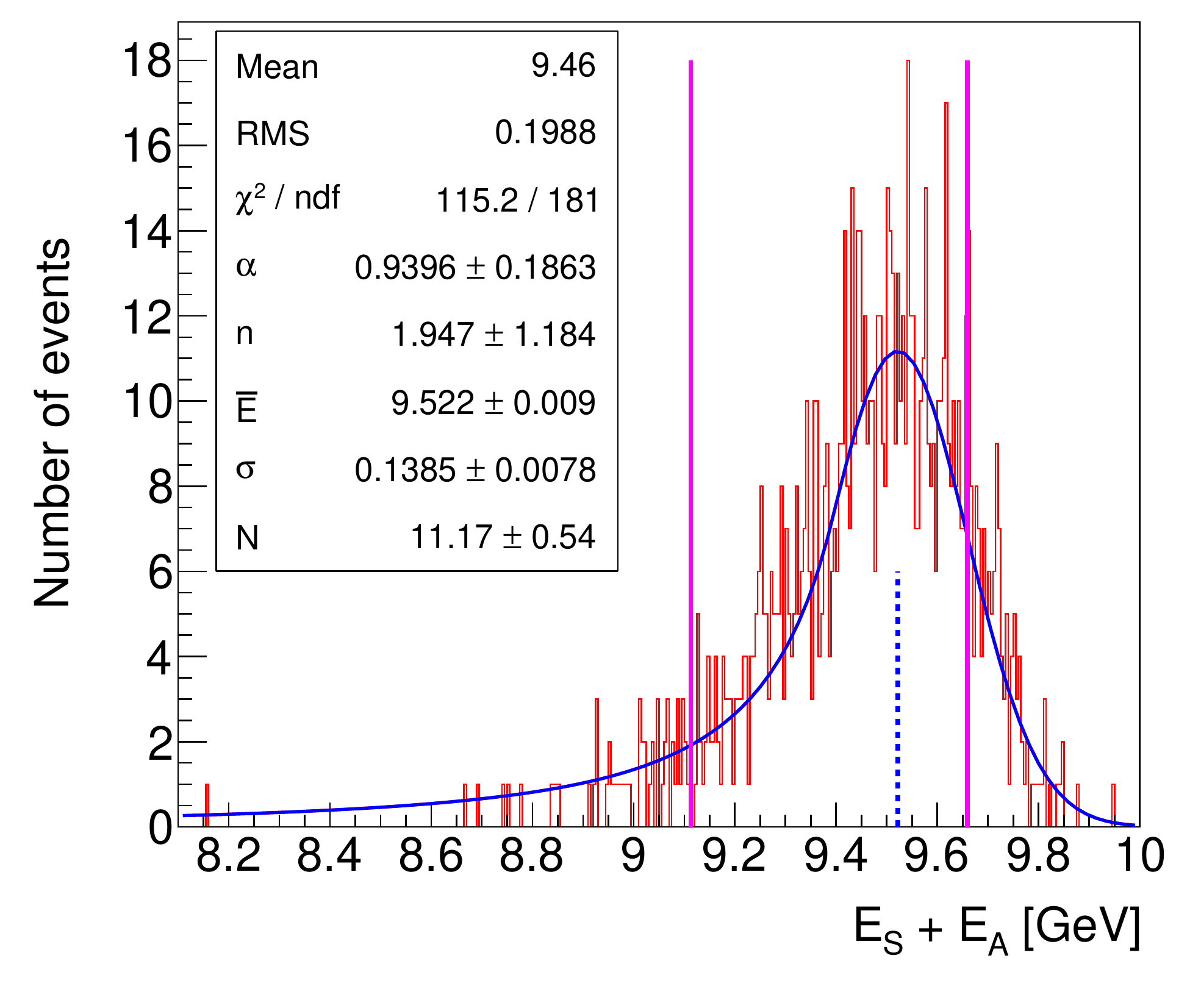}&
\includegraphics[width=0.45\textwidth]{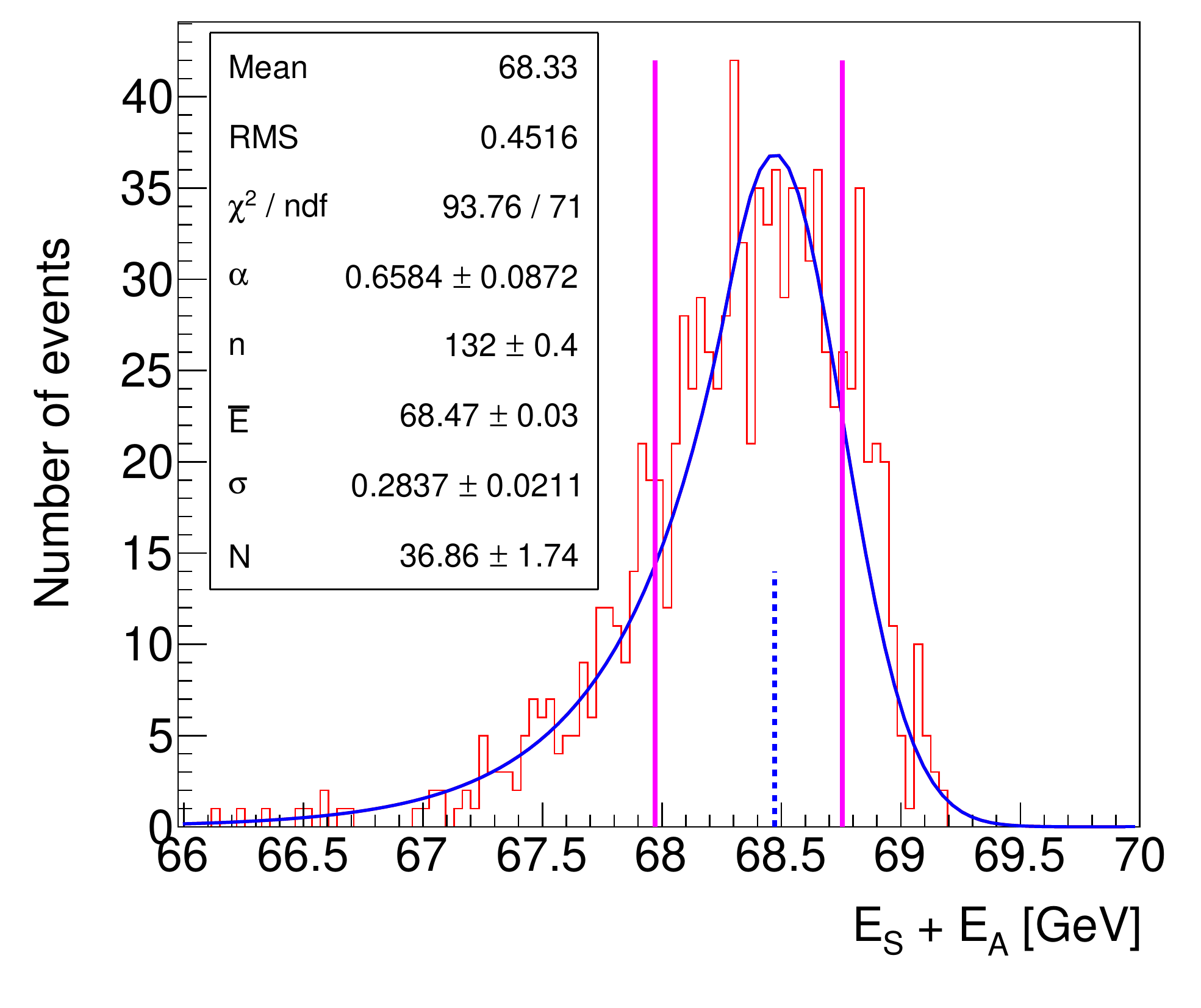}\\
\includegraphics[width=0.45\textwidth]{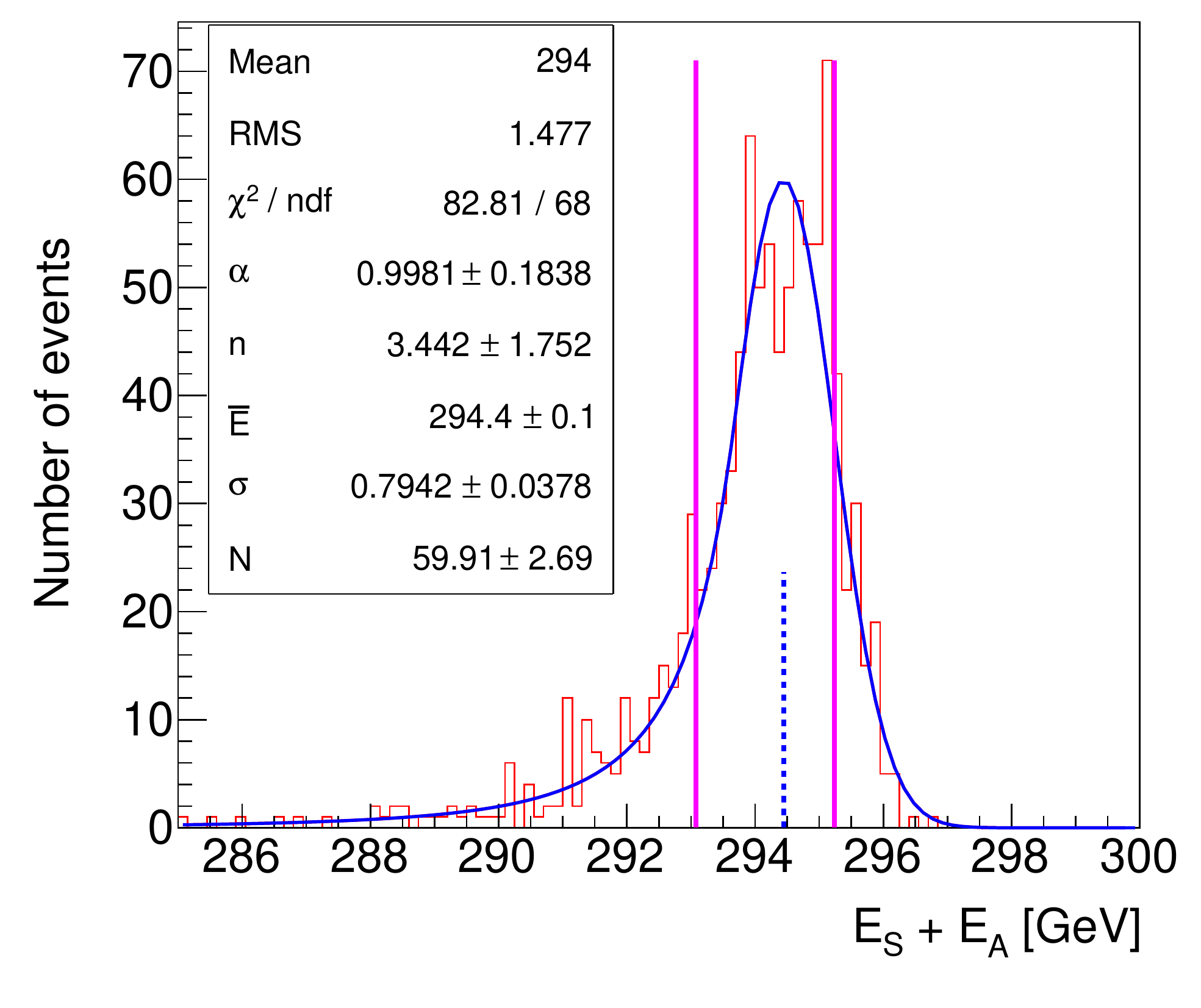}&
\includegraphics[width=0.45\textwidth]{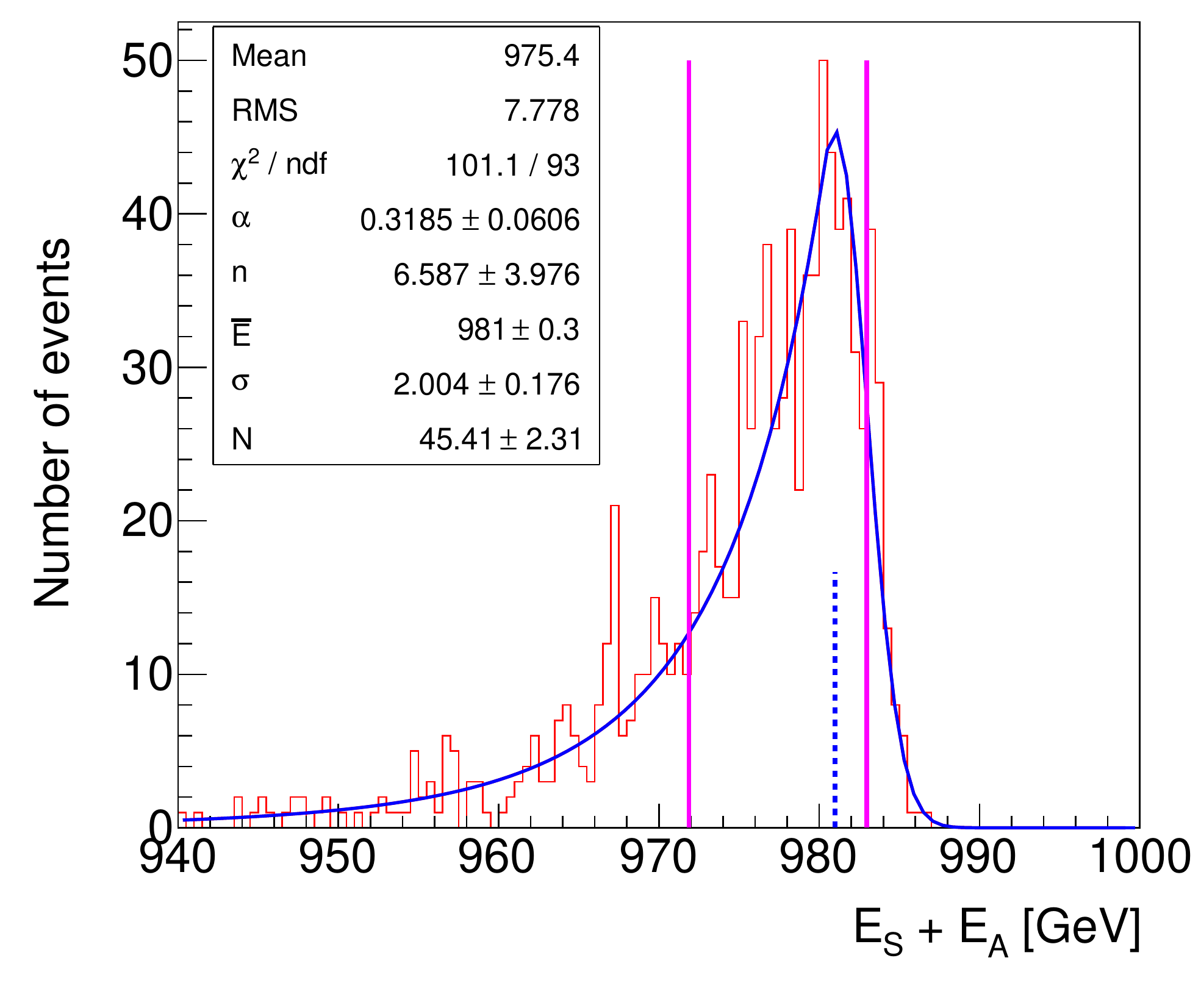}
  \end{tabular}\end{center}  
\caption{The distribution of E$_{S}$+E$_{A}$ for 10 GeV electrons on the top-left; 
         70 GeV electrons on top-right; 300 GeV electrons on bottom-left and 
         1000 GeV electrons on bottom-right. 
         The blue curves show the Crystal Ball fit to the distributions. 
         The pink lines show the $\sigma_{68}$ which is the 68$\%$ interval 
         constructed as discussed in the text and the
         blue dotted lines show the fitted Crystal Ball mean.}
\label{fig:eTot_el10501003005001000gev}
\end{figure}
\item[Sampling fluctuation:] It is estimated for each energy point exactly 
the same way as described in the previous sub-section. This is done for a 28 
layer configuration. The distribution of  $\sigma/\bar{E}$ is plotted as a function of 
energy and fitted with a function of type $p_{0}/\sqrt{E}$.
Fitted value of $p_{0}$ comes out to be 0.104 $\pm$ 0.001 .
\item[Photo-statistics:] To estimate the contribution due to photo-statistics, the energy collected in 
all the scintillator  layers is converted to the number of photo-electrons 
(p.e.) in the photo-detector. 
The total number of p.e. from the scintillator is
\begin{equation}
\label{eqn:EtoPEconv}
N_{pe} = \Sigma E_{i}~\times ~LE ~\times ~LY
\end{equation}
where $E_{i}$ is the energy deposit in i'th layer, $LE$ is the light collection efficiency 
and $LY$ is the light yield. The light yield is different for different materials~\cite{bib:crystal_calo}. 
The photo-statistics contribution can vary depending on the light yield of the material. 
Average light collection efficiency (LE) is taken to be 0.5\% [14] for each layer. 
(Only a small fraction of scintillation photons is collected by the fibers which go through 
the holes of each layer and thus leads to such a low efficiency). 
The distribution of p.e. follows Poisson distribution and hence the fluctuation has a $\sqrt{LY}$ dependence. 
The distributions of $\sigma/\bar{E}$ in Figure \ref{fig:photostat_LY} are fitted with a function of type $p_0/\sqrt{LY}$. 
The fit yields $p_0$ to be 0.1068 $\pm$ 0.0001 for 100 GeV electron beam energy. 
Figure \ref{fig:photostat_E} shows the distribution of $\sigma/\bar{E}$ as a function of incident energy for different values of $LY$. 
Fits of these distributions to functions of the the type $p_0/\sqrt{E}$ yield $p_0$ to be 0.0171 $\pm$ 0.0001 for light yield value of 4000 p.e./MeV.
\begin{figure}[h!]
\begin{minipage}{.48\textwidth}
  \centering
  \includegraphics[width=.99\linewidth]{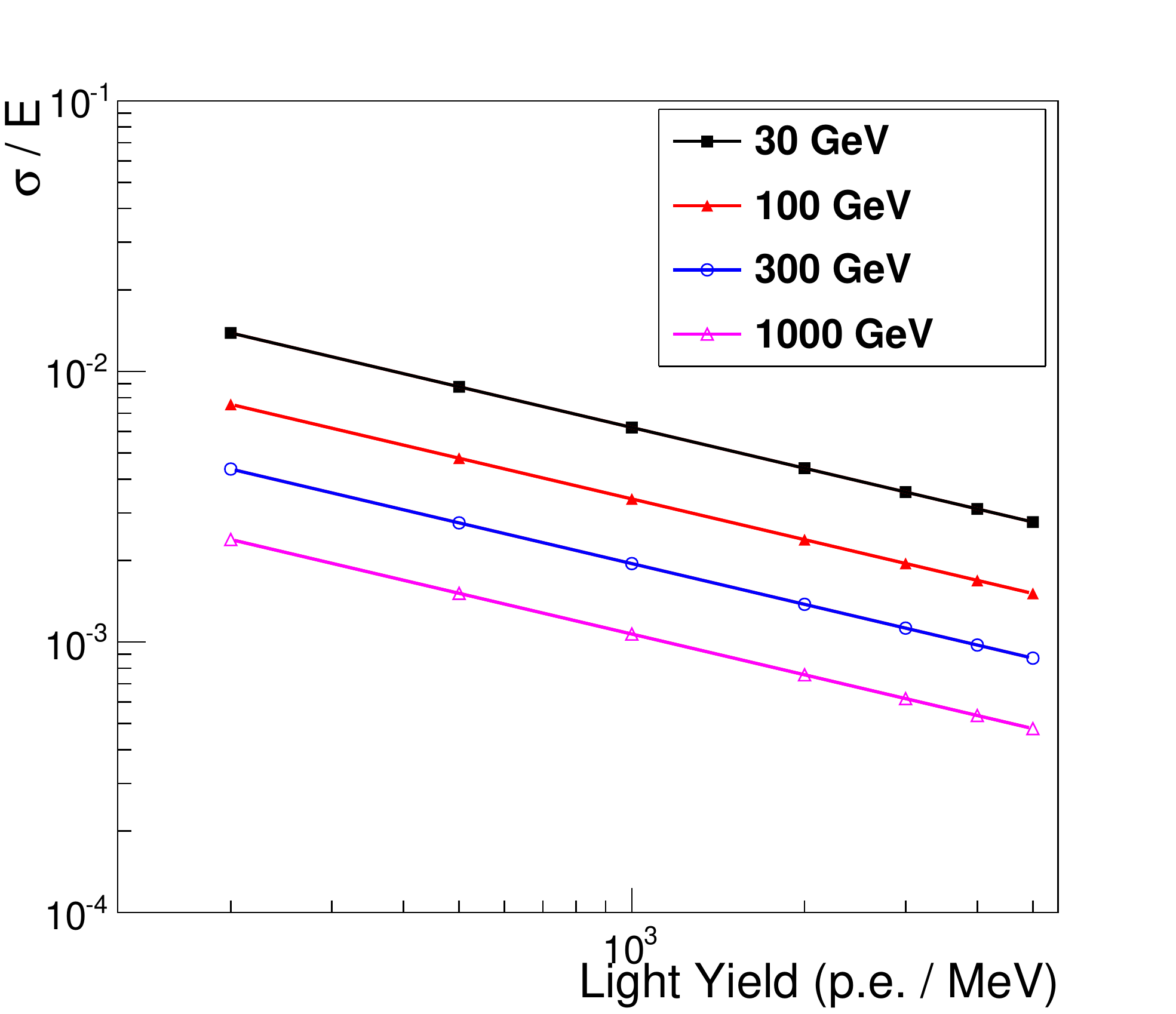}
  \caption{
   Photo-statistics contribution as a function of light yield (LY) for different beam energies}
  \label{fig:photostat_LY}
\end{minipage}\hfill
\begin{minipage}{.48\textwidth}
  \centering
  \includegraphics[width=.99\linewidth]{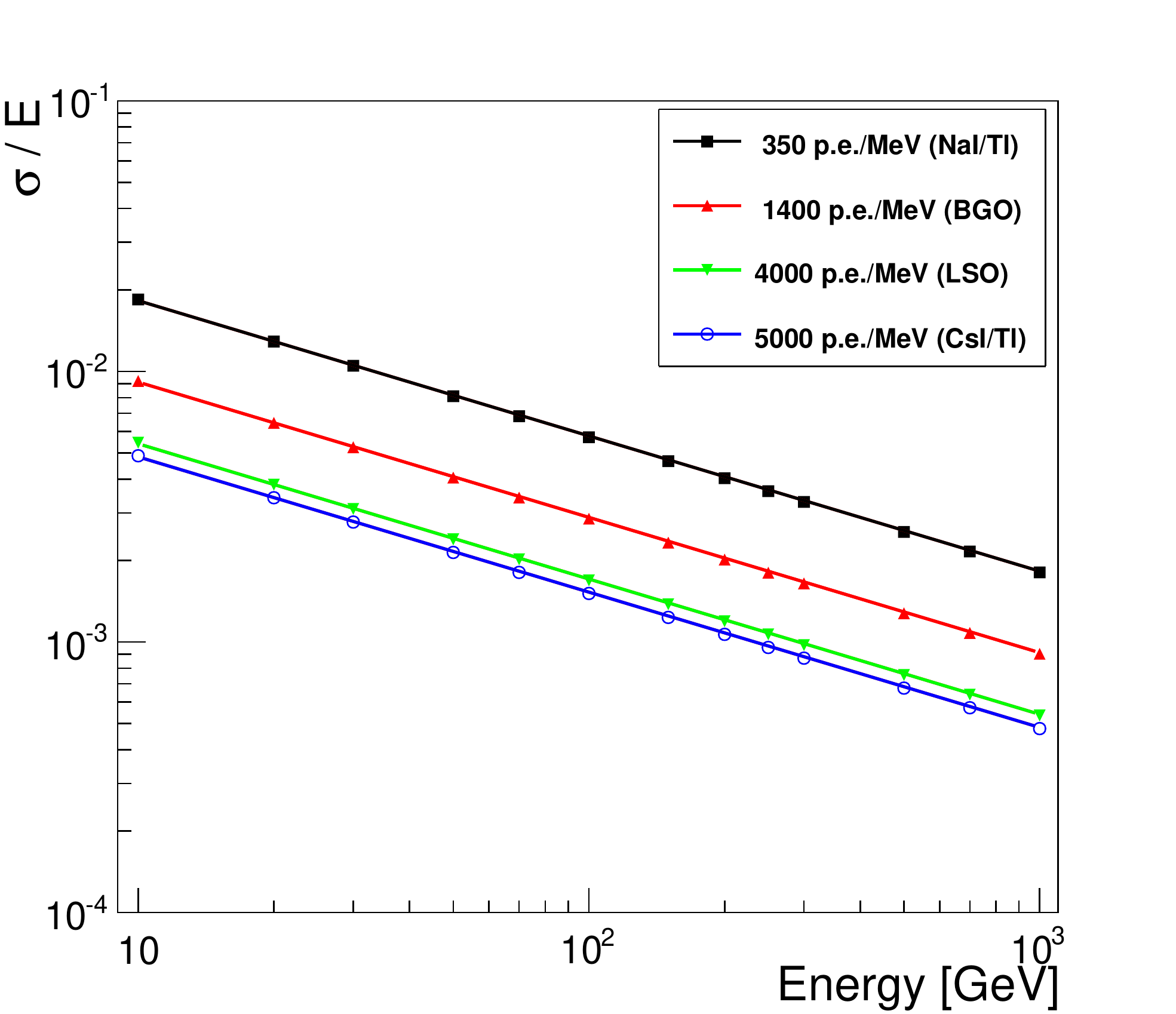}
  \caption{
   Photo-statistics contribution as a function of energy for different light yields. 
  }
  \label{fig:photostat_E}
\end{minipage}
\end{figure}
\item[Electronic noise:] 
 Contribution of electronic noise in the photo-detectors to the energy resolution depends on the fluctuation 
in the number of photo electrons contributing to the noise. It varies depending on the read-out scheme. 
Figure \ref{fig:noise_pe} shows the noise distribution as a function of mean number of p.e. corresponding to electronic noise 
for different beam energies. 
Fits to the energy resolution $\sigma/\bar{E}$ to a function of the type $p_0 \times p.e.$ yield a value 
of $p_0$ to be $(2.8 \pm 0.1) \times 10^{-6}$ for 100 GeV electron beam energy.
\begin{figure}[h!]
\begin{minipage}{.48\textwidth}
  \centering
  \includegraphics[width=.99\linewidth]{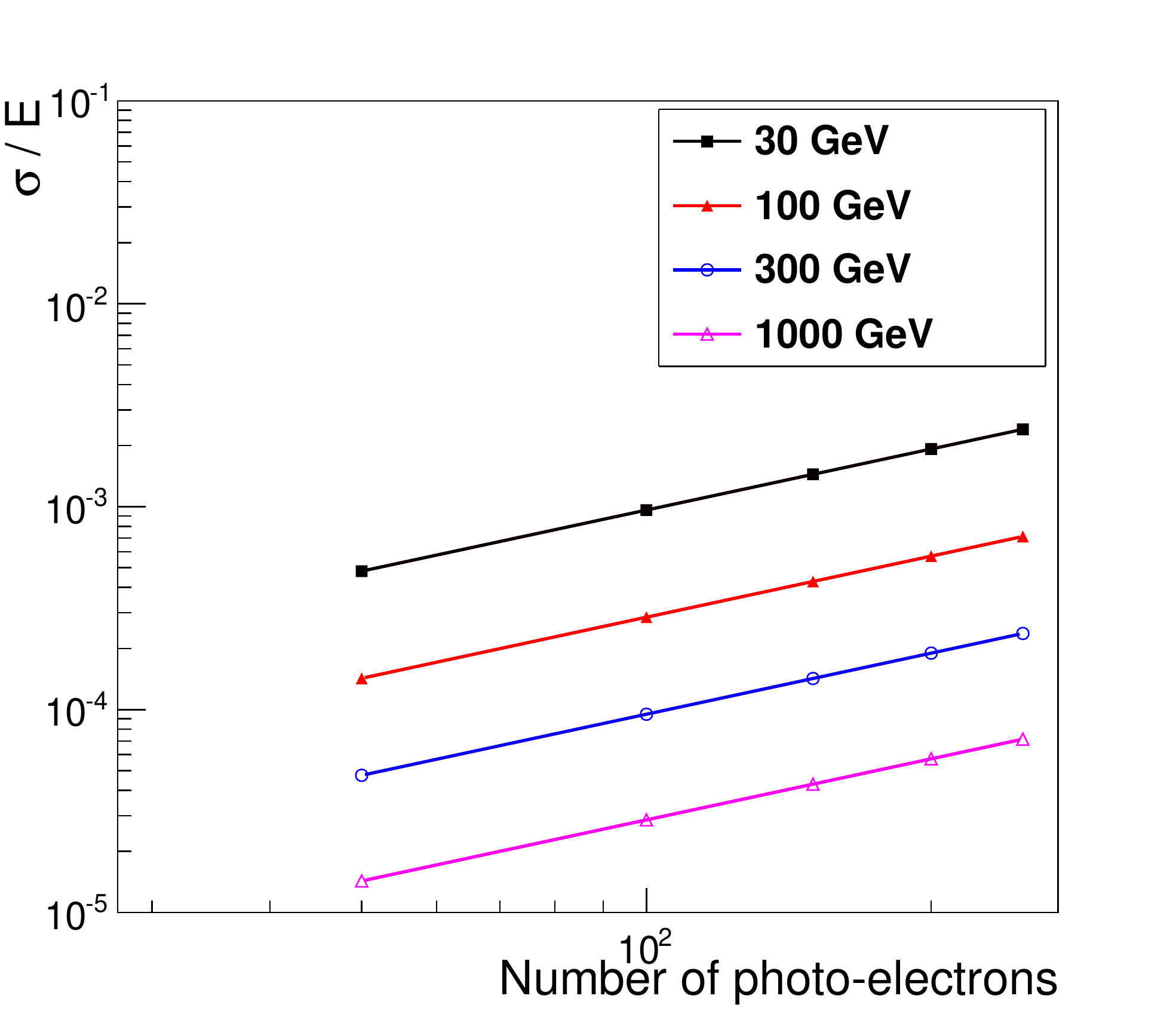}
  \caption{Noise contribution as a function of photo-electrons for different beam energies.}
  \label{fig:noise_pe}
\end{minipage}\hfill
\begin{minipage}{.48\textwidth}
  \centering
  \includegraphics[width=.99\linewidth]{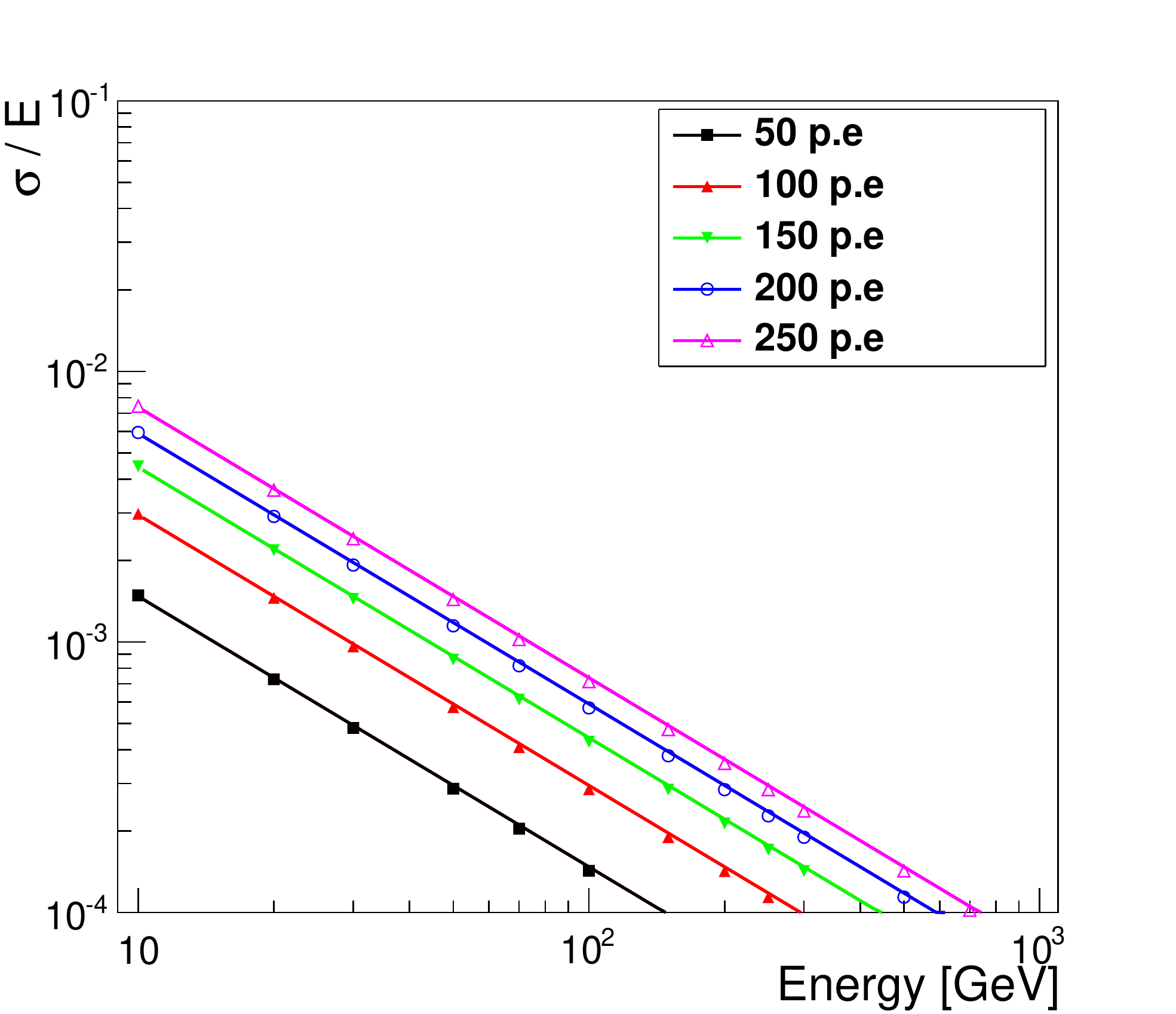}
  \caption{Noise contribution as a function of energy for different number of photo-electrons.}
  \label{fig:noise_E}
\end{minipage}
\end{figure}
Figure \ref{fig:noise_E} shows the distributions of  $\sigma/\bar{E}$ as 
a function of energy and are fitted with functions of the type $p_{0}/E$ for for different numbers of mean photo-electrons.
Fitted value of $p_{0}$ comes out to be 
0.0442 $\pm$ 0.0002 for 150 as mean number of photo-electrons.
\item[Total energy resolution:] Total energy resolution is the sum (in 
quadrature) of the above four terms. 
The distribution of $\sigma/\bar{E}$ is plotted as a function of beam energy in Figure \ref{fig:totreso} 
for light yield value of  4000 p.e./MeV and mean noise of 150 p.e. 
The energy resolution is fitted with a function of the type \\

$\sqrt{\left(p_{0}/\sqrt{E}\right)^{2} + \left(p_{1}/E\right)^{2} + \left(p_{2} + p_{3} \times \ln E + p_{4} \times (\ln E)^{2} + p_{5} \times (\ln E)^{3} \right)^{2}}$.

and the fitted parameters are shown in Table \ref{tab:fittot}. 
Figure \ref{fig:totreso} also shows contributions of each of the term as given above.
It can be seen that the parameters of the total fit are in agreement with the individual 
parameters obtained by fitting all the terms ({\it i.e.} leakage, sampling, 
statistics and noise) of the resolution individually. The parameter, $p_{0}$ 
has contribution from both the terms, sampling and statistics. But the major 
contribution comes from sampling and hence the fitted parameter $p_{0}$ is 
closer to the fitted value of the sampling term. 
\begin{table}[htbp]
\begin{center}
\begin{tabular}{|c|c|}
\hline
parameter & Fitted value            \\ \hline
$p_{0}$ & 0.103 $\pm$ 0.006         \\ \hline
$p_{1}$ & 0.087 $\pm$ 0.056         \\ \hline
$p_{2}$ & 0.118 $\pm$ 0.002         \\ \hline
$p_{3}$ & -0.058 $\pm$ 0.001       \\ \hline
$p_{4}$ & $(9.8 \pm 0.1)~\times 10^{-3}$    \\ \hline
$p_{5}$ & $(-5.3 \pm 0.1)~\times 10^{-4}$     \\ \hline
\end{tabular}
\end{center}
\caption{Table showing fitted values of the parameters when $\sigma/\bar{E}$
         of total energy resolution is is fitted with function $ \sqrt{ \left(
         p_{0}/\sqrt{E}\right)^{2} + \left(p_{1}/E\right)^{2} + \left(p_{2} 
         + p_{3} \times \ln E + p_{4} \times (\ln E)^{2} + p_{5} \times 
         (\ln E)^{3}\right)^{2}}$.}
\label{tab:fittot}
\end{table}
\begin{figure}[htbp]
\centering
\includegraphics[width=0.6\textwidth]{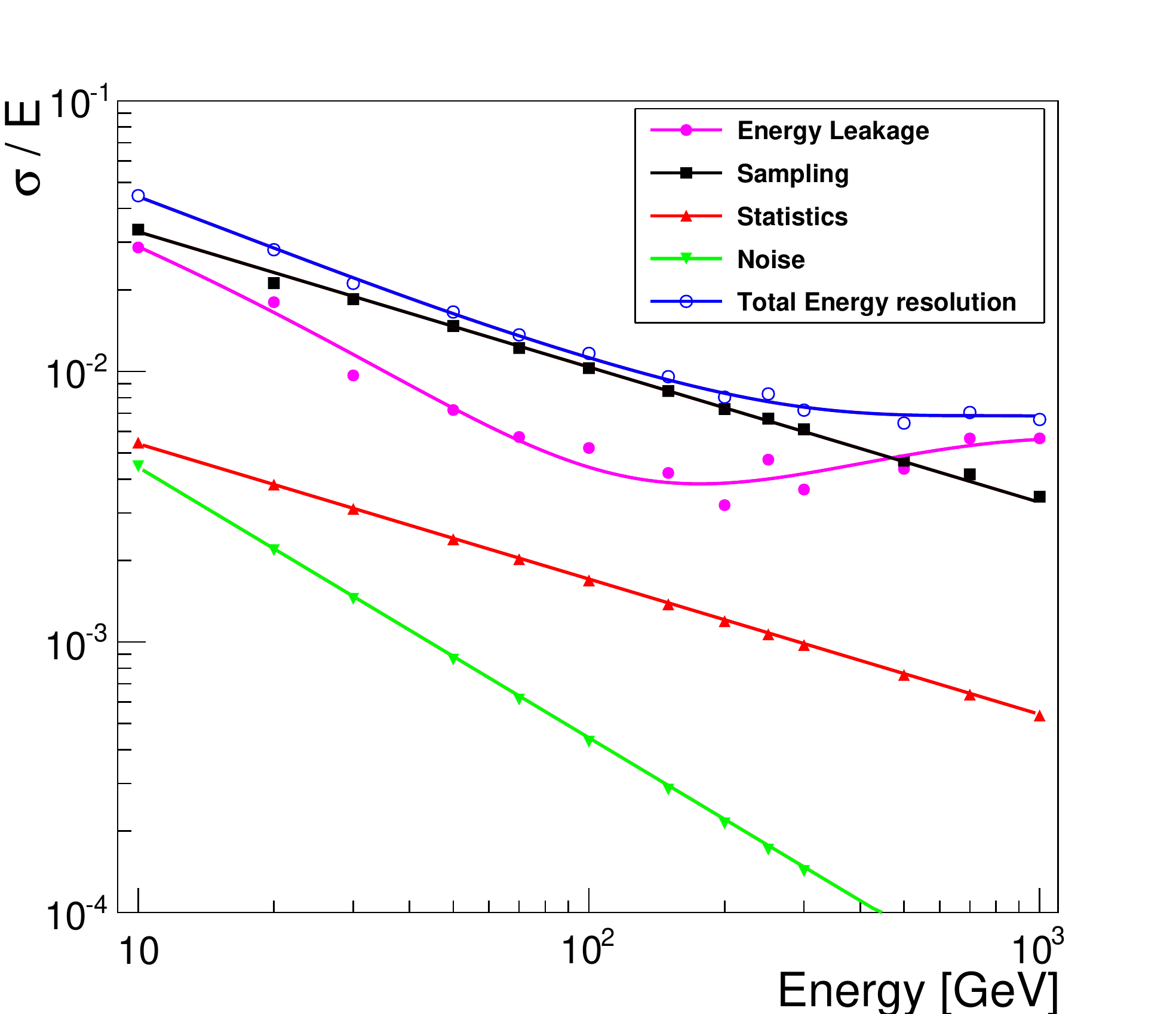}
\caption{Contributions of each term to the total resolution. Points are the 
         actual points as a function of energy and lines are the fitted 
         functions.}
\label{fig:totreso}
\end{figure}
\end{description}
Finally, to check the effect of this detector resolution on the mass resolution 
of the Higgs boson, 10000 events are generated where the Higgs boson is 
produced via gluon gluon fusion, and decays to a photon pair. Higgs mass is 
taken to be 125 GeV. In order to mimic the effect of the Shashlik detector, 
energy of each photon was smeared with the parameters as given 
in Table \ref{tab:fittot} in the following way:
\begin{enumerate}
\item a resolution term, $\sigma/$E is calculated for each of the 
photons using the parameters of Table \ref{tab:fittot} and the 
expression given in there;
\item a Gaussian random number is generated with mean 1 and 
      $\sigma_{gaus} = \sigma/$E;
\item energy and momentum of each photon is multiplied with the above 
      random number;
\item the diphoton mass is fitted with a Gaussian function. The $\sigma$ from
      the fit is taken as the estimate of resolution of Higgs mass. 
\end{enumerate}
%
%
\begin{figure}[htbp]
  \begin{center}
    \includegraphics[width=0.6\textwidth]{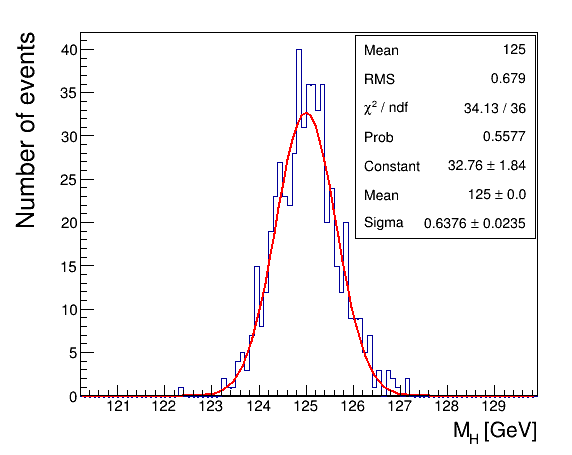}
  \end{center}
\caption{Resolution on Higgs mass with a Shashlik ECAL and with both the 
         photons in the endcap electromagnetic calorimeter.}
\label{fig:higgsreso}
\end{figure}
Figure \ref{fig:higgsreso} shows the Gaussian fit to Higgs mass,  
when both the photons are in the endcap electromagnetic calorimeter, obtained 
using the above procedure. 
The fitted $\sigma$ is estimated to be $0.64 \pm 0.02$. 
\section{Position resolution} \label{sec:position}
When a particle hits the electromagnetic calorimeter, its deposited energy gets 
distributed in the hit tower as well as the towers around it. The hit position 
can be estimated from the weighted mean of the position of the towers in which 
energy has been deposited, where the weights 
are proportional to the energy deposited in the towers
(so that higher the energy, more is the weight and hence more likely that the 
particle has hit that tower). This method of estimating the position 
is called the center of gravity (COG) method. The equations used to estimated 
coordinates in this way are given in Equation \ref{eqn:linearweight}. 
\begin{eqnarray}
x_{meas}  &= &\Sigma x_{i} \times E_{i}/\Sigma E_{i} \nonumber\\
y_{meas}  &= &\Sigma y_{i} \times E_{i}/\Sigma E_{i}
\label{eqn:linearweight}
\end{eqnarray}

\begin{figure}[htbp]
  \begin{center} \begin{tabular}{cc}
\includegraphics[width=0.45\textwidth]{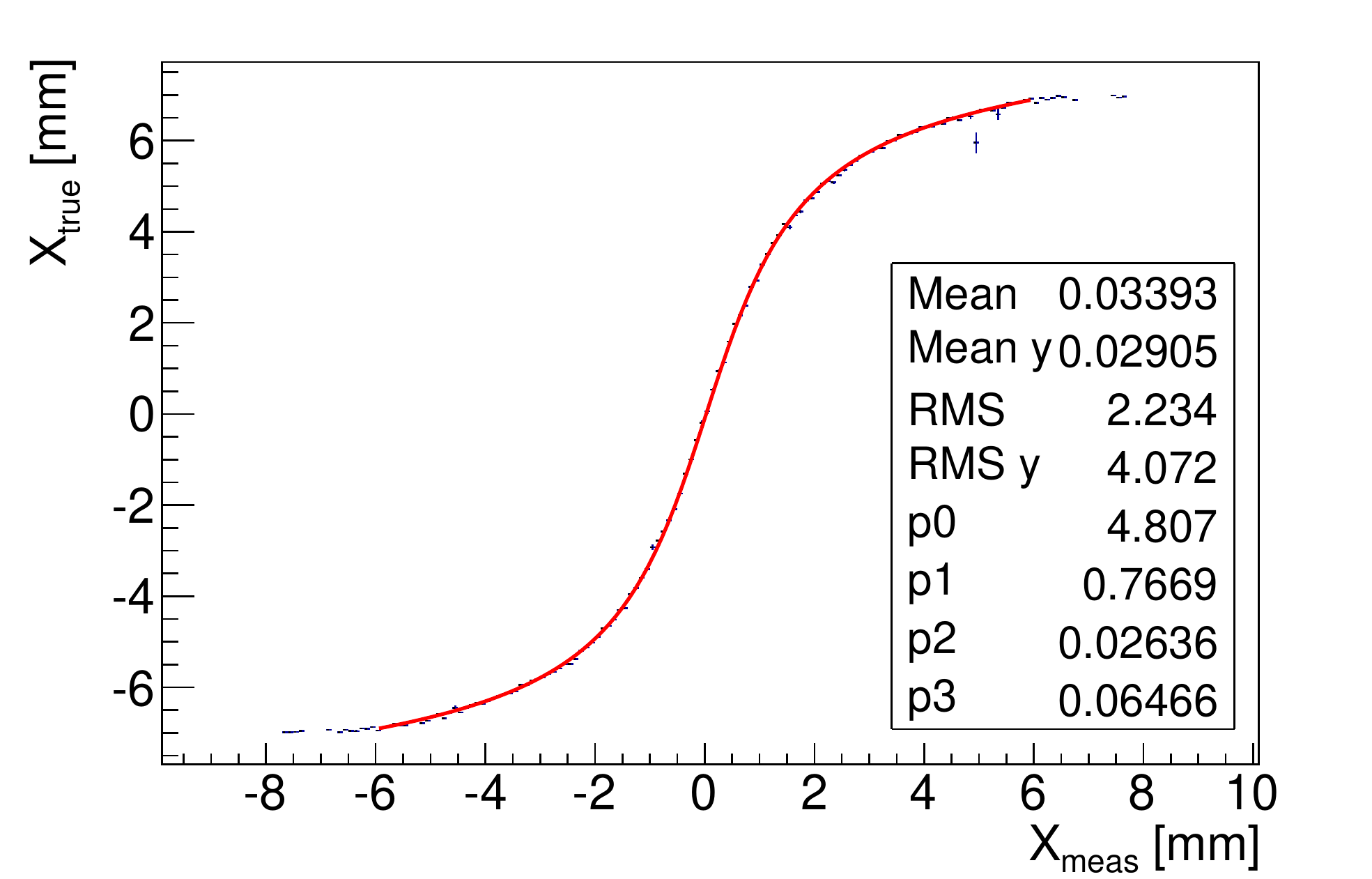}&
\includegraphics[width=0.45\textwidth]{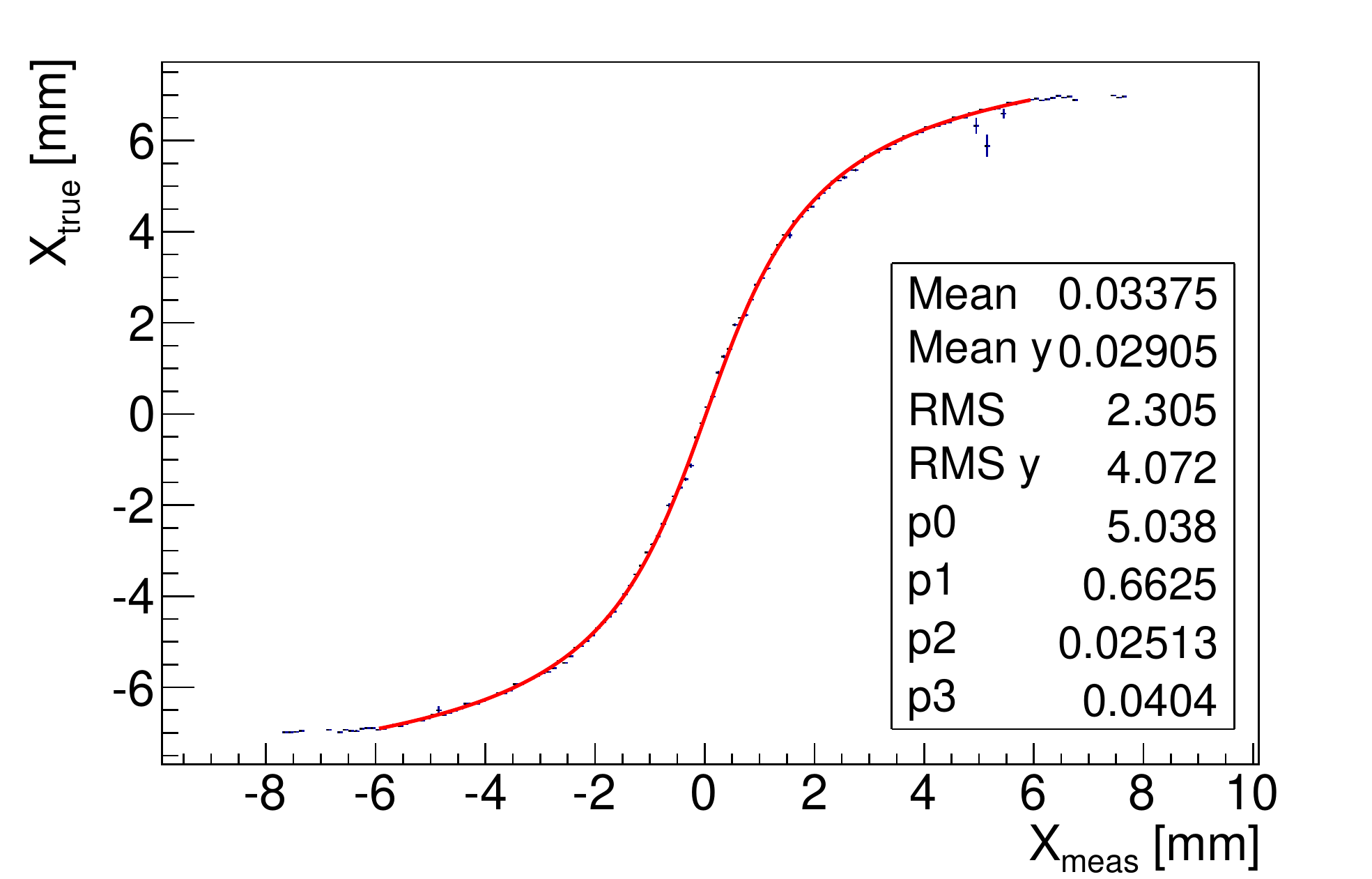}\\
\includegraphics[width=0.45\textwidth]{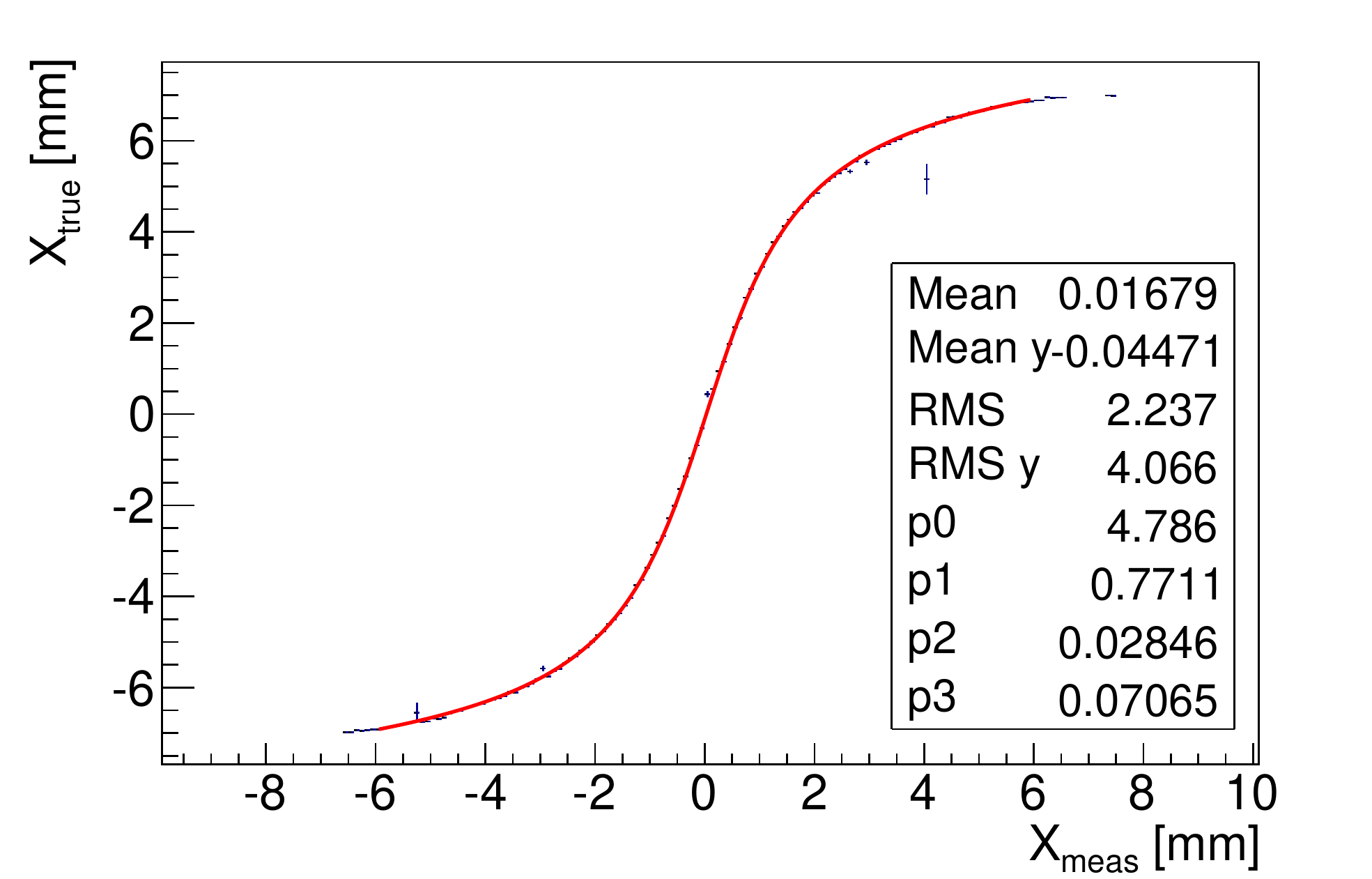}&
\includegraphics[width=0.45\textwidth]{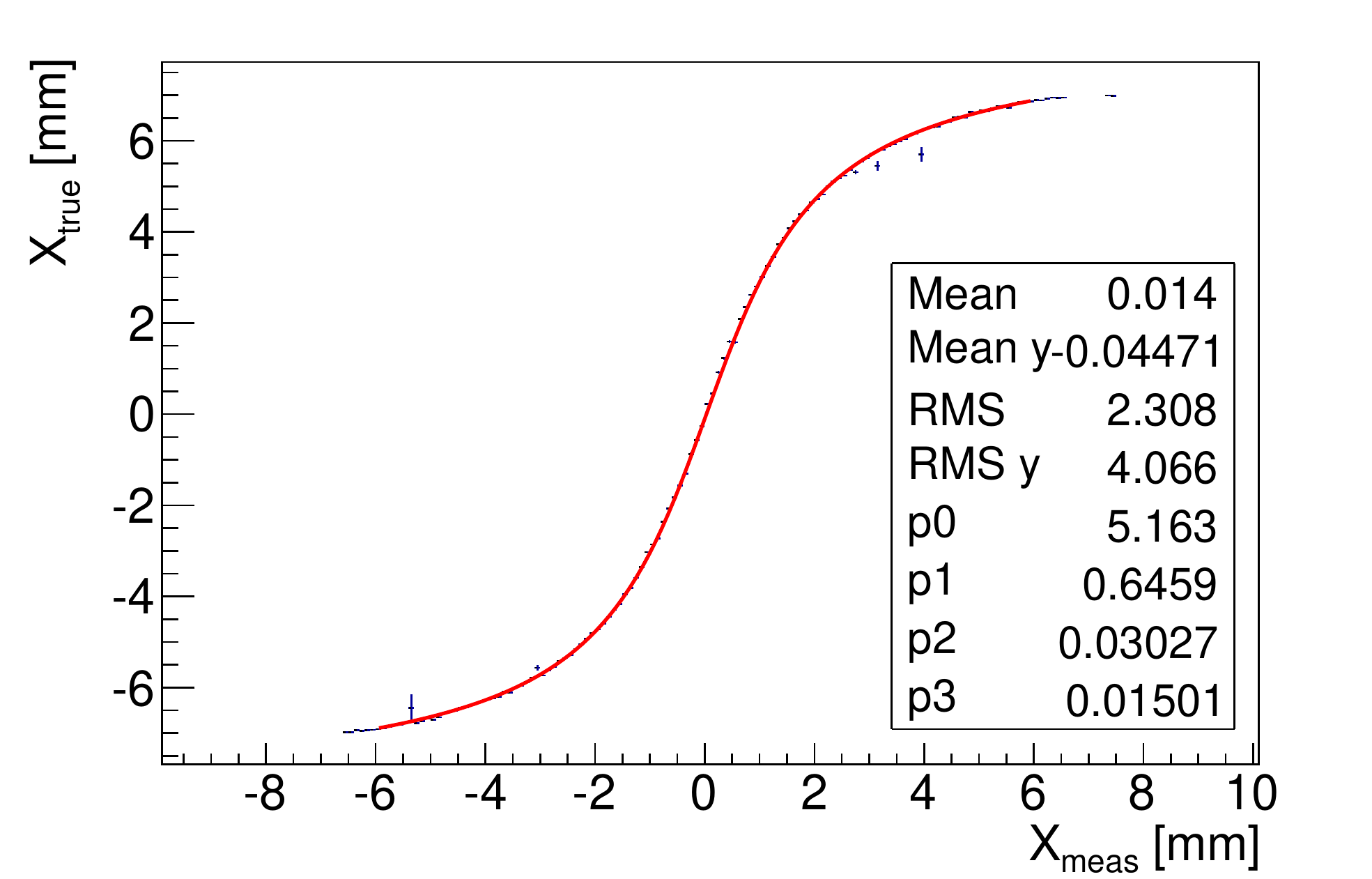}
  \end{tabular} \end{center}
\caption{2-D distribution of x$_{true}$ versus x$_{meas}$ for photons of energy 
         50 GeV (top plots) and 150 GeV (bottom plots) when linear weights are 
         used to estimate the COG. The relation between x$_{meas}$  and 
         x$_{true}$ is fitted with a S-shaped curve as parametrized in the
         Equation \ref{eqn:sshapefit} and shown by the red curve in the figure. 
         The left (right) figures refer to cases when combined (individual) 
         fiber information is used.}
\label{fig:xsshape_50150}
\end{figure}

\begin{figure}[htbp]
  \begin{center} \begin{tabular}{cc}
\includegraphics[width=0.45\textwidth]{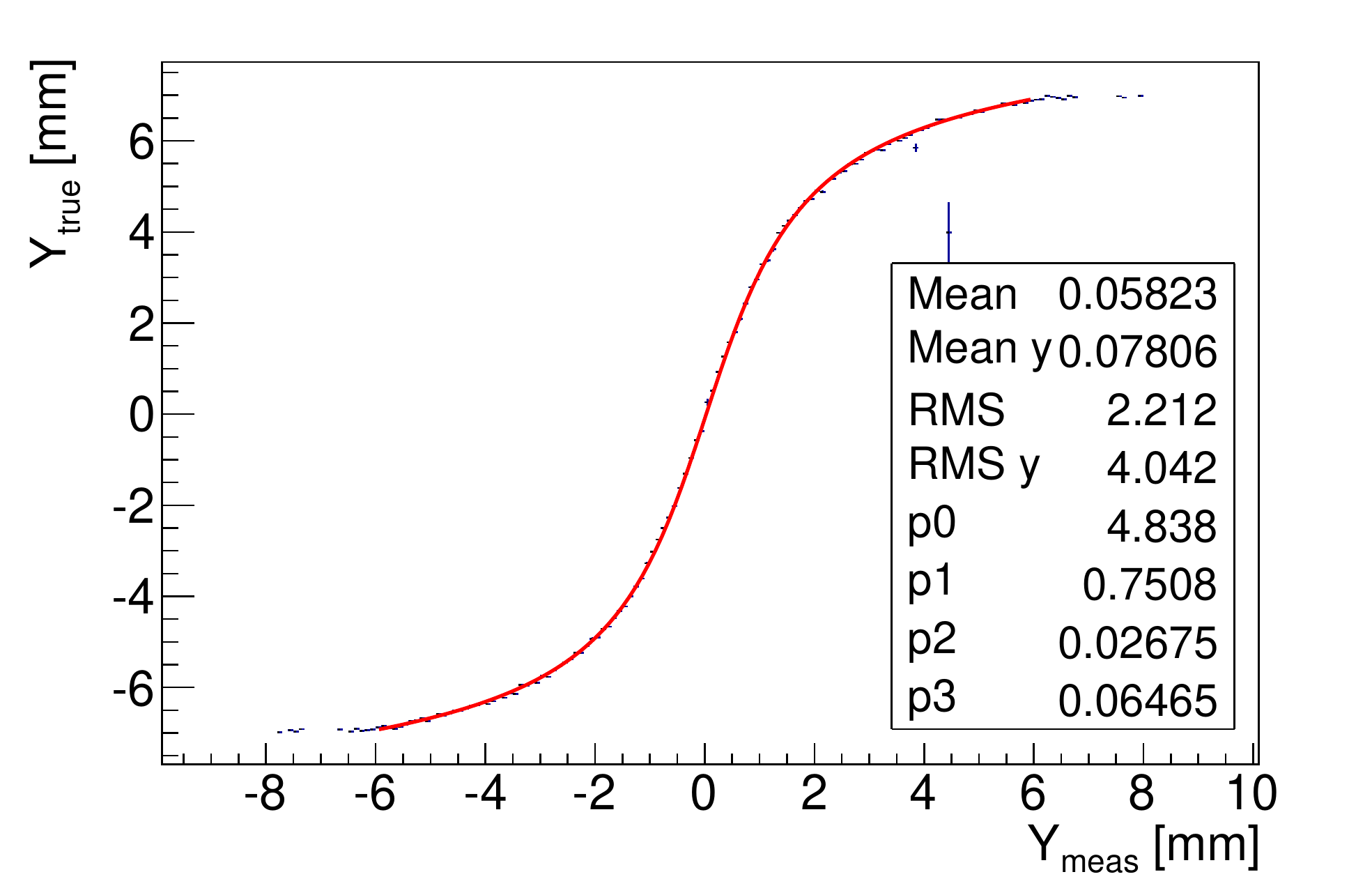}&
\includegraphics[width=0.45\textwidth]{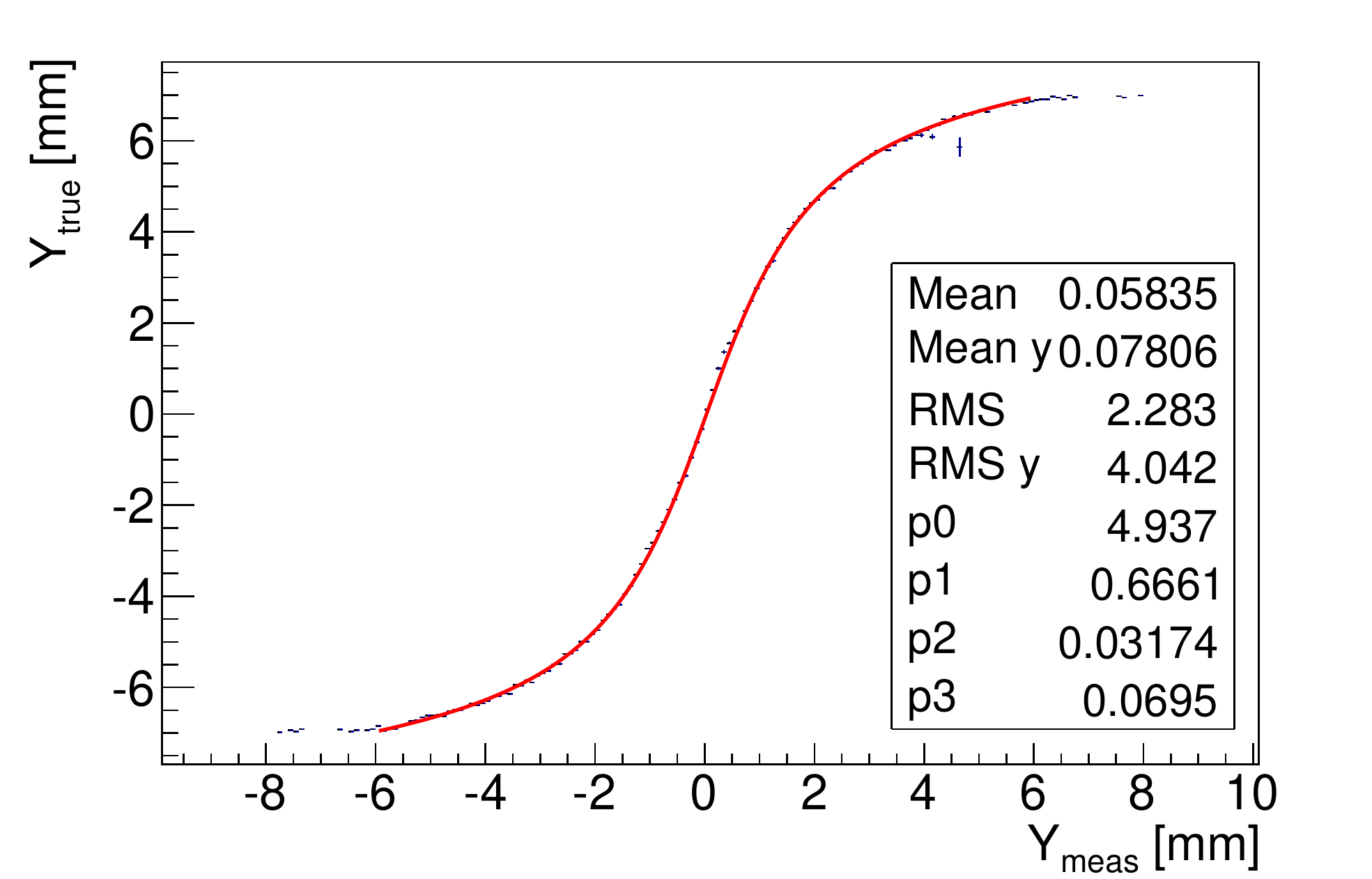}\\
\includegraphics[width=0.45\textwidth]{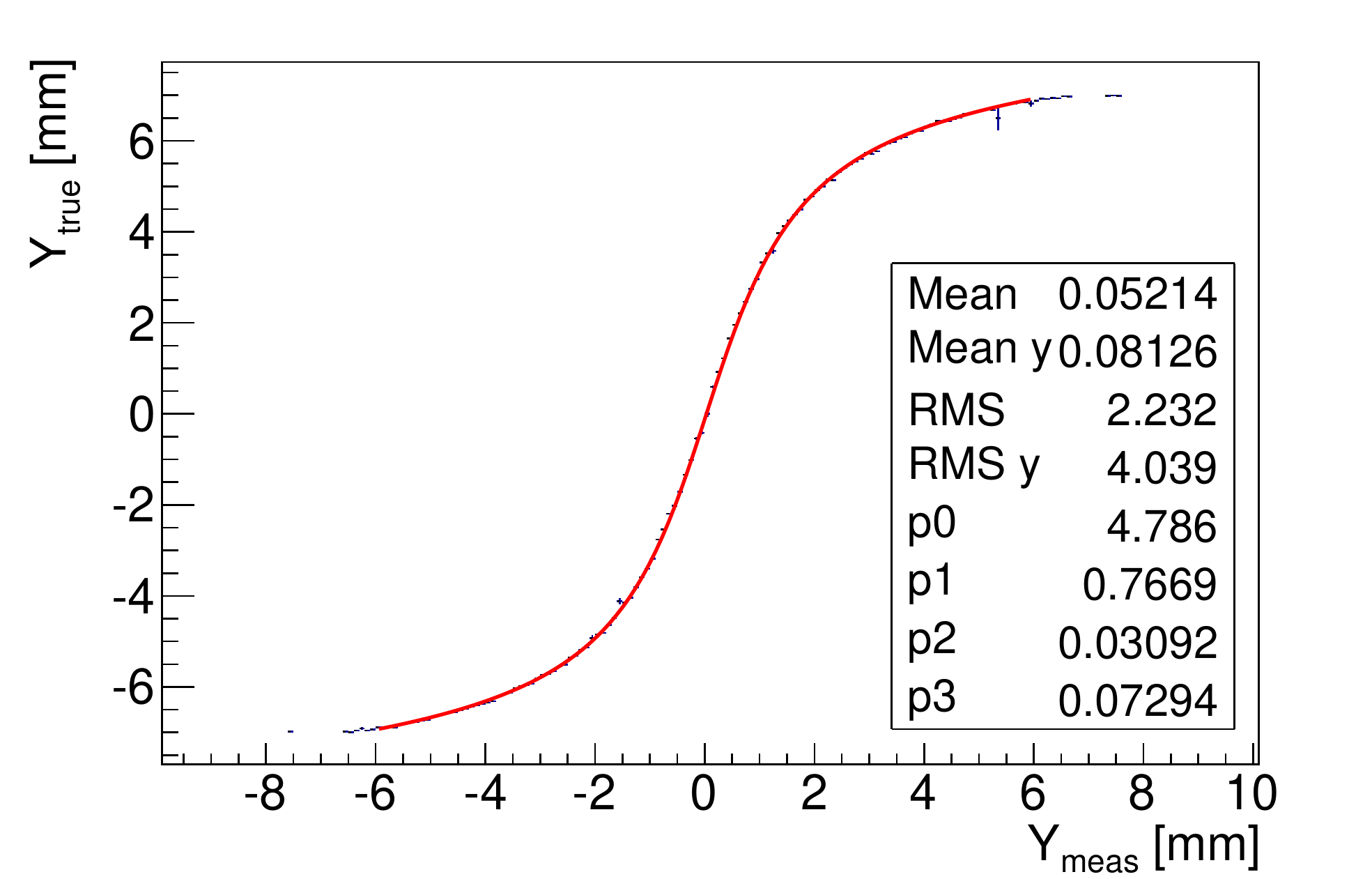}&
\includegraphics[width=0.45\textwidth]{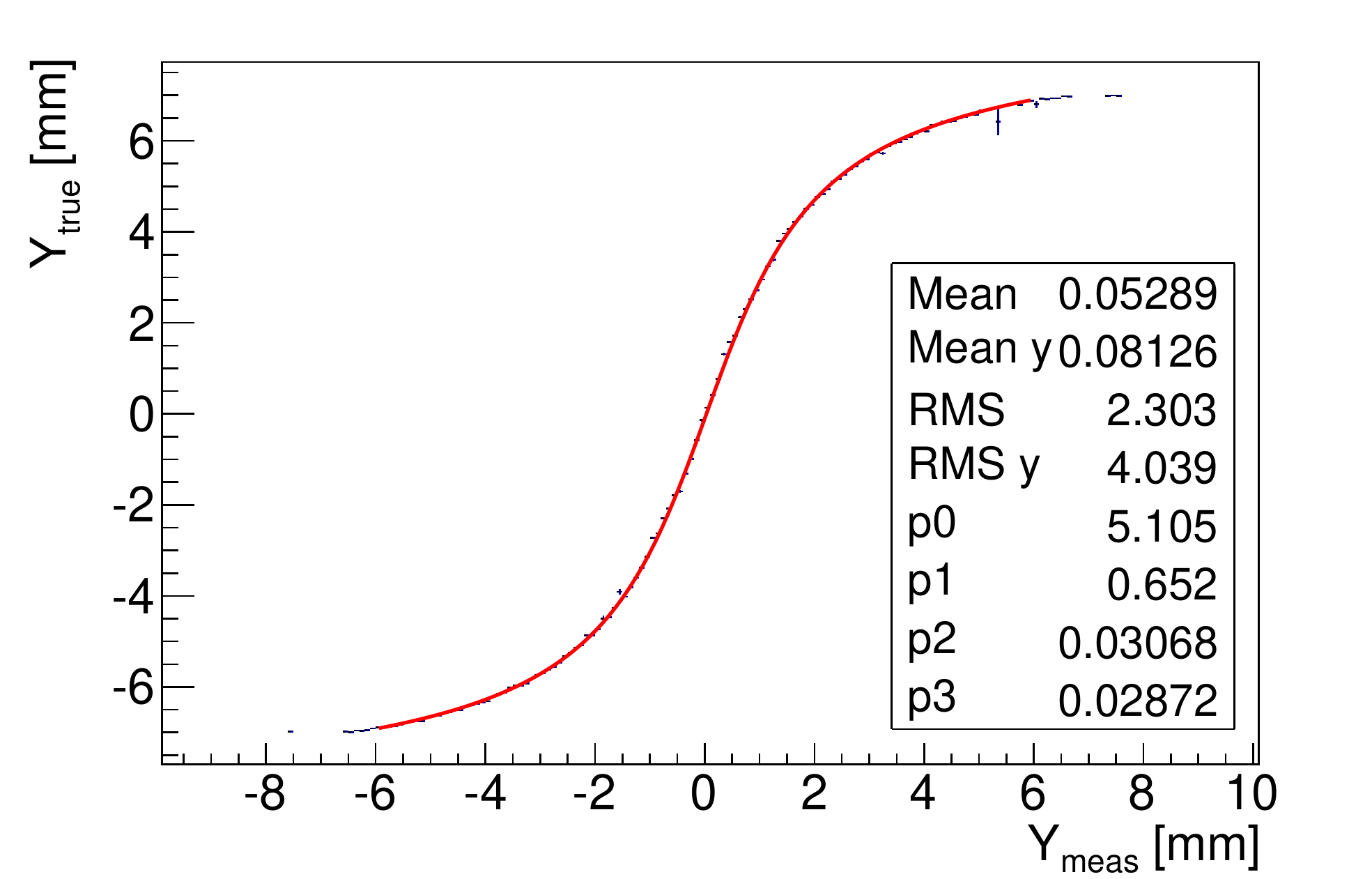}
  \end{tabular}\end{center}
\caption{2-D distribution of y$_{true}$ versus y$_{meas}$ for photons of energy 
         50 GeV (top plots) and 150 GeV (bottom plots) when linear weights are 
         used to estimate the COG. The relation between y$_{meas}$  and 
         y$_{true}$ is fitted with a S-shaped curve as parametrized in the
         Equation \ref{eqn:sshapefit} and shown by the red curve in the figure. 
         The left (right) figures refer to cases when combined (individual) 
         fiber information is used.}
\label{fig:ysshape_50150}
\end{figure}

Here, the sum is over the 3 $\times$ 3 array of towers, if a combined
signal from the four fibers are read out for each tower, or it is over
3 $\times$ 3 $\times$ 4 array of fibers, if individual fiber information is 
used. Figures \ref{fig:xsshape_50150} and \ref{fig:ysshape_50150} show the 
true impact point $x_{true}$ and $y_{true}$ as a function of the measured 
coordinates $x_{meas}$  or $y_{meas}$  for 50 GeV and 150 GeV photons.
The resulting distributions show deviations from linearity and roughly
follows an S-shape. This feature is observed for combined as well as 
individual fiber readouts.

\begin{figure}[htbp]
  \begin{center} \begin{tabular}{cc}
\includegraphics[width=0.45\textwidth]{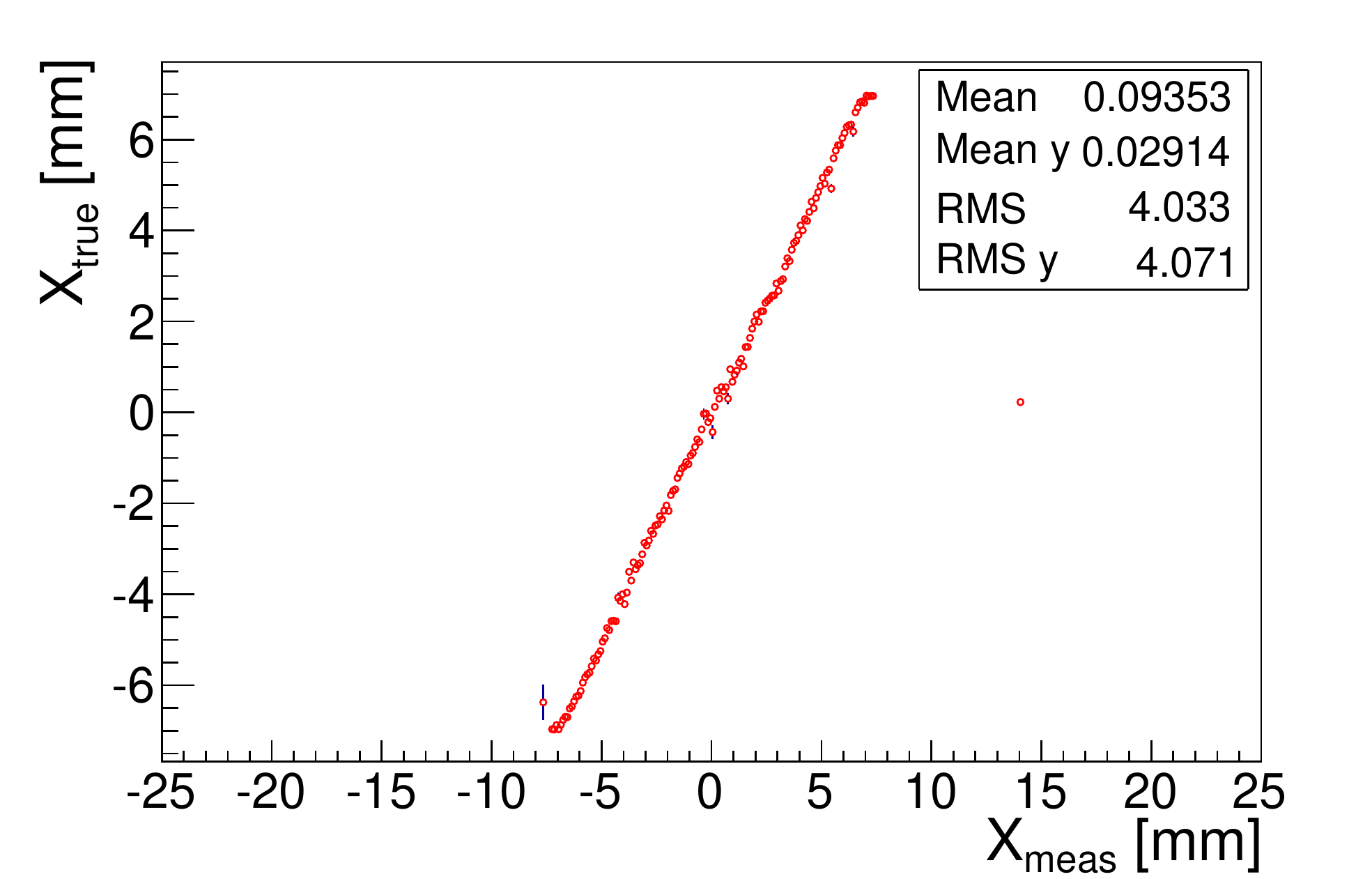}&
\includegraphics[width=0.45\textwidth]{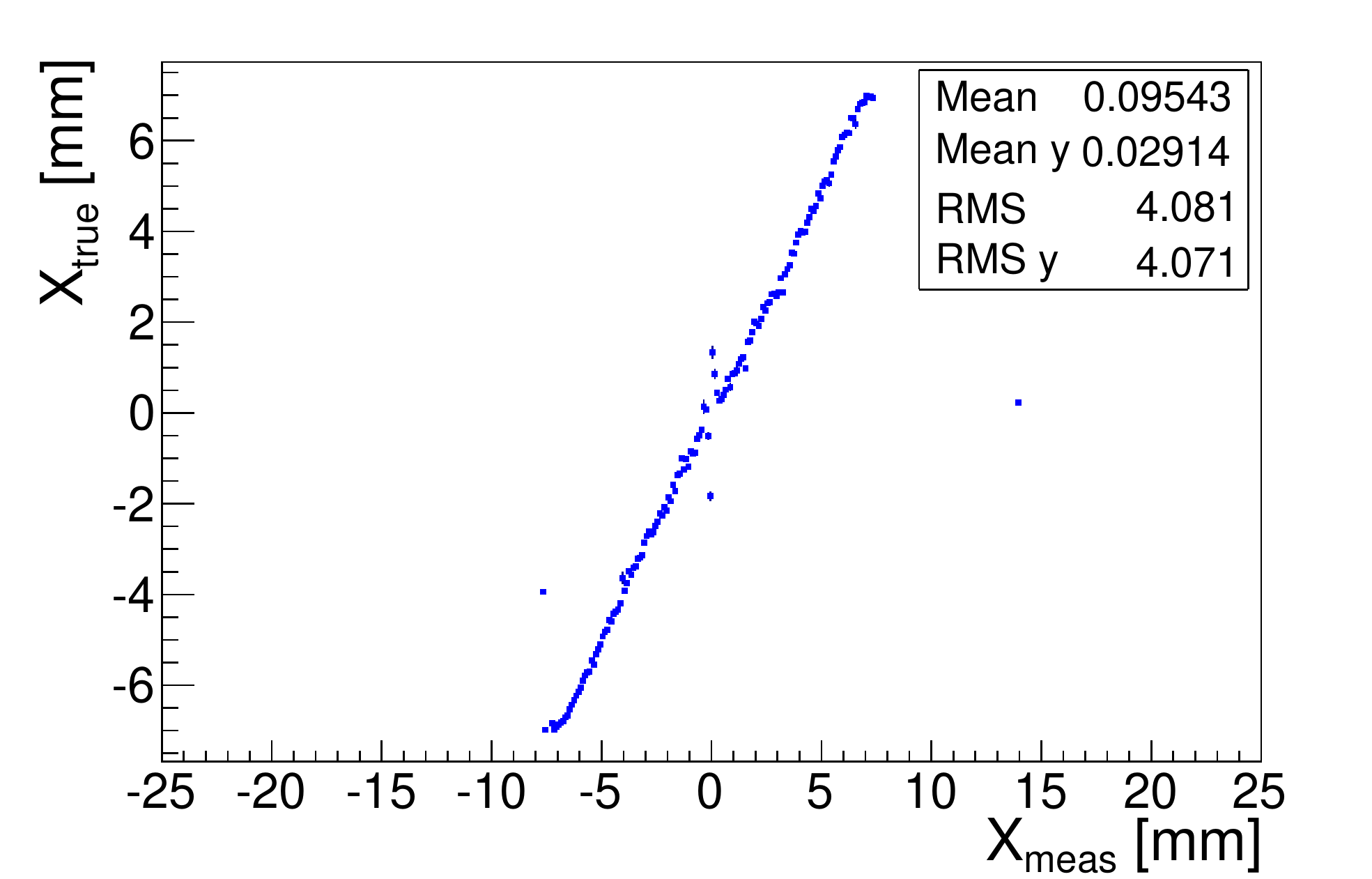}\\
\includegraphics[width=0.45\textwidth]{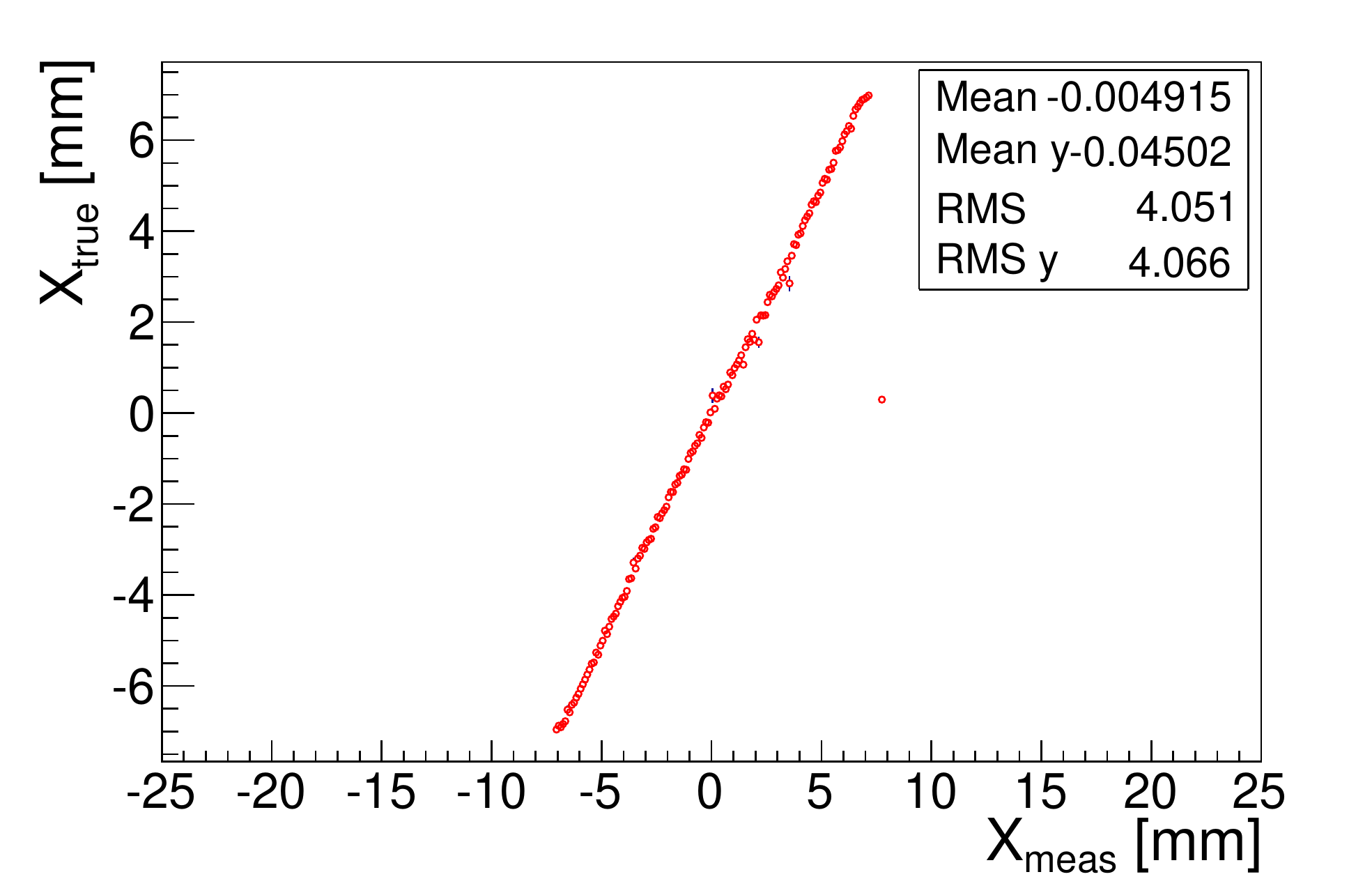}&
\includegraphics[width=0.45\textwidth]{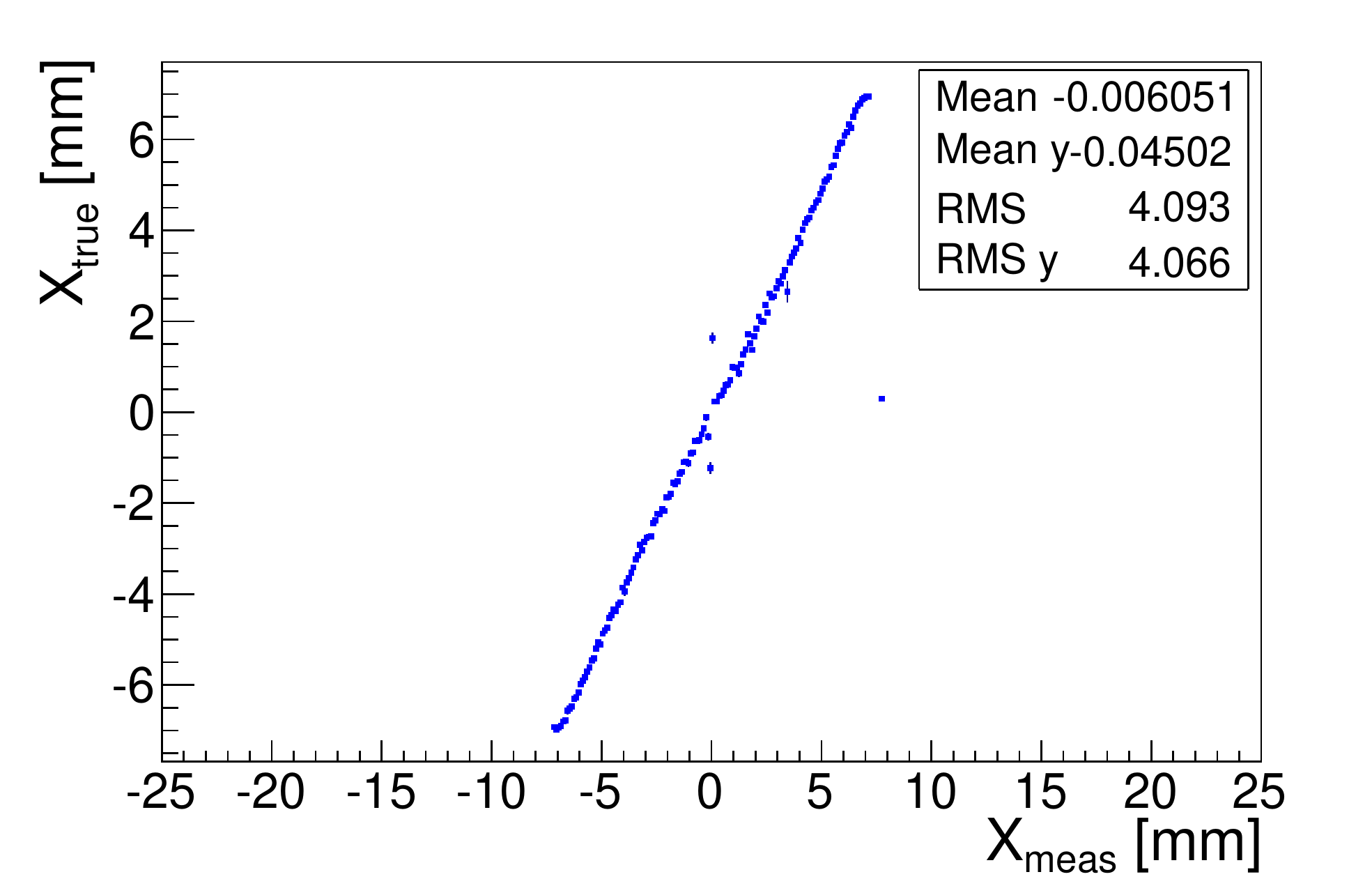}
  \end{tabular}\end{center}
\caption{2-D distribution of x$_{true}$ versus x$_{meas}$ for photons of energy 
         50 GeV and 150 GeV when log weights are used to estimate the COG as 
         given in Equation \ref{eqn:logweight}. The left
         (right) figures refer to cases when combined (individual) fiber
         information is used.}
\label{fig:logweightx_50150}
\end{figure}

\begin{figure}[htbp]
  \begin{center} \begin{tabular}{cc}
\includegraphics[width=0.45\textwidth]{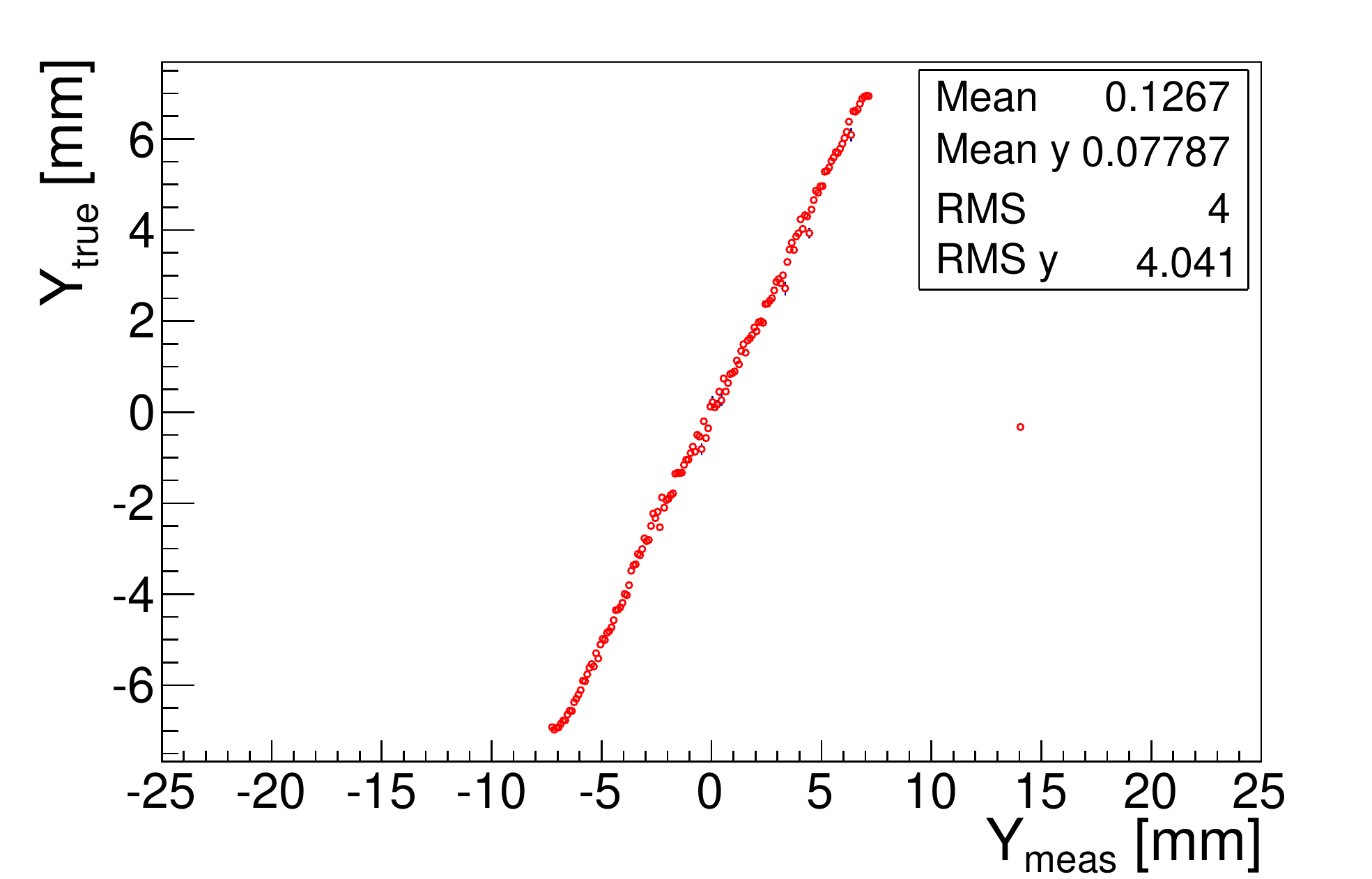}&
\includegraphics[width=0.45\textwidth]{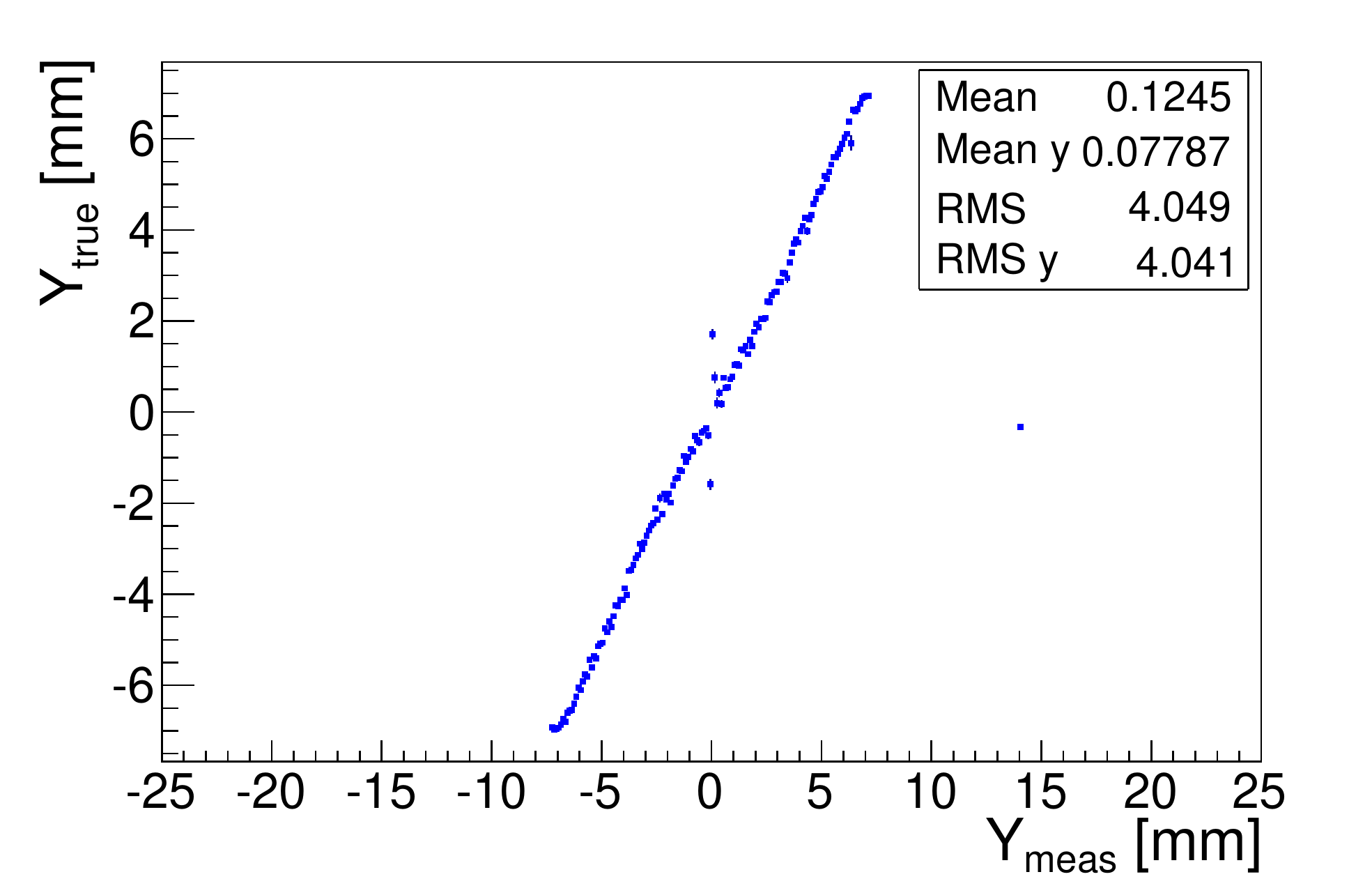}\\
\includegraphics[width=0.45\textwidth]{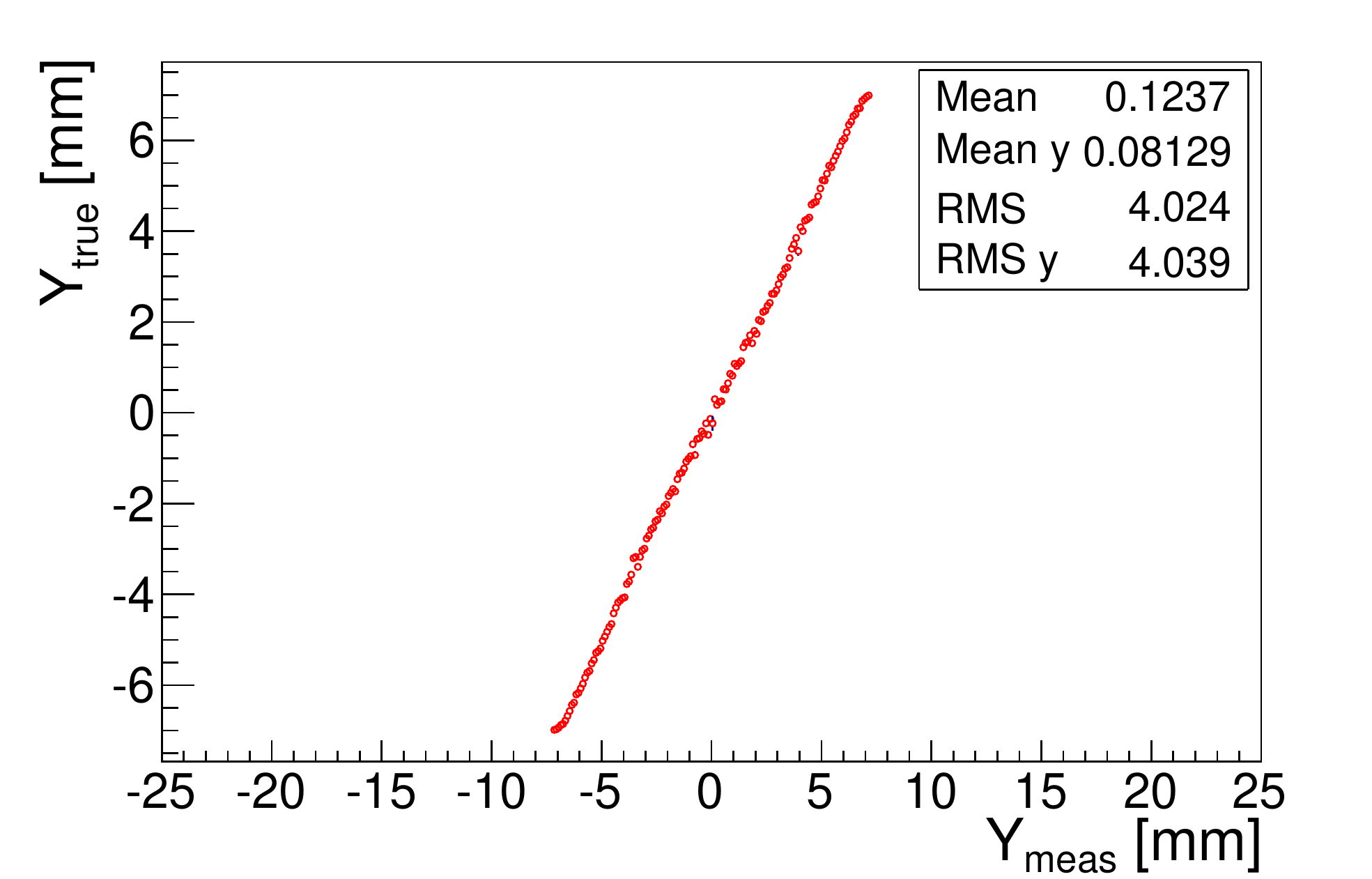}&
\includegraphics[width=0.45\textwidth]{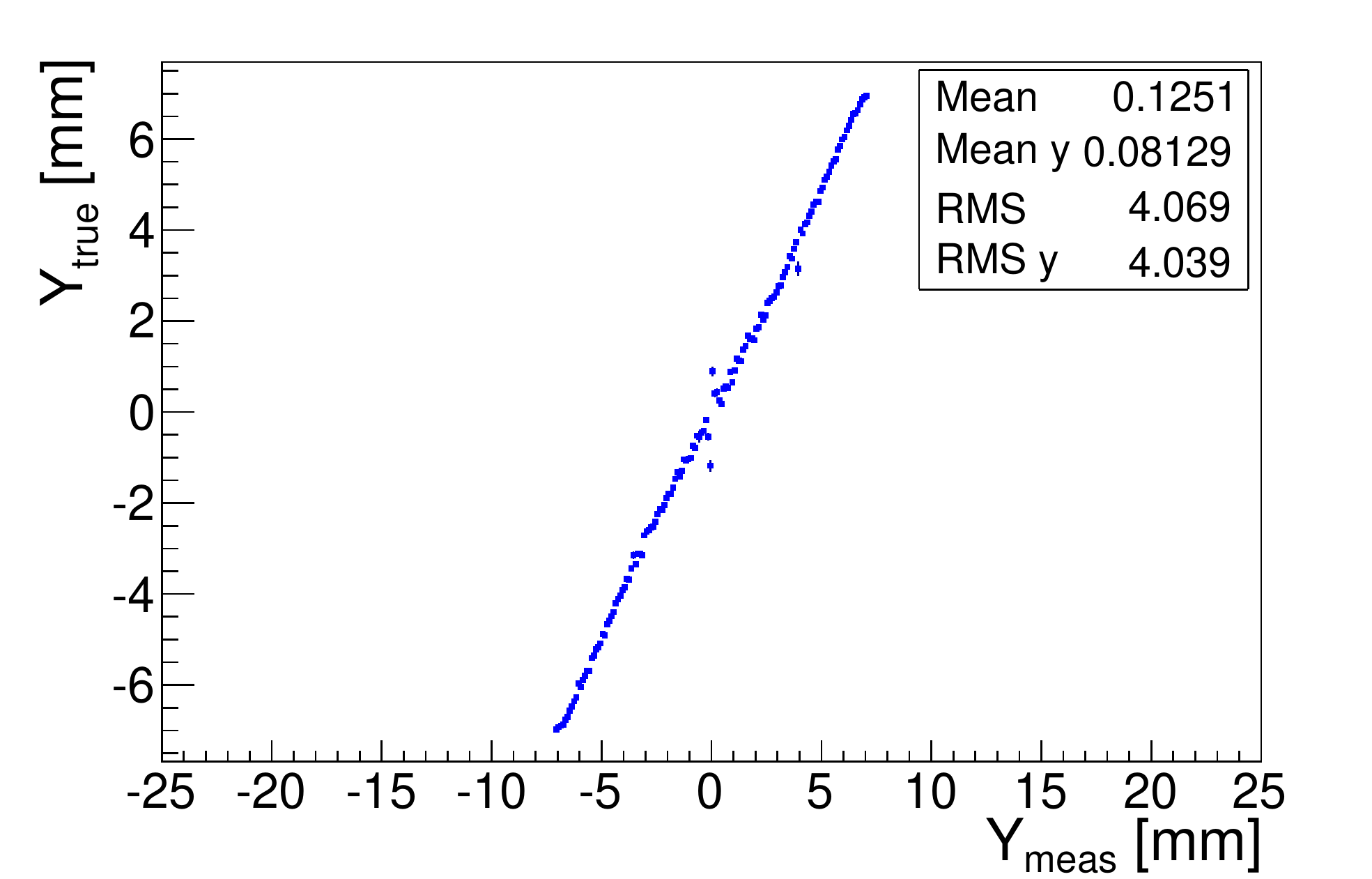}
  \end{tabular}\end{center}
\caption{2-D distribution of y$_{true}$ versus y$_{meas}$ for photons of energy 
         50 GeV and 150 GeV when log weights are used to estimate the COG as 
         given in Equation \ref{eqn:logweight}. The left
         (right) figures refer to cases when combined (individual) fiber
         information is used.}
\label{fig:logweighty_50150}
\end{figure}

Some features of the S-shape curve are summarized below: 
\begin{enumerate}
\item At the center, x$_{true}$ (y$_{true}$) = x$_{meas}$ (y$_{meas}$). This is 
because the deposited energy is mostly contained in the central tower. 
\item On moving away from the center in either direction, 
x$_{true}$ (y$_{true}$) $>$ x$_{meas}$ (y$_{meas}$). There is an exponential fall 
in the spread of the energy for other towers. So linear weights E$_{i}$ give 
more weight to the hit tower and hence the position is not correctly 
determined. 
\item At the edge (in this case at $\pm$6~mm), the energy is distributed 
equally in the adjacent towers and roughly equal weight is given to 
them and hence x$_{true}$ (y$_{true}$) again becomes x$_{meas}$ (y$_{meas}$). 
\end{enumerate}
Above three points essentially summarize why the S-shape arises. 
Instead of linear weights, log weights of energy fraction are also 
tried. Since the energy falls off as an exponential, the log weights 
compensate the exponential decrease and hence the estimated position is 
closer to the true one.
Equation \ref{eqn:logweight} shows the relations used to estimate 
the coordinates of the hit point with log weights. This equation 
depends highly on the value of $w_{0}$. The optimum value of $w_{0}$
depends on whether individual or combined fiber information is used.
Figures \ref{fig:logweightx_50150} and  \ref{fig:logweighty_50150}  
show the 2-D distribution of $x_{true}$ ($y_{true}$) VS $x_{meas}$ ($y_{meas}$) 
for 50 GeV and 150 GeV  photons when log weights are used for the 
two cases of using combined or individual fiber information.
\begin{eqnarray}
x_{meas} &= &\Sigma x_{i} \times w_{i}/\Sigma w_{i}\nonumber\\
y_{meas} &= &\Sigma y_{i} \times w_{i}/\Sigma w_{i}\nonumber\\
w_{i}    &= &Max(0, w0 + \ln(E_{i}/E_{T}))
\label{eqn:logweight}
\end{eqnarray}
where $w_0$ $=$ 4.7 for combined fiber information and $w_0$ $=$ 6 for 
individual fiber information.
%
\begin{figure}[htbp]
  \begin{center} \begin{tabular}{cc}
\includegraphics[width=0.45\textwidth]{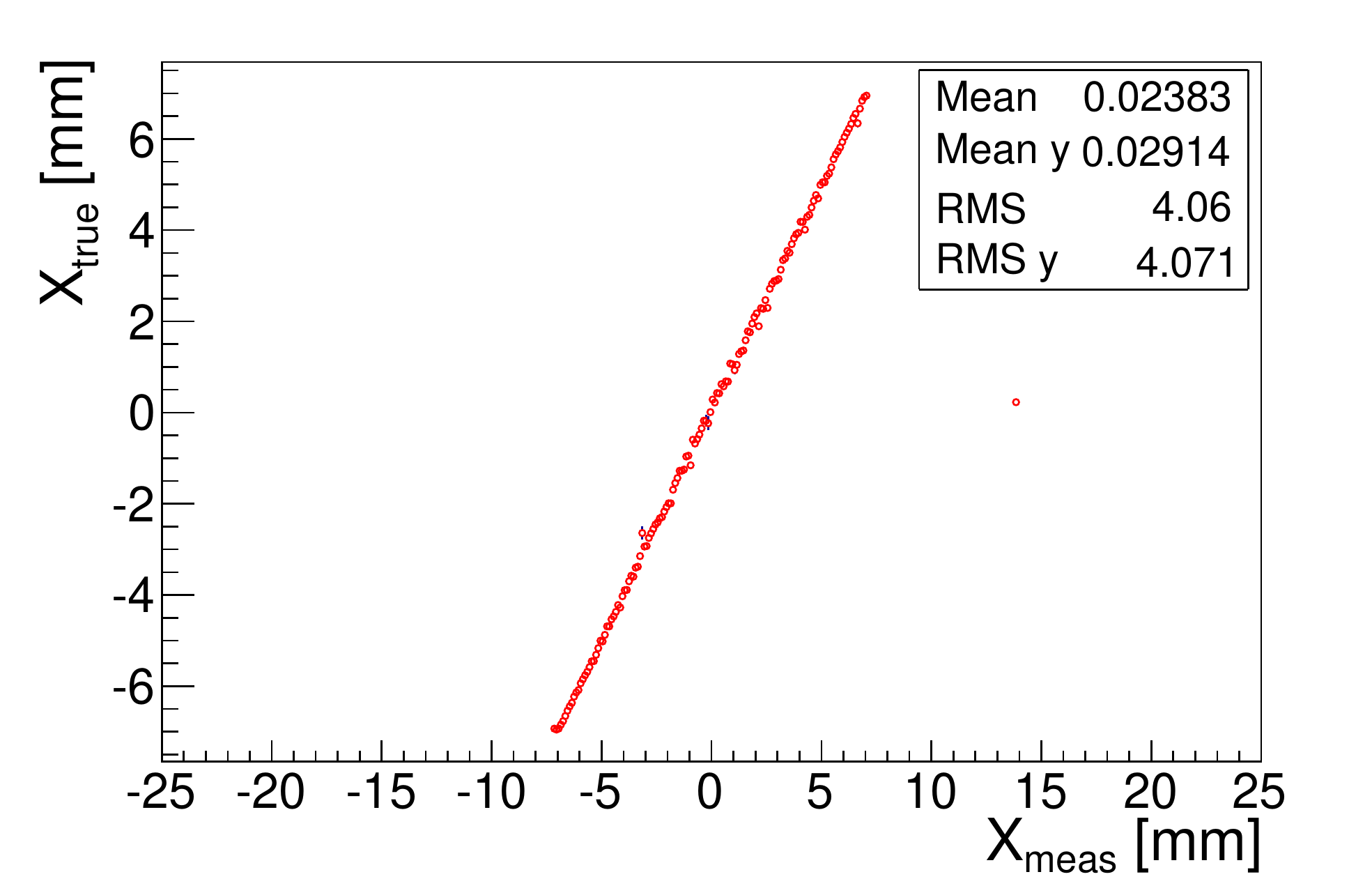}&
\includegraphics[width=0.45\textwidth]{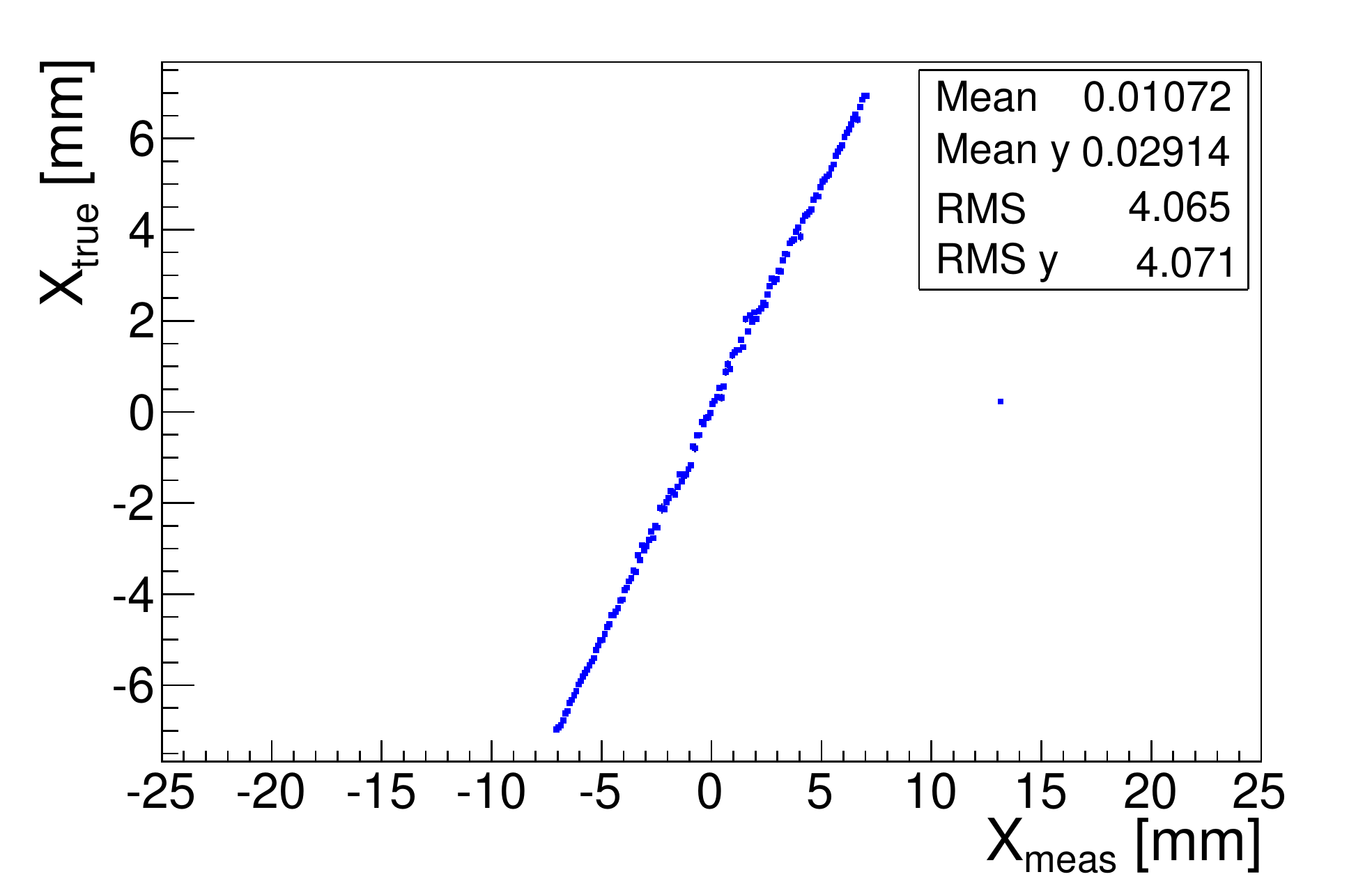}\\
\includegraphics[width=0.45\textwidth]{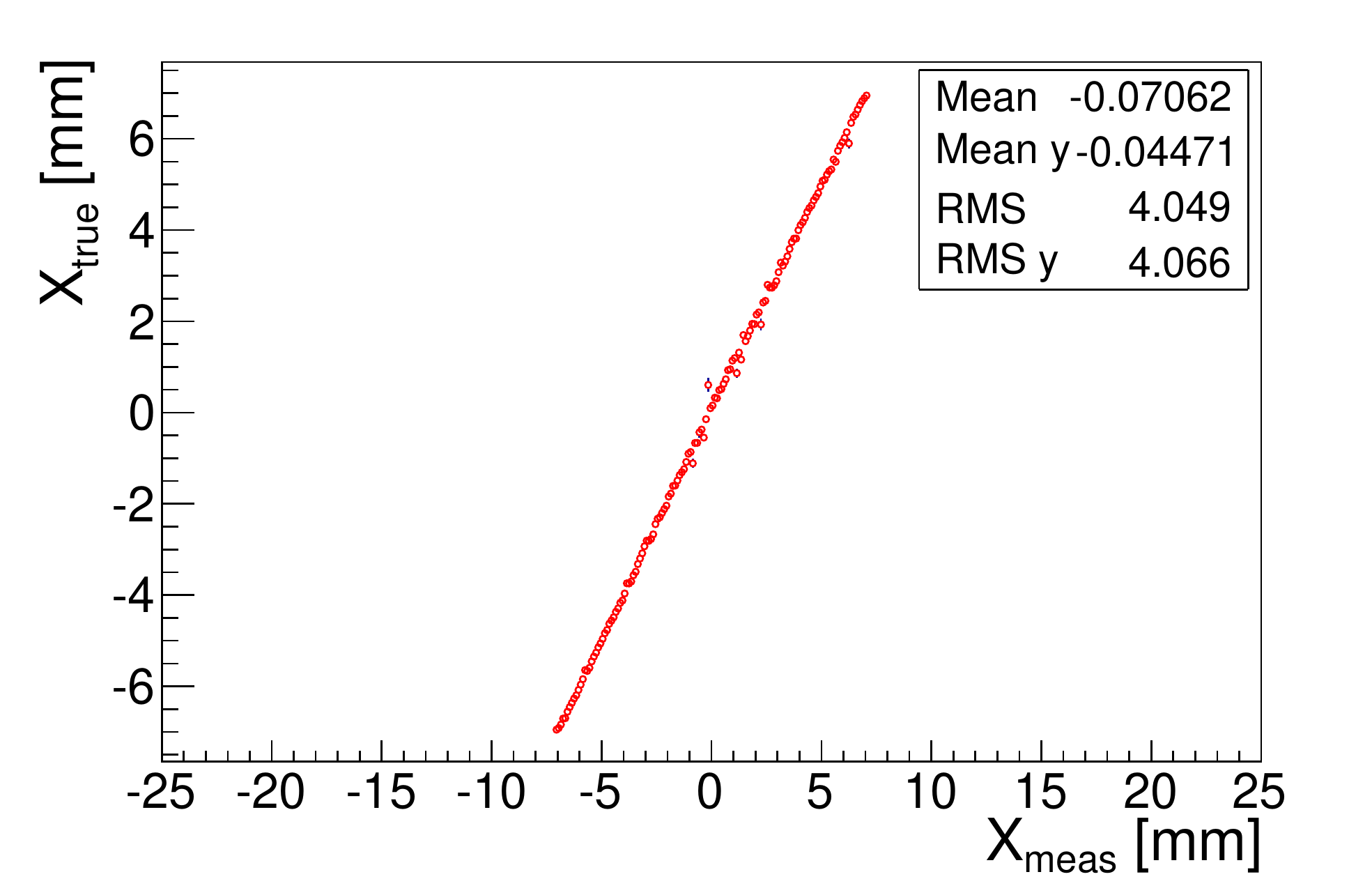}&
\includegraphics[width=0.45\textwidth]{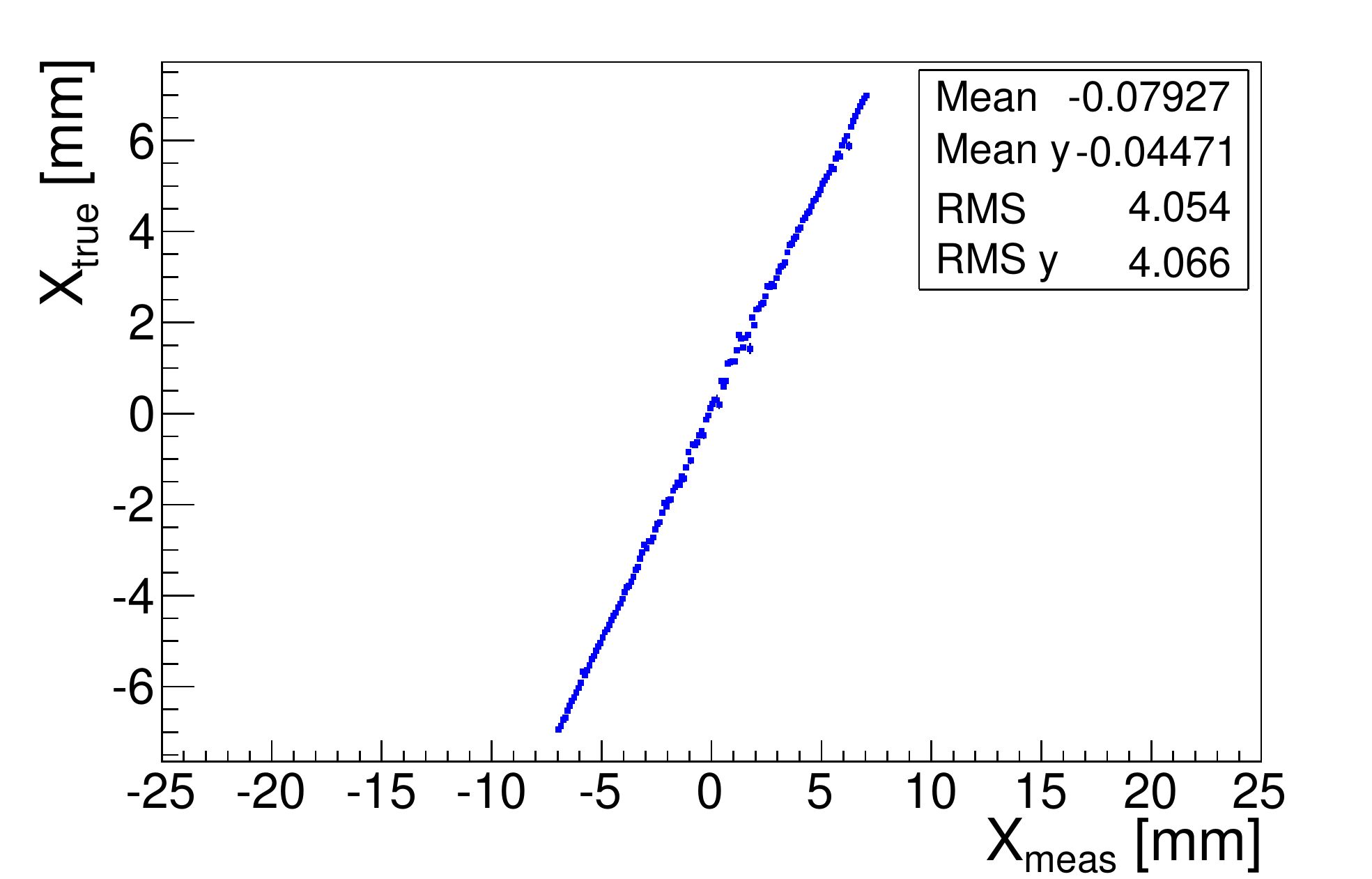}
  \end{tabular}\end{center}
\caption{2-D distributions of x$_{true}$ versus S-shape corrected x$_{meas}$
         for photons of energy 50 GeV (top) and 150 GeV (bottom)
         when linear weights are used to estimate the COG as given in 
         Equation \ref{eqn:linearweight}. The left
         (right) figures refer to cases when combined (individual) fiber
         information is used.}
\label{fig:xsshapecorr_50150pho}
\end{figure}

\begin{figure}[htbp]
  \begin{center} \begin{tabular}{cc}
\includegraphics[width=0.45\textwidth]{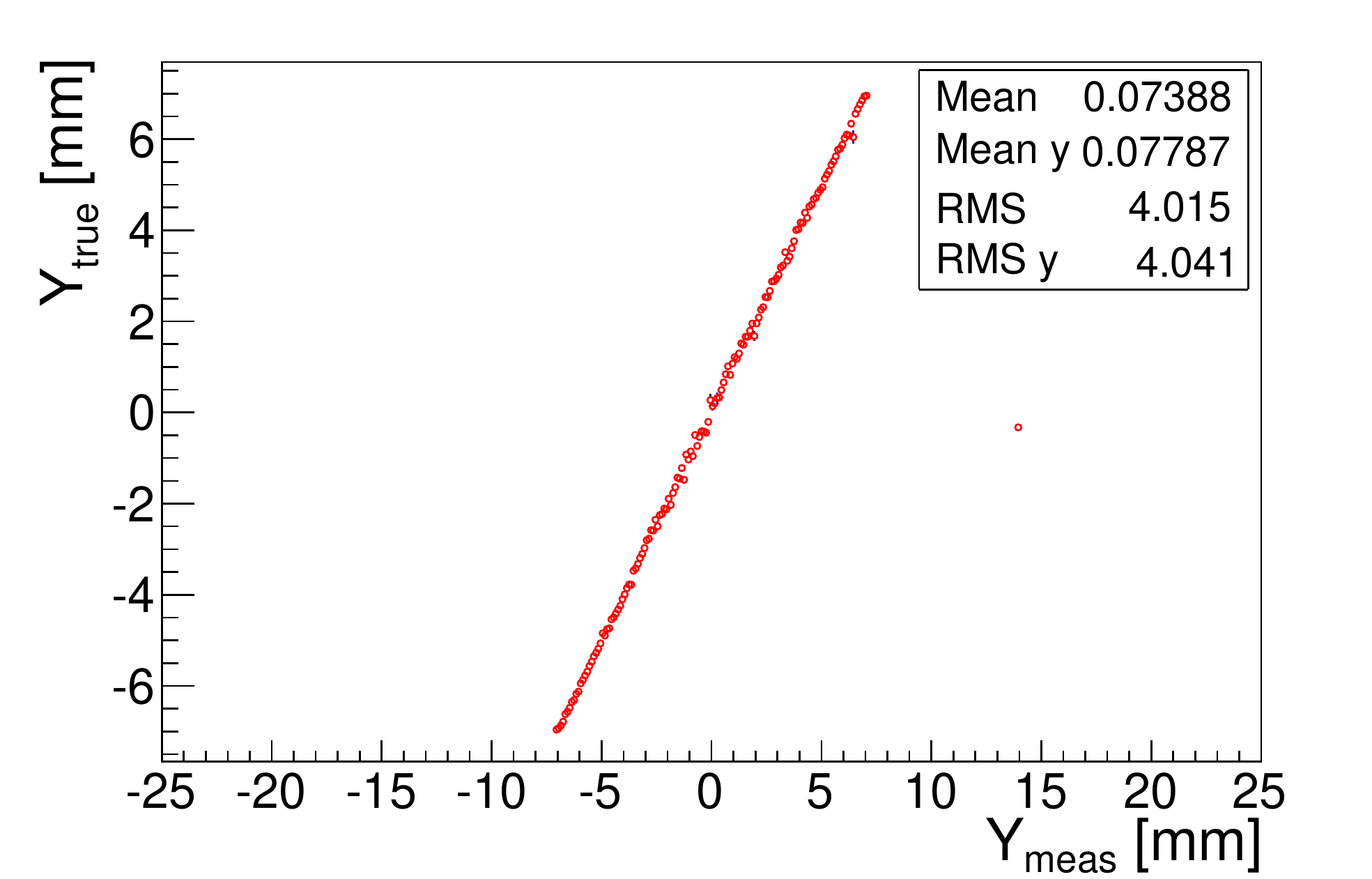}&
\includegraphics[width=0.45\textwidth]{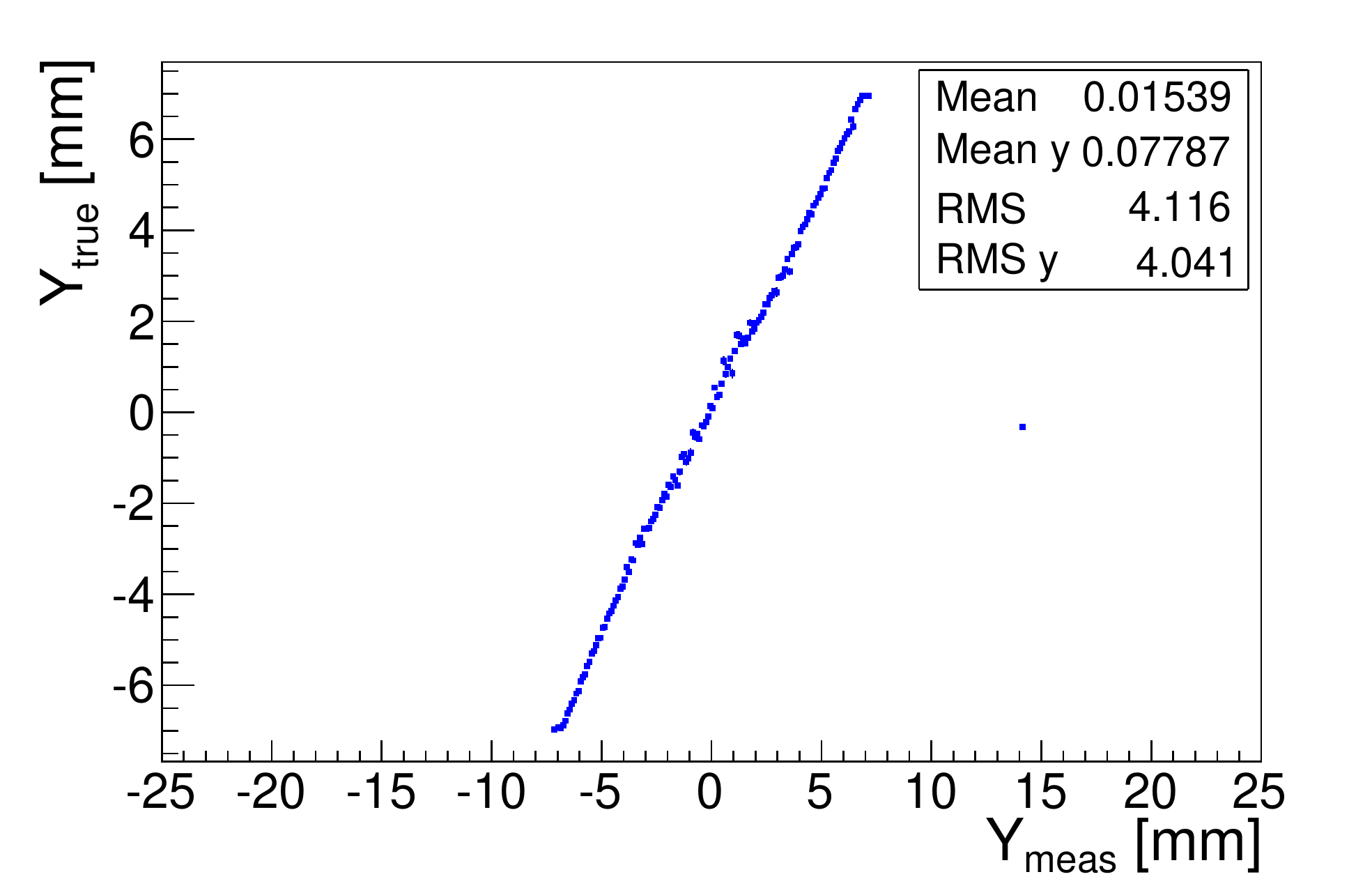}\\
\includegraphics[width=0.45\textwidth]{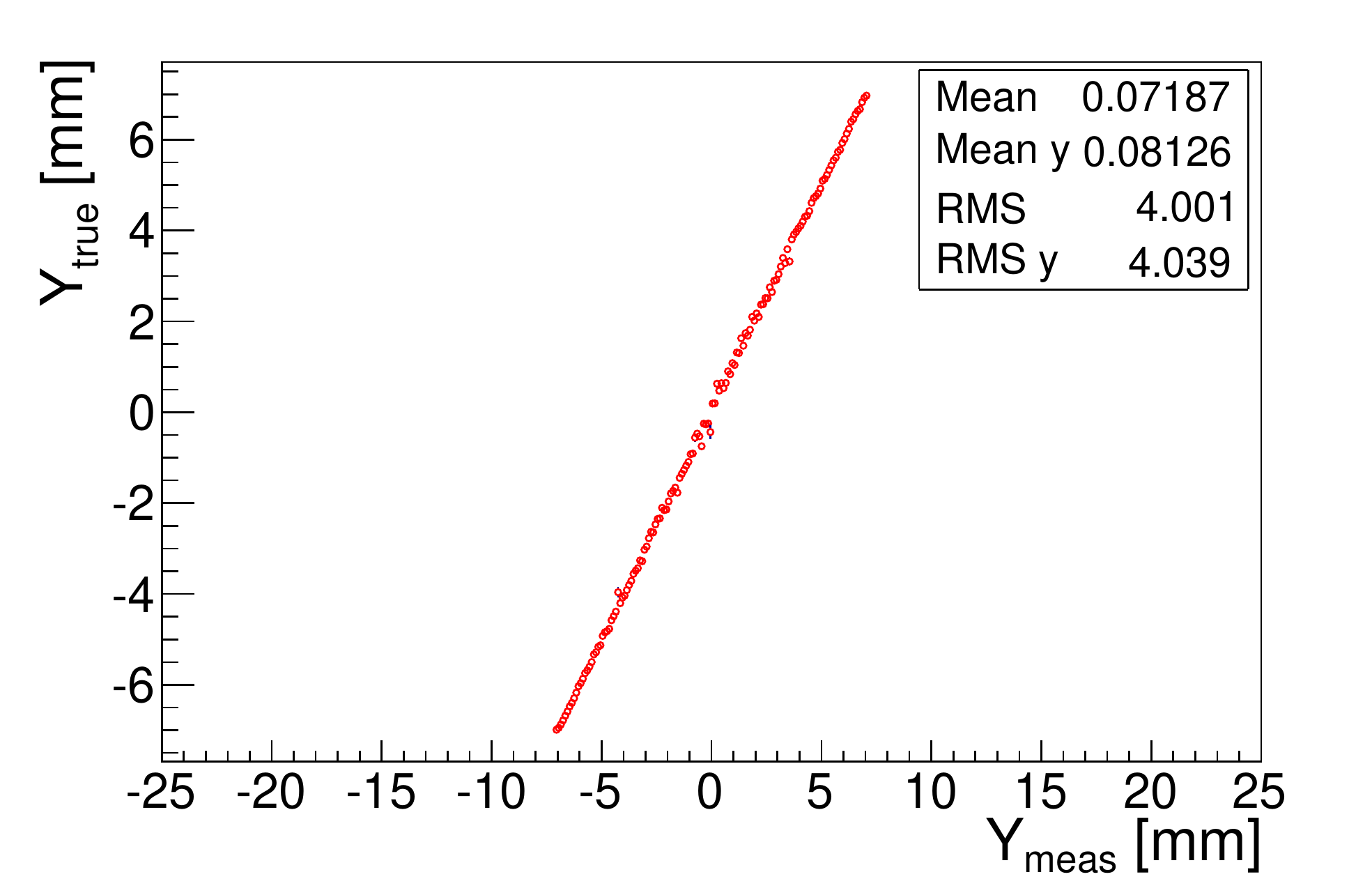}&
\includegraphics[width=0.45\textwidth]{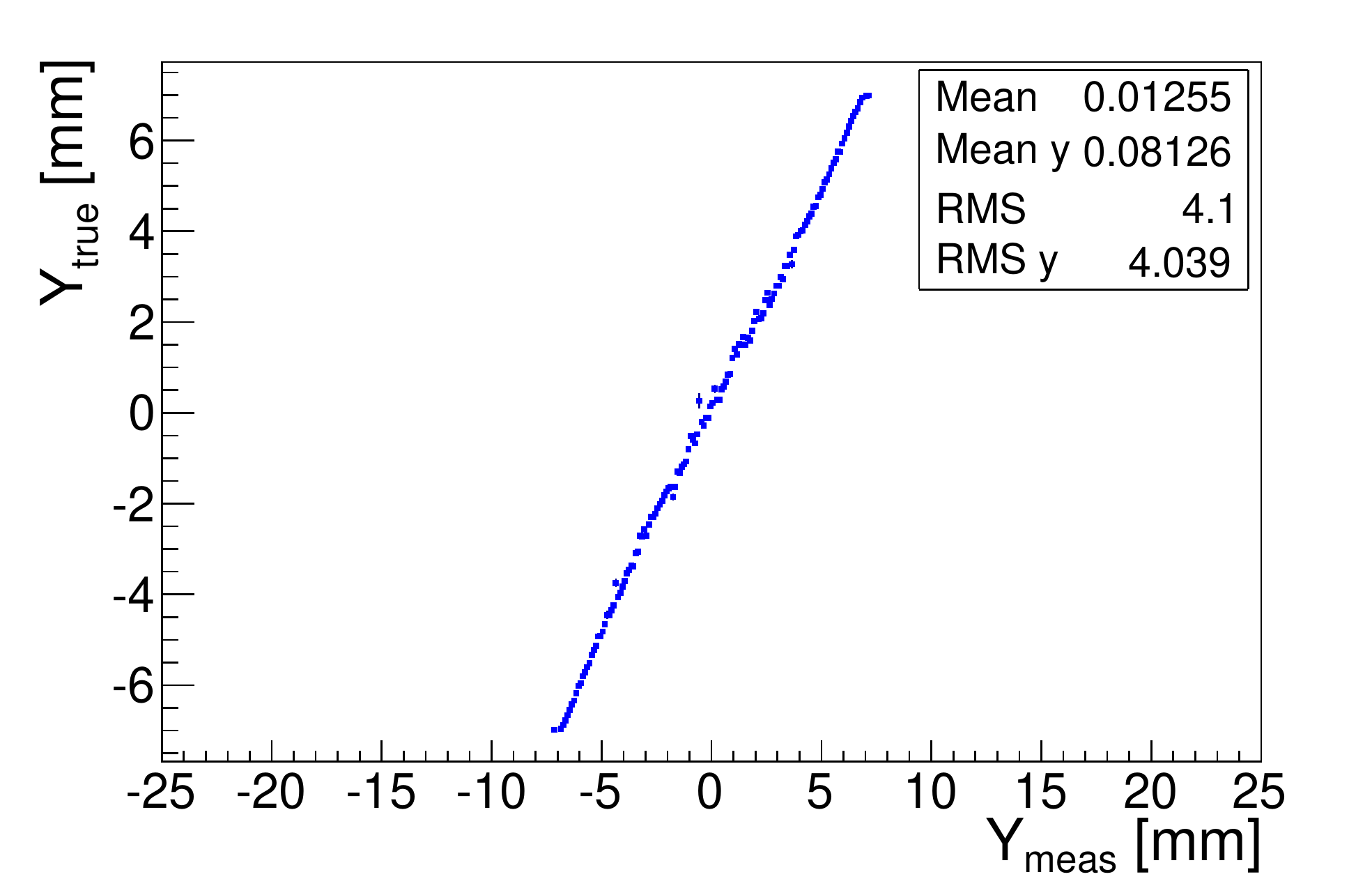}
  \end{tabular}\end{center}
\caption{2-D distributions of y$_{true}$ versus S-shape corrected y$_{meas}$
         for photons of energy 50 GeV (top) and 150 GeV (bottom)
         when linear weights are used to estimate the COG as given in 
         Equation \ref{eqn:linearweight}. The left
         (right) figures refer to cases when combined (individual) fiber
         information is used.}
\label{fig:ysshapecorr_50150pho}
\end{figure}

For this analysis, linear weights are used to estimate the position of COG.
These S-shape curves are then fitted with a function of the form 

\begin{eqnarray}
x_{true}  &= &P_{0} \times \tan^{-1} {(P_{1} \times (x_{meas} - P_{2}))} + P_{3} \times {(x_{meas} - P_{2})} \nonumber \\
y_{true}  &= &P_{4} \times \tan^{-1} {(P_{5} \times (y_{meas} - P_{6}))} + P_{7} \times {(y_{meas} - P_{6})}
\label{eqn:sshapefit}
\end{eqnarray}

Figures \ref{fig:xsshape_50150} and \ref{fig:ysshape_50150}  show how the 
fitted functions for X and Y look like for 50 GeV and 150 GeV photons.
Using these fitted functions, the $x_{meas}$ ($y_{meas}$) are corrected so that 
the measured position coordinates are nearer to the true ones, i.e., $x_{true}$
($y_{true}$). Fitted parameters differ slightly depending on the energy of the 
photon. For simplicity fitted parameters from 50 GeV photons 
are used to fit all energy particles and to obtain S-shape corrected 
$\overline{x}$ and $\overline{y}$. These S-shaped corrected positions are 
then used in Equation \ref{eqn:covquant} as $\overline{x}$ and $\overline{y}$. 
Figure \ref{fig:xsshapecorr_50150pho} shows the S-shape corrected X 
position of 50 GeV and 150 GeV photons for both the cases of combined and 
individual fiber information. Similarly, Figure \ref{fig:ysshapecorr_50150pho} 
shows the S-shape corrected Y position of 50 GeV and 150 GeV photons.

\begin{figure}[htbp]
 \begin{center} 
  \includegraphics[width=0.325\textwidth]{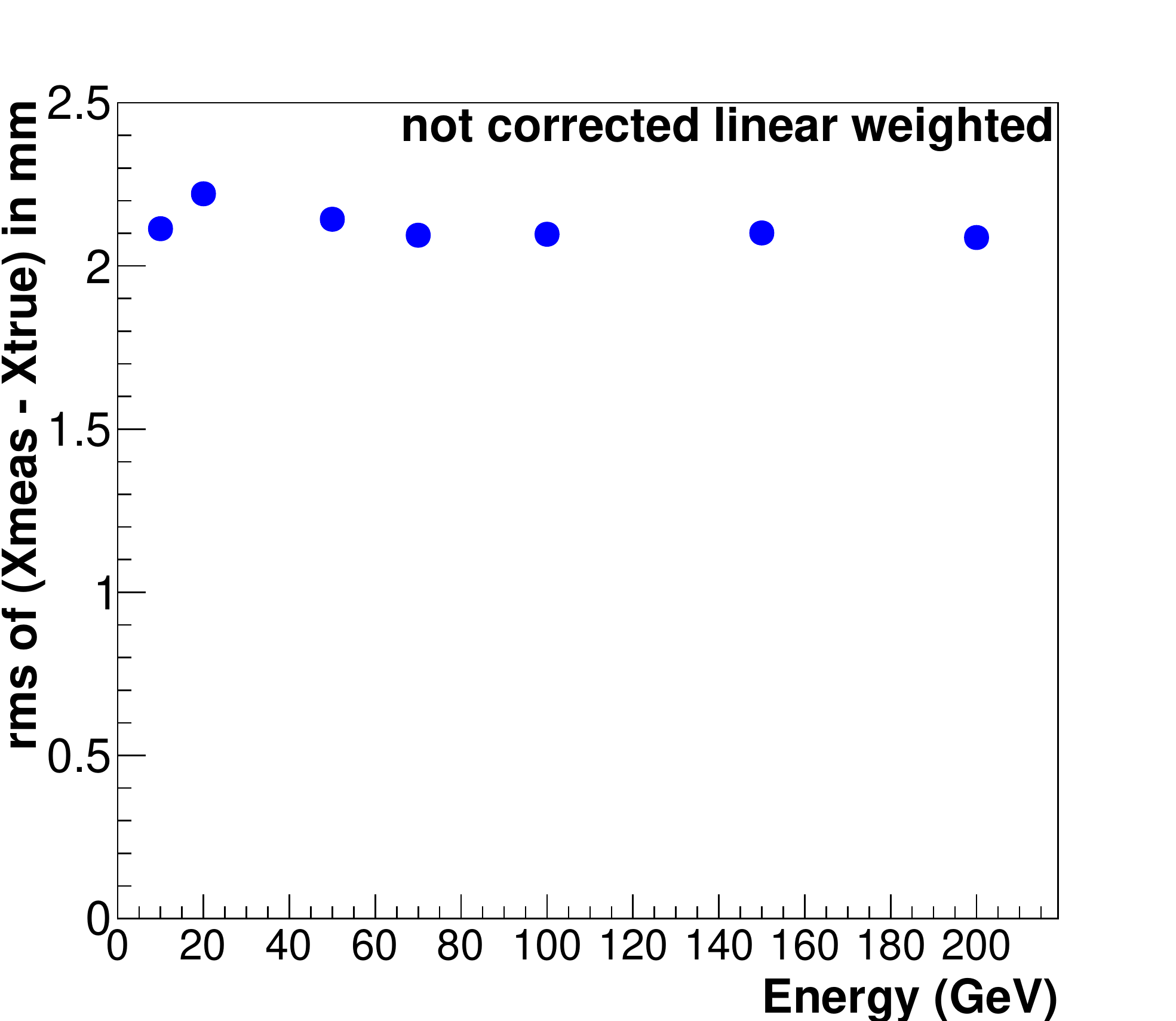}
  \includegraphics[width=0.325\textwidth]{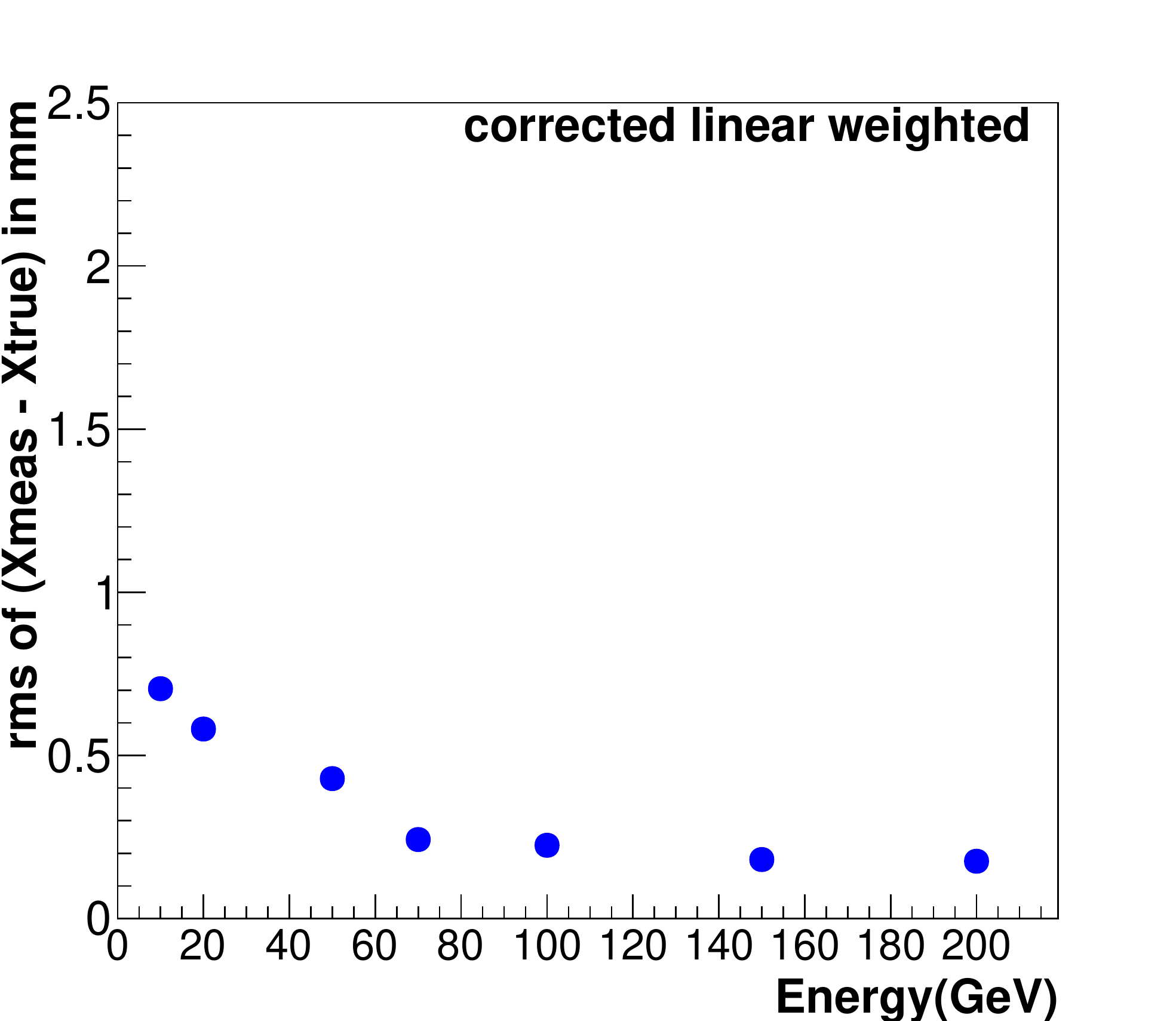}
  \includegraphics[width=0.325\textwidth]{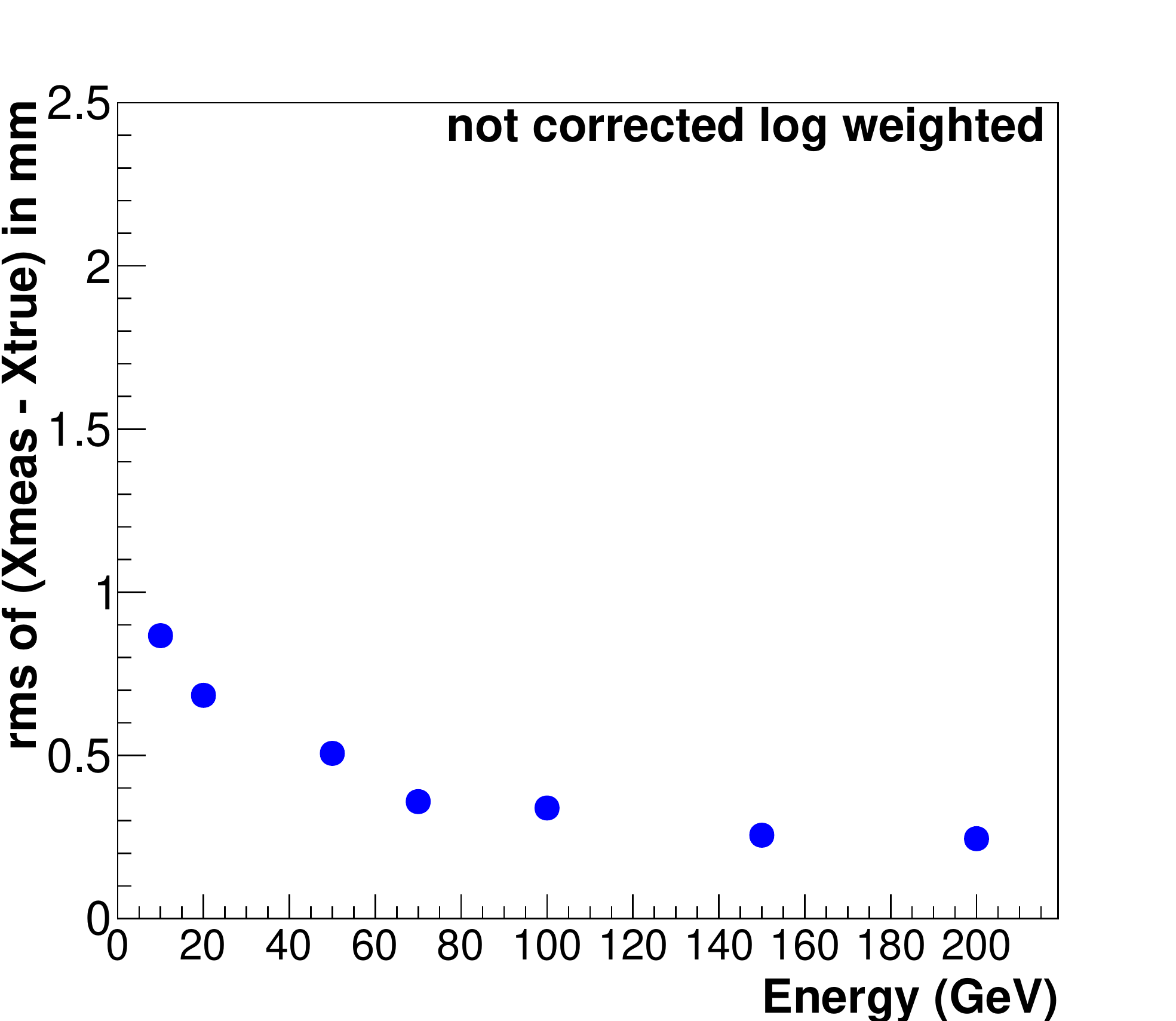}
 \end{center}
\caption{Position resolution for photons in a Shashlik detector as a function
         of photon energy with position reconstructed using linear weighting in 
         energy (a); S-shape corrected position with linear weighting in energy 
         (b); and position with logarithmic weighting (c).}
 \label{fig:posresol}
\end{figure}

Position resolution is studied from samples of photons produced with impact 
points randomly distributed on the front face of the module. The difference 
between the true and measured position along X and Y directions are plotted.
These distributions follow roughly a Gaussian shape. The RMS of these
distributions is used to estimate the position resolution of these photons.
Figure \ref{fig:posresol} shows position resolution as a function of photon
energy for three different scenario: (a) position with linear weighting in
energy; (b) S-shape corrected position with linear weighting in energy; and
(c) position with logarithmic weighting. Position resolution improves as a 
function of the photon energy in all three cases. 

Position resolution is quite large and is related to the lateral size of
the tower if one simply
uses linear weighting in energy in determining the impact point and without
any further correction. With S-shape correction, the resolution improves
significantly. The resolution is 0.70~mm for photons at 10 GeV and it improves
with energy becoming 0.22~mm at 200 GeV. Logarithmic weighting takes care of
the correction to some extent and even without any further correction the 
resolution is 0.87~mm at 10 GeV and 0.34~mm at 200 GeV. 
A precise measurement of the impact position is extremely useful for $\pi^{0}$/$\gamma$ separation.

\section{$\pi^{0}$/$\gamma$ separation} \label{sec:Shashlikpi0gammaStudy}
An important measure of the performance of an electromagnetic calorimeter used in 
a high energy physics experiment is its ability to separate between photons 
and $\pi^0$s. In high energy collisions 
any final state with photons has a background contribution 
from jets which fake photons. This is because $\pi^{0}$'s in jets decay 
to 2$\gamma$'s almost 99.9\% of the time. 
For decays of a high energy $\pi^0$, the angle between the two photons can become 
comparable with or smaller than the granularity of the calorimeter. 
It is very difficult to separate the photons from such a decay
from photons which are coming either from the interaction vertex or from
radiation off charged leptons.

In this study, the idea of exploiting the information from the
four fibres for $\pi^{0}$/$\gamma$ separation has been investigated.
The idea behind using information from all the four fibers 
individually is that a larger fraction of the deposited energy from a single 
photon will be collected by the fiber which is closest to the impact point, 
while $\pi^{0}$, decaying to a pair of photons will have two impact points
on the Shashlik detector and the sharing of light among the fibers will
significantly increase.
\subsection{Shower shapes} \label{sec:showershapes}
In general, the lateral shower profile tends to be broader for photons coming 
from $\pi^{0}$ compared to that of prompt photons. This holds true for lower 
energy $\pi^{0}$'s (less than 100 GeV). 
Therefore shower shape variables are useful for discriminating between $\pi^0$'s and photons. 
For all shower shape variables, the tower with maximum energy deposit is first 
identified and then the shape parameters are formed around it. The following
shape variables are considered:

\begin{description}
\item[$S1/S9$:] This ratio makes use of $S1$, 
the maximum energy deposited in a tower, 
and $S9$, the energy deposited in 3$\times$3 array around 
the maximum energy deposited tower. Figure \ref{fig:s1s9_105070150} shows the array of 
3$\times$3 towers formed around the maximum energy deposited tower and the 
distribution of $S1/S9$ for 50 GeV, 70 GeV and 150 GeV photons and $\pi^{0}$'s. 

\begin{figure}[htbp]
  \begin{center} \begin{tabular}{cc}
\hspace{20pt}{\raisebox{15pt}{
\includegraphics[width=4cm,height=3cm]{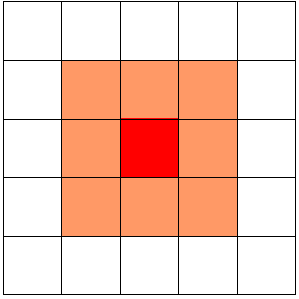}}} & 
\includegraphics[width=0.45\textwidth]{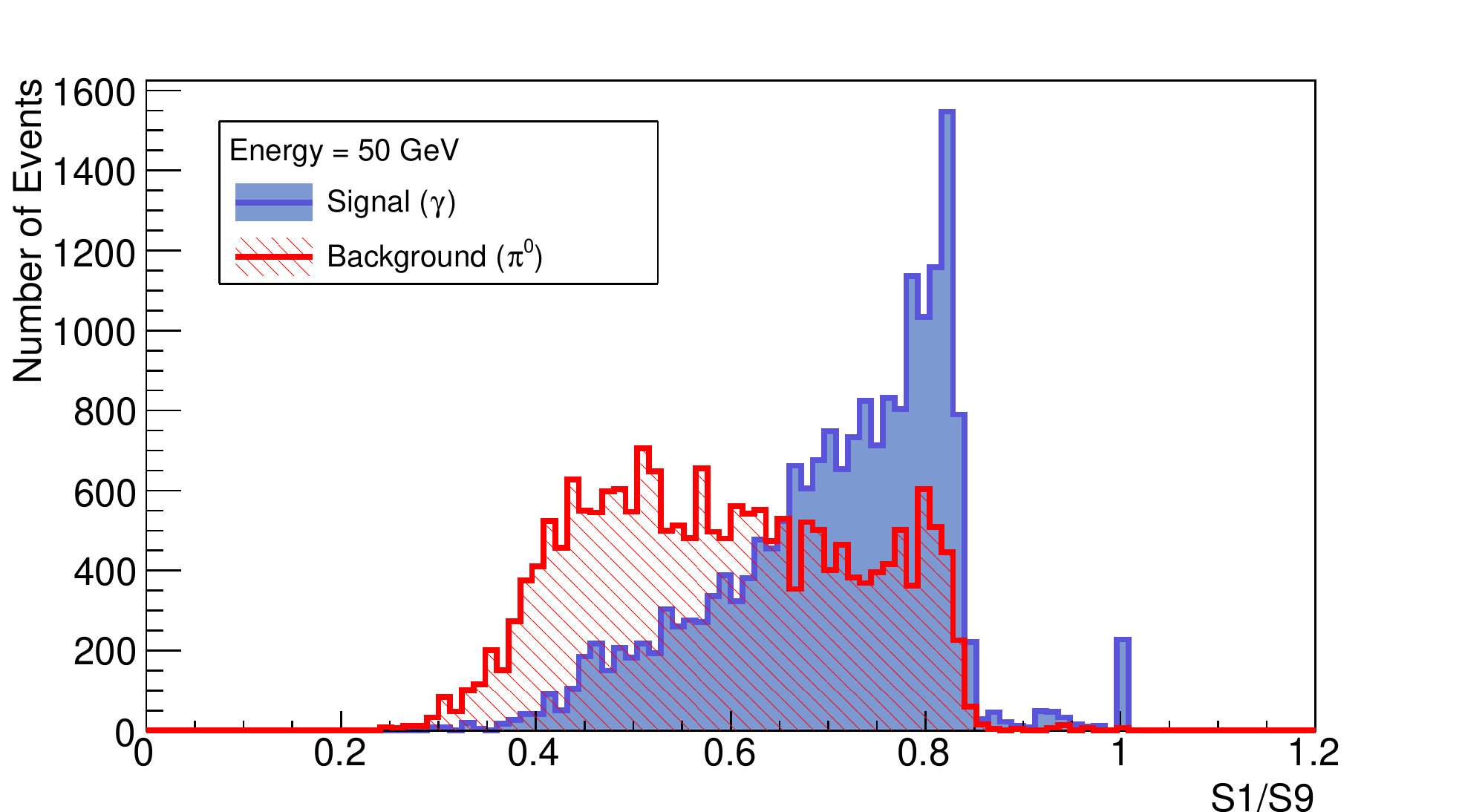}\\
\includegraphics[width=0.45\textwidth]{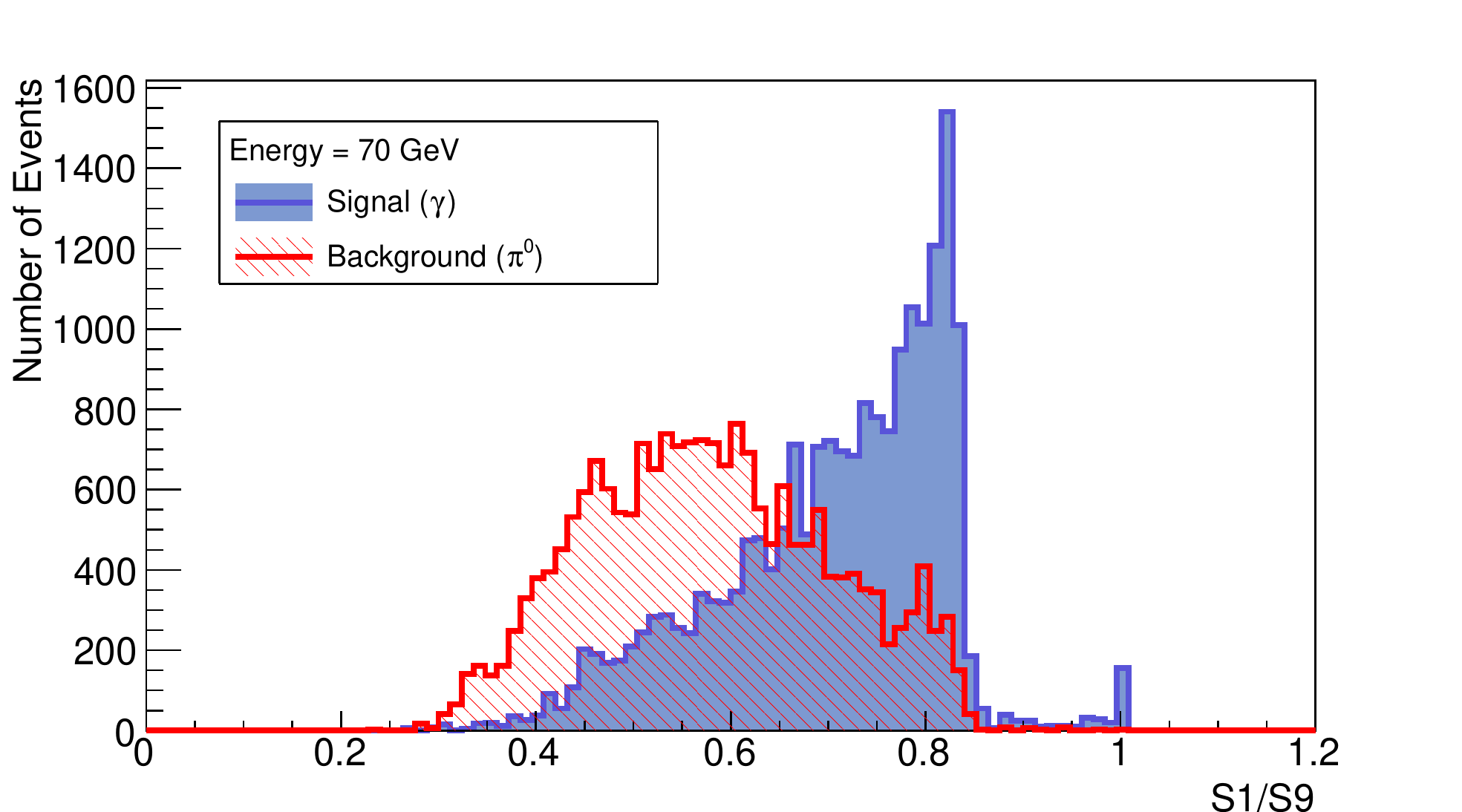}&
\includegraphics[width=0.45\textwidth]{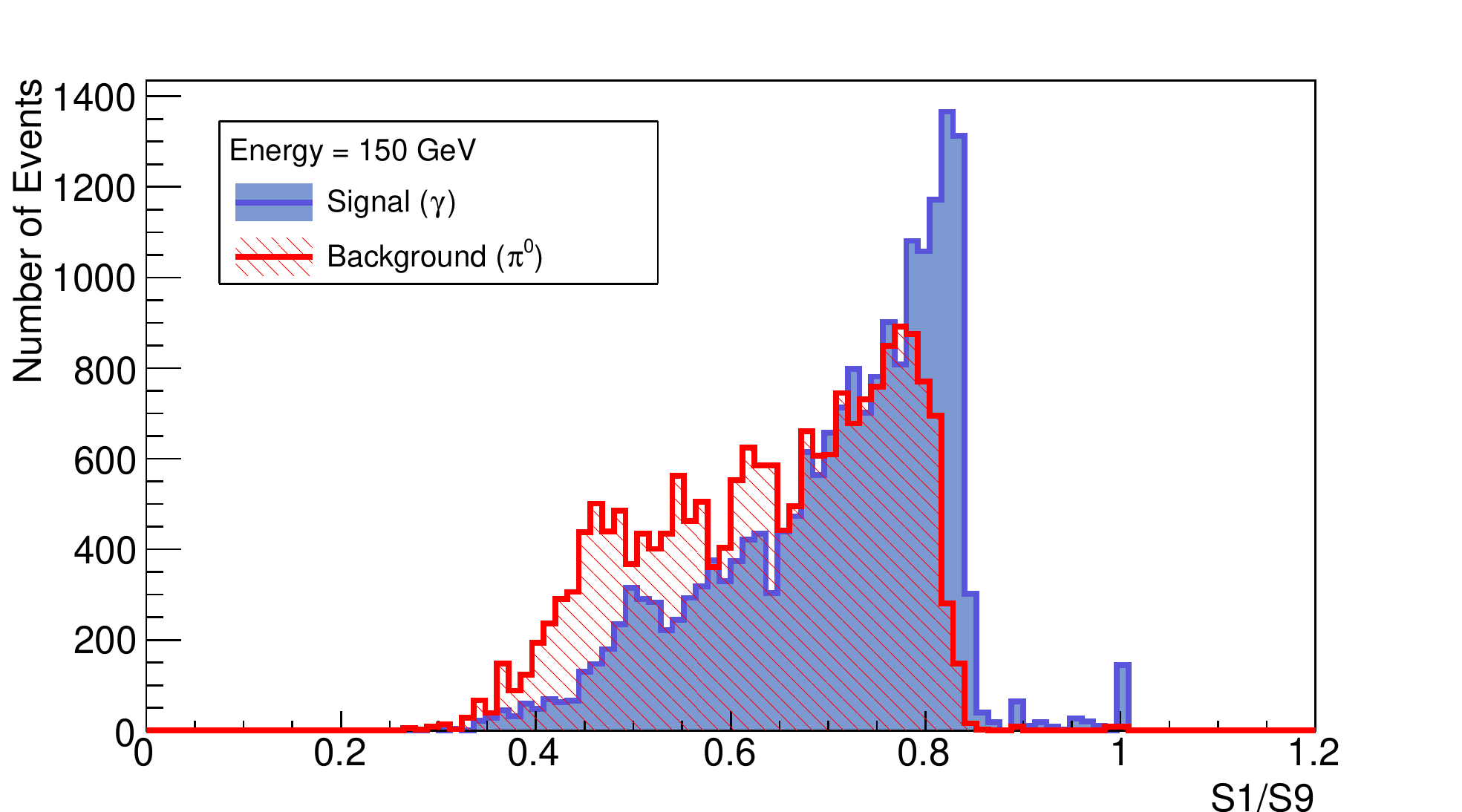}
  \end{tabular} \end{center}
\caption{The top left diagram shows 3$\times$3 array of towers in light 
         orange color and the maximum energy deposited tower in red color. The top-right 
         figure shows the distribution of S1/S9 for 50 GeV photons and 
         $\pi^{0}$'s. Similar distribution for 70 GeV photons and $\pi^{0}$'s 
         is on the bottom left; and 150 GeV photons and $\pi^{0}$'s is on the 
         bottom right. The blue hatched histogram is for photons and the red
         hatched histogram is for $\pi^{0}$'s \cite{ref:confpaper}.}
\label{fig:s1s9_105070150}
\end{figure}
\item[$S1/S4$:] This ratio uses $S4$, the energy deposited in 2$\times$2 array
including the maximum energy tower. 
Four possible 2$\times$2 arrays are 
possible which include the maximum energy tower. The combination which 
corresponds to the largest sum total energy is used in determining the ratio.
Figure \ref{fig:s1s4_105070150} shows the four possible combinations of 
2$\times$2 array of towers and the distributions of $S1/S4$ for 50, 70 and 150
GeV photons and $\pi^{0}$'s.

\begin{figure}[htbp]
  \begin{center} \begin{tabular}{cc}
\hspace{20pt}{\raisebox{15pt}{
\includegraphics[width=4.5cm,height=3cm]{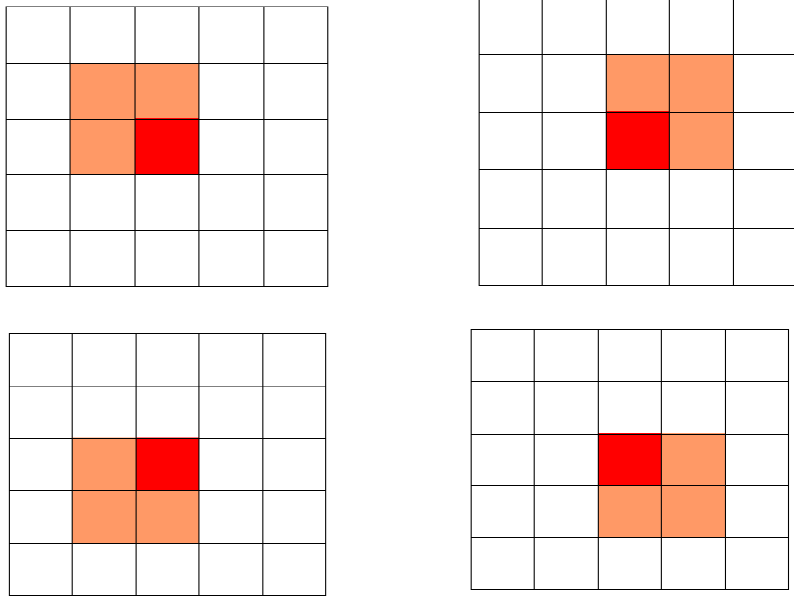}}}&
\includegraphics[width=0.45\textwidth]{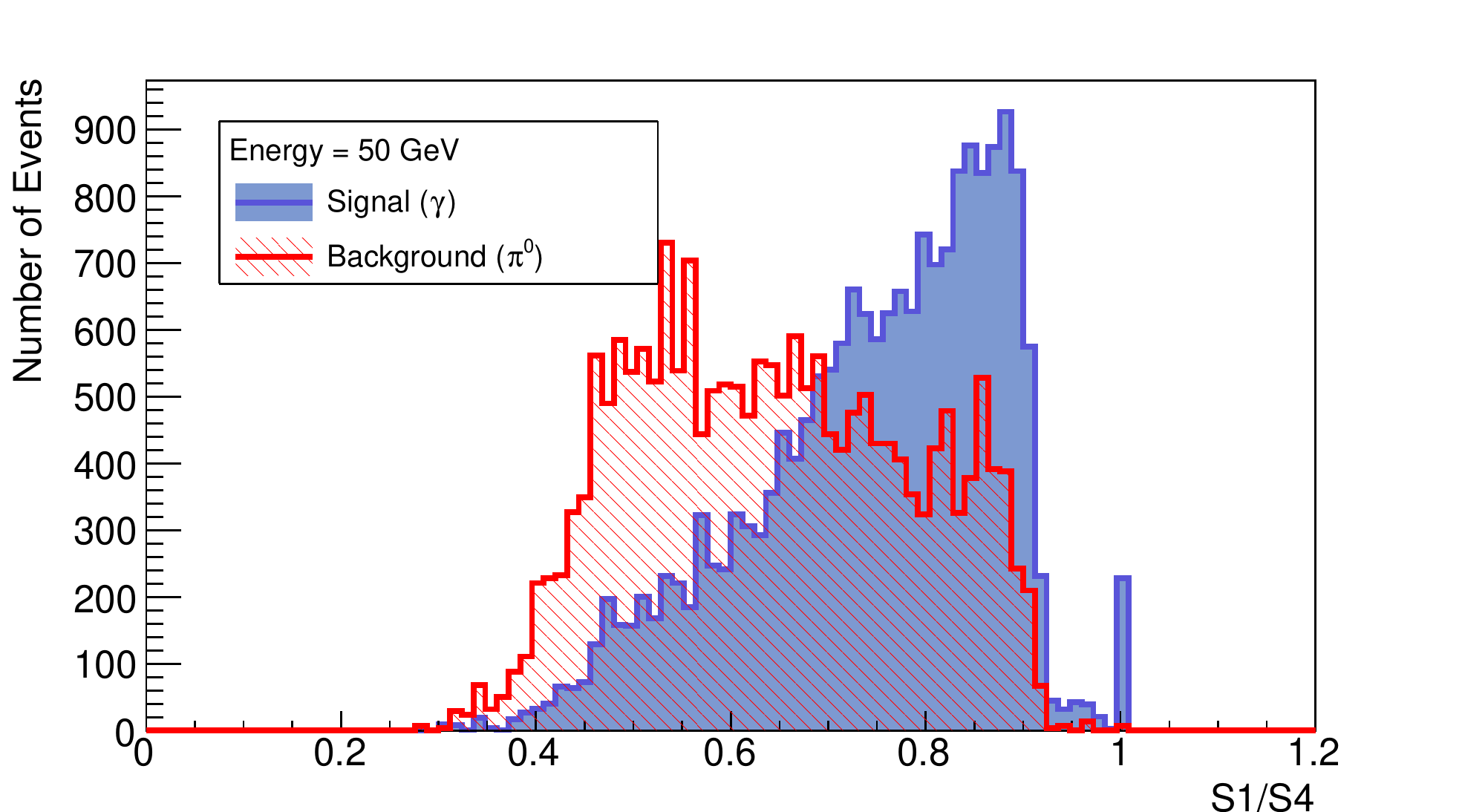}\\
\includegraphics[width=0.45\textwidth]{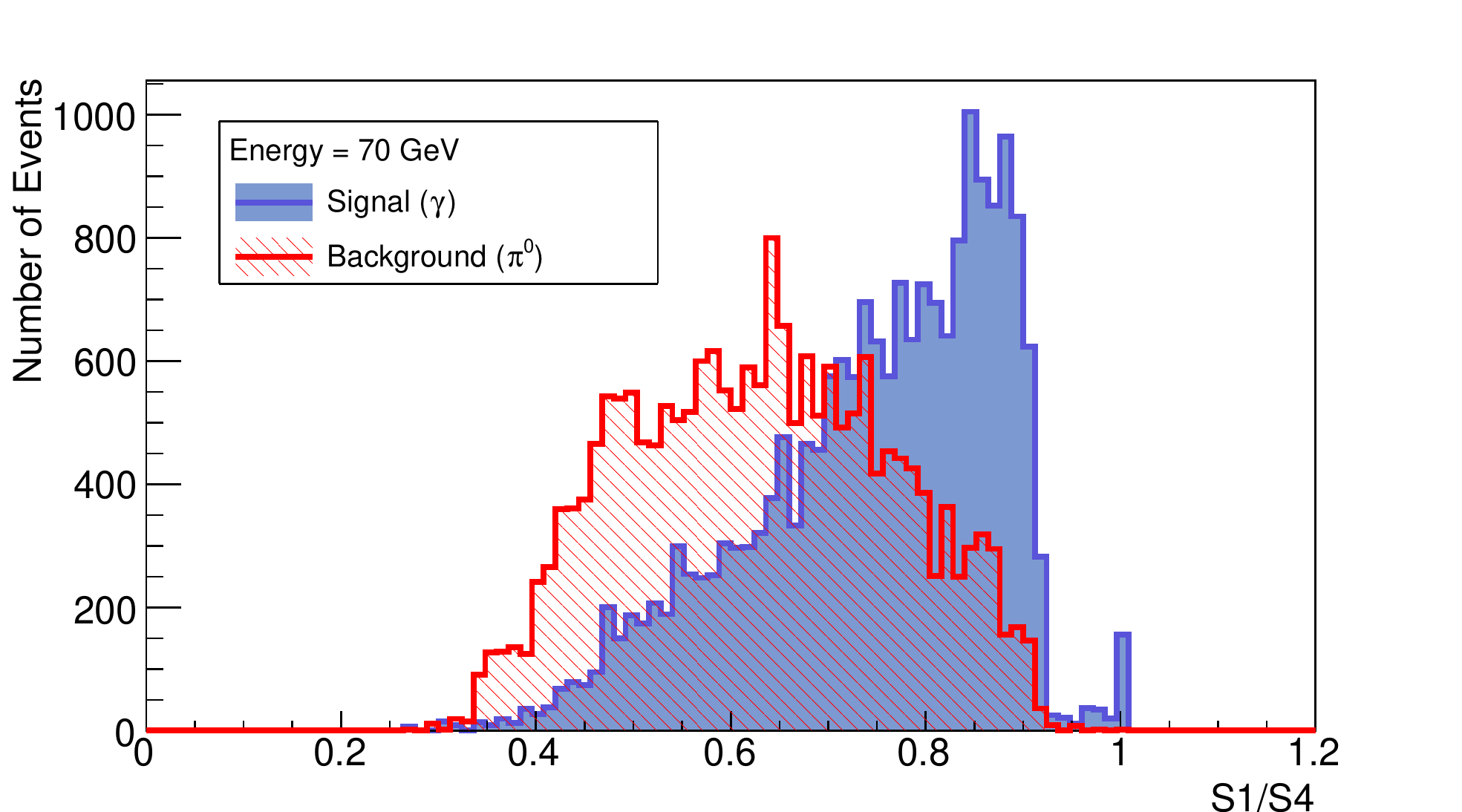}&
\includegraphics[width=0.45\textwidth]{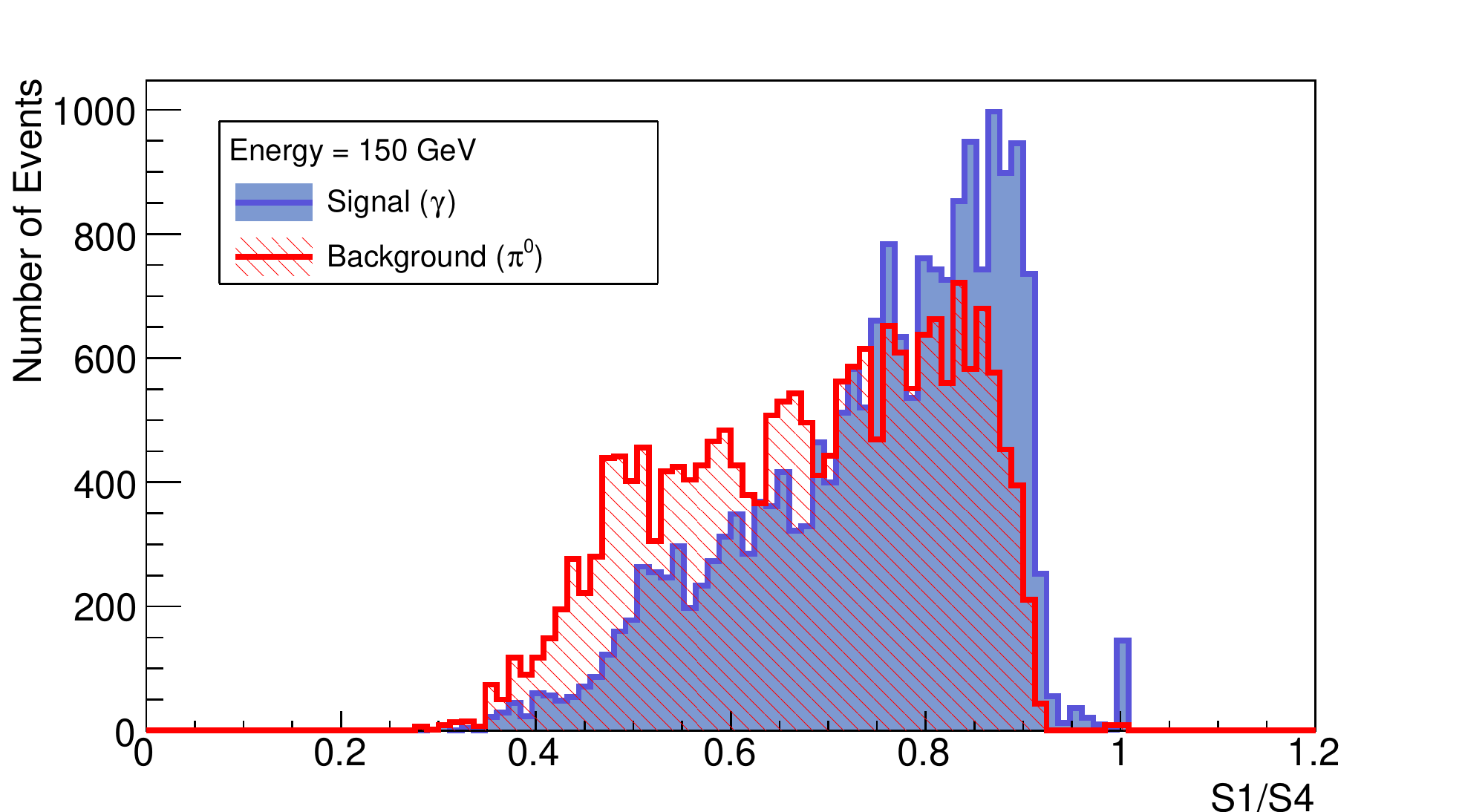}
  \end{tabular} \end{center}
\caption{The top left diagram shows the four possible combinations of 2$\times$2
         arrays which can be formed including the maximum energy 
         tower. The top right figure shows the distribution of S1/S4 variable 
         for 50 GeV photons and $\pi^{0}$'s. The bottom left and right plots
         correspond to 70 GeV and 150 GeV photons and $\pi^{0}$'s. The blue 
         hatched histogram is for photons and the red hatched histogram is 
         for $\pi^{0}$'s \cite{ref:confpaper}.}
\label{fig:s1s4_105070150}
\end{figure}

\item[2-D distribution of $F_{16}$ vs $F_{9}$:] The variables, $F_{9}$ and 
$F_{16}$, are defined through equation \ref{eqn:eqnf9f16}.
\begin{eqnarray}
\label{eqn:eqnf9f16}
F_{9}  & =& \frac{S9 - S1}{S9}\nonumber \\
F_{16} & =& \frac{S16 - S4}{S16}
\end{eqnarray}

where $S16$ is the energy deposited in the 4$\times$4 array of towers 
that is centered on the 2$\times$2 array of towers with the maximum energy, 
among the four possible combinations as explained above. 
Figure \ref{fig:F9F16_105070150} 
shows the diagrammatic view of 4$\times$4 array of towers  around the 
2$\times$2 array of towers and the 2-D distribution of F$_{16}$  and F$_{9}$
for 50, 70 and 150 GeV photons and $\pi^{0}$'s. 

\begin{figure}[htbp]
  \begin{center} \begin{tabular}{cc}
   \hspace{20pt}{\raisebox{15pt}{
       \includegraphics[width=4cm,height=3cm]{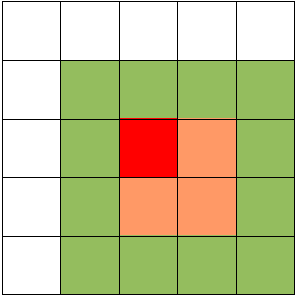}}} &
       \includegraphics[width=0.45\textwidth]{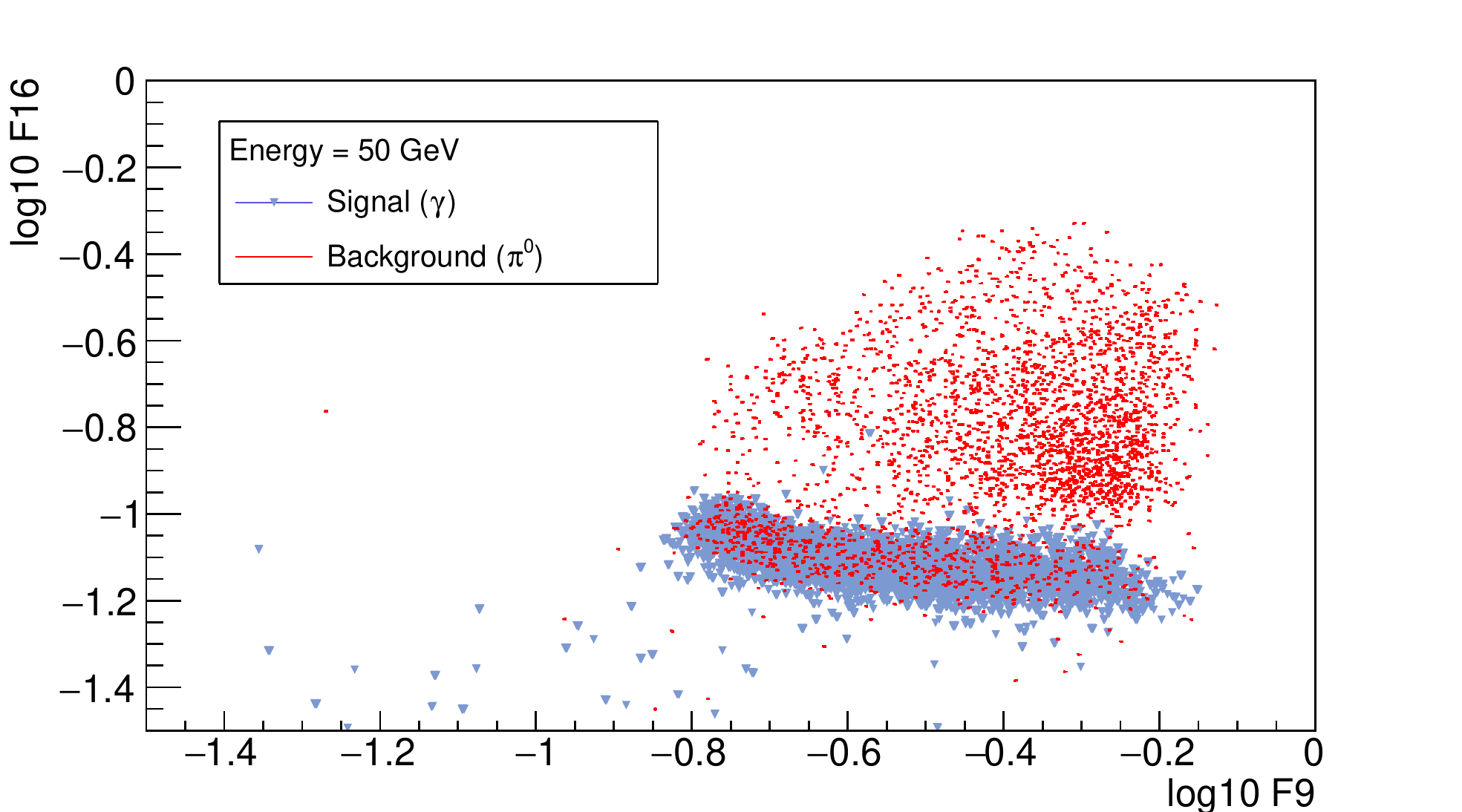}\\
       \includegraphics[width=0.45\textwidth]{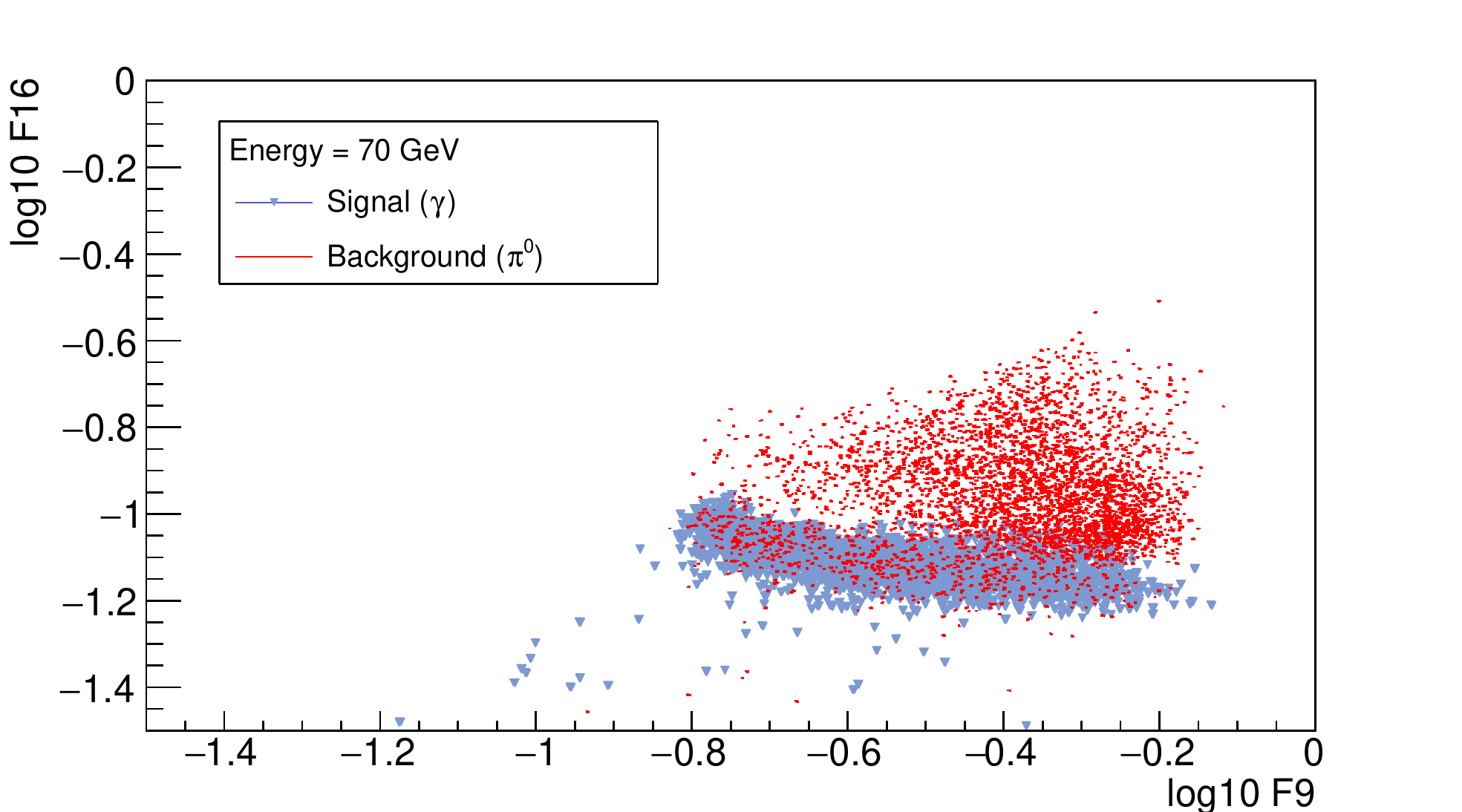}&
       \includegraphics[width=0.45\textwidth]{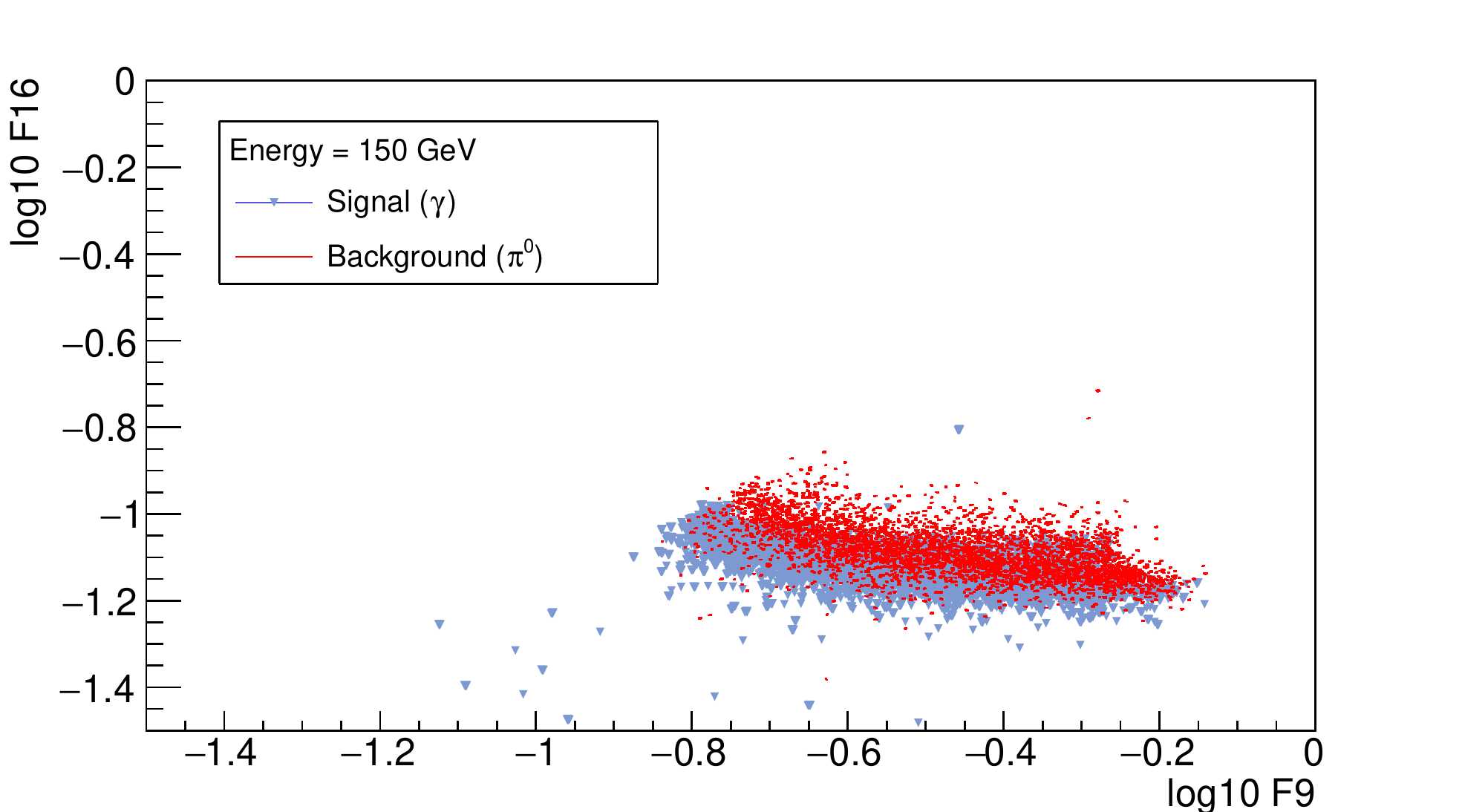}
  \end{tabular}\end{center}
\caption{The top left diagram shows the 4$\times$4 arrays of towers formed 
         around that array of 2$\times$2 towers which has maximum energy of 
         the four possible combinations as explained in the text. The 
         top-right plot shows the 2-D distribution of F$_{16}$ along the 
         Y axis and F$_{9}$ along the X axis for 50 GeV photons and $\pi^{0}$'s.
         Similar plots at 70 GeV and 150 GeV are at the bottom left and at
         the bottom right respectively. The blue points are for photons and the
         red points are for $\pi^{0}$'s \cite{ref:confpaper}.}
\label{fig:F9F16_105070150}
\end{figure}
\end{description}
The performance of the variables $S1/S9$, $S1/S4$ and $F_{16}$ vs $F_{9}$ are
summarized below:
\begin{itemize}
\item Shower shape variables $S1/S9$ and $S1/S4$ lose the sensitivity for 
$\gamma$/$\pi^{0}$ separation at energies above 70 GeV;
\item The 2-D distribution of $F_{16}$ vs $F_{9}$ performs better for 70 GeV 
$\gamma$/$\pi^{0}$ discrimination compared to $S1/S9$ and $S1/S4$. But again 
it loses power for discrimination at energies above 70 GeV. 
\end{itemize}

\subsection{Moment analysis} \label{sec:momentana}
\begin{figure}[htbp]
\centering
\includegraphics[width=0.8\textwidth]{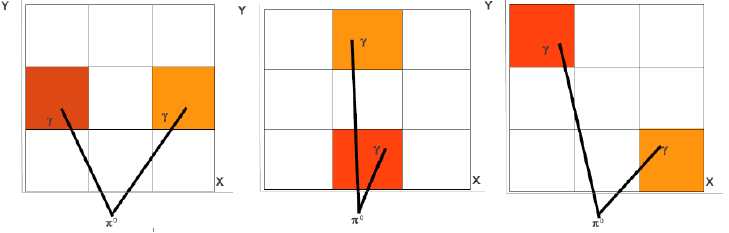}
\caption{Diagram showing three different topologies of a $\pi^{0}$ decay. }
\label{fig:phodecayplane}
\end{figure}

This analysis is based on the consideration that when the $\pi^{0}$ decays to 
two $\gamma$'s, 
the shower tends to be elliptical, as shown in Figure \ref{fig:phodecayplane}, 
whereas for a prompt photon the spread in X and Y will tend to be similar, 
because the shower spreads uniformly in all the directions. 
The above holds true for photons not converted before they reach the front 
face of the tower and when there is no magnetic field. Early conversion and
passage through magnetic field make the decay topology more complicated. 

The ratio of major axis of the ellipse
to its minor axis or vice-versa is
utilized to distinguish between photons and $\pi^{0}$'s. For $\pi^{0}$'s, 
the ratio between the two axes is expected to be away 
from 1, whereas, for photons, it is expected to be close to 1. 
To get the values of the length of both of the axis of the shower, 
a covariance matrix can be formed with the quantities as given in 
Equation \ref{eqn:covquant}.  The eigenvalues of this matrix is 
related to the length of the shower axis. 
The spread is calculated from the point of impact of the photon or the 
$\pi^{0}$. The point of impact is estimated using the COG method and
correcting for the S-shape as discussed in Section \ref{sec:position}.
The 2$\times$2 matrix is formed in each event as follows:

\begin{eqnarray*}
 M & = & \left| \begin{array}{cc}
                 \sigma^{2}_{x} &  \sigma^{2}_{xy}\\
                 \sigma^{2}_{xy} & \sigma^{2}_{y}\\
                \end{array} \right| 
\end{eqnarray*}
The definition of each of the terms of the above matrix is given below
\begin{eqnarray}
\label{eqn:covquant}
\sigma^{2}_{x} &= &\Sigma(x_{i}-\overline{x})^{2} \times w_{i}/\Sigma w_{i}\nonumber\\
\sigma^{2}_{y} &= &\Sigma(y_{i}-\overline{y})^{2} \times w_{i}/\Sigma w_{i}\nonumber\\
\sigma^{2}_{xy} &= &\Sigma(x_{i}-\overline{x}) \times (y_{i}-\overline{y}) \times w_{i}/\Sigma w_{i}, 
\end{eqnarray}
where $\overline{x}$ and $\overline{y}$ are the S-shape 
corrected Center of Gravity (COG) positions of the shower in X and Y, 
and $w_{i}$ are the weights associated with each contribution. 
This matrix is then diagonalized to extract the eigenvalues, $\lambda_{+}$ and 
$\lambda_{-}$. The ratio of eigenvalues of this matrix, 
$\lambda_{-}$/$\lambda_{+}$ or $\lambda_{+}$/$\lambda_{-}$ measures the 
eccentricity of the energy distribution and is used to discriminate direct
$\gamma$ from $\pi^0$.

There are two ways in which the weights, $w_{i}$ are formed:
\begin{description}
\item[Linear weights, $w_{i}$:]
Here the weights, $w_{i}$ are given by Equation \ref{eqn:linearweight0}:
\begin{eqnarray}
\label{eqn:linearweight0}
x_{meas}  &= &\Sigma x_{i} \times w_{i},\nonumber\\
y_{meas}  &= &\Sigma y_{i} \times w_{i},\nonumber\\
w_{i} &= &E_{i}/\Sigma E_{i}.
\end{eqnarray}
\item[Logarithmic weights, $w_{i}$:]
In this case, the weights, $w_{i}$ are given by the Equation 
\ref{eqn:logweight0}
\begin{eqnarray}
\label{eqn:logweight0}
x_{meas}  &= &\Sigma x_{i} \times w_{i}/\Sigma w_{i},\nonumber\\
y_{meas}  &= &\Sigma y_{i} \times w_{i}/\Sigma w_{i},\nonumber\\
w_{i} &= &Max(0, w_{0} + \ln(E_{i}/E_{T})), 
\end{eqnarray}
where $w_{0}$ is related to the threshold for the towers below which the towers 
are not included in the sum, and $E_{T}$ is the total energy deposited in the 
array.
\end{description}
There are two ways in which the covariance matrix can be constructed:
\begin{enumerate}
\item use combined information of all four fibers in a given tower [Coarse grain information];
\item use information from individual fiber in a given tower [Fine grain information].  
\end{enumerate}

\subsubsection{Coarse grain Information} \label{sec:result_momana_crys}
\begin{figure}[htbp]
\centering
\begin{tabular}{cc}
\includegraphics[width=0.45\textwidth]{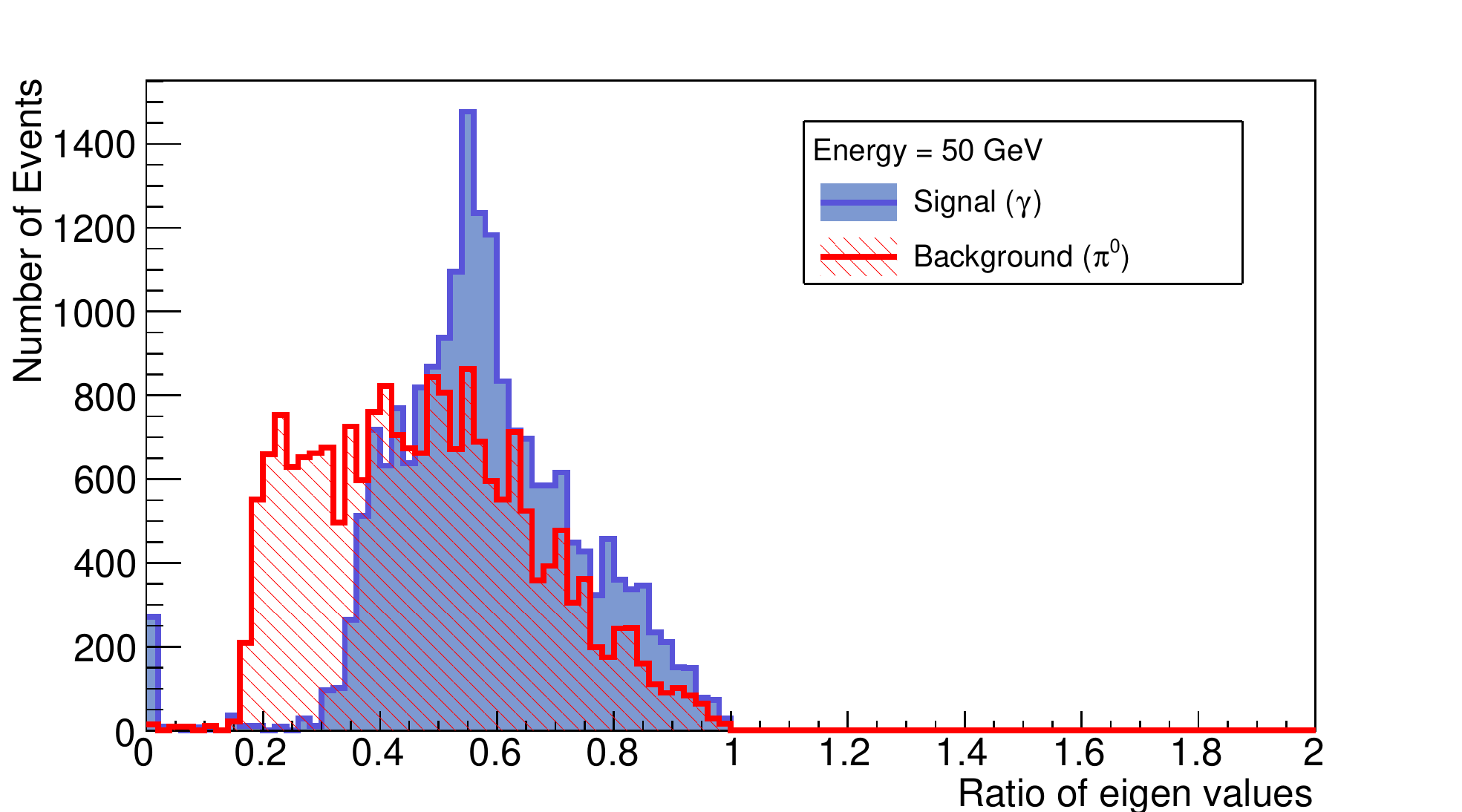}&
\includegraphics[width=0.45\textwidth]{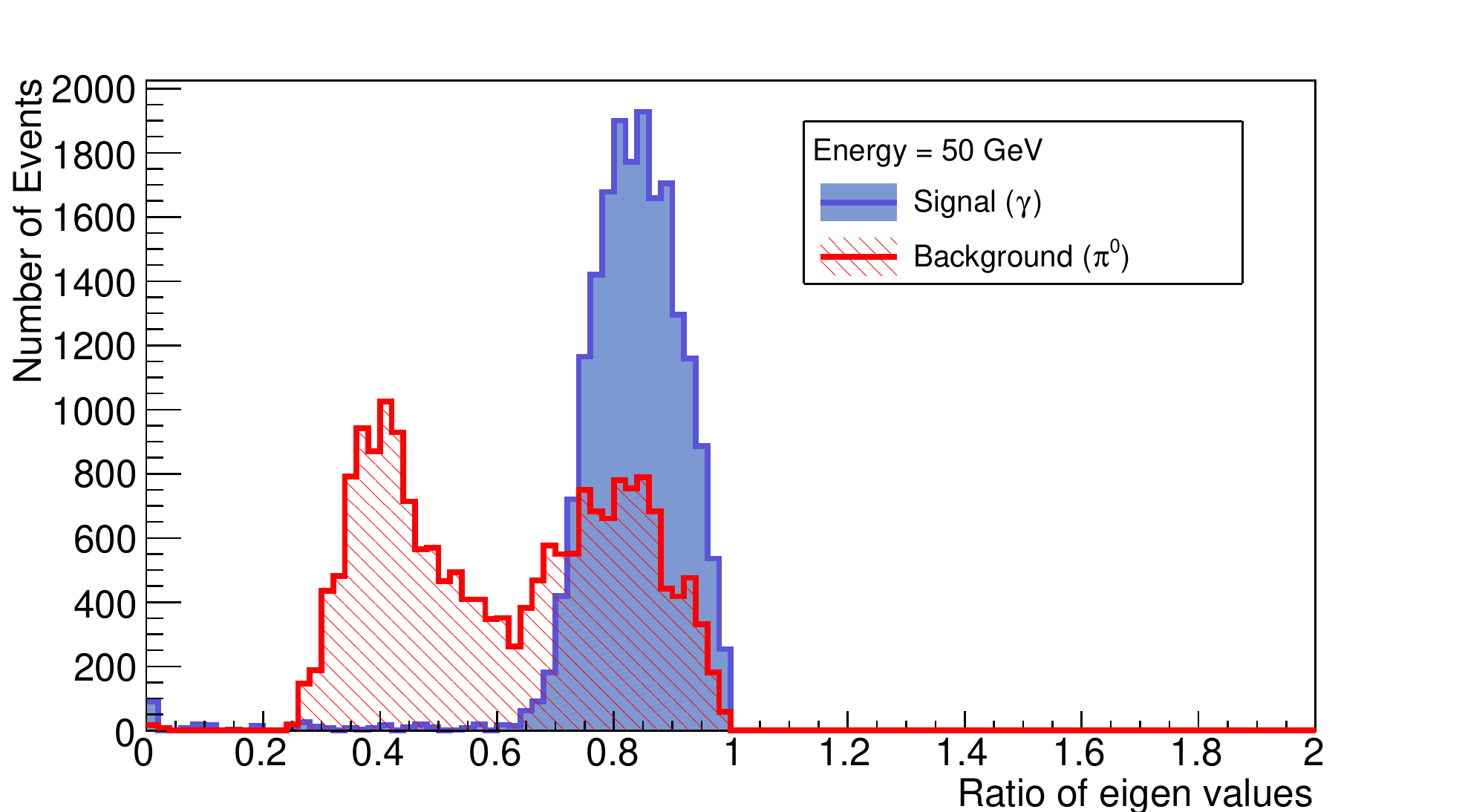}\\
\end{tabular}
\caption{Distribution of $\lambda_{-}$/$\lambda_{+}$ for photons and $\pi^{0}$'s
         of 50 GeV for the case of coarse grain information (no information
         from individual fibers used). The plot on the left is done with 
         linear weights for determining the impact point while the plot
         on the right is done with the log weights.}
\label{fig:ratioplots_50}
\end{figure}

\begin{figure}[htbp]
\centering
\begin{tabular}{cc}
\includegraphics[width=0.45\textwidth]{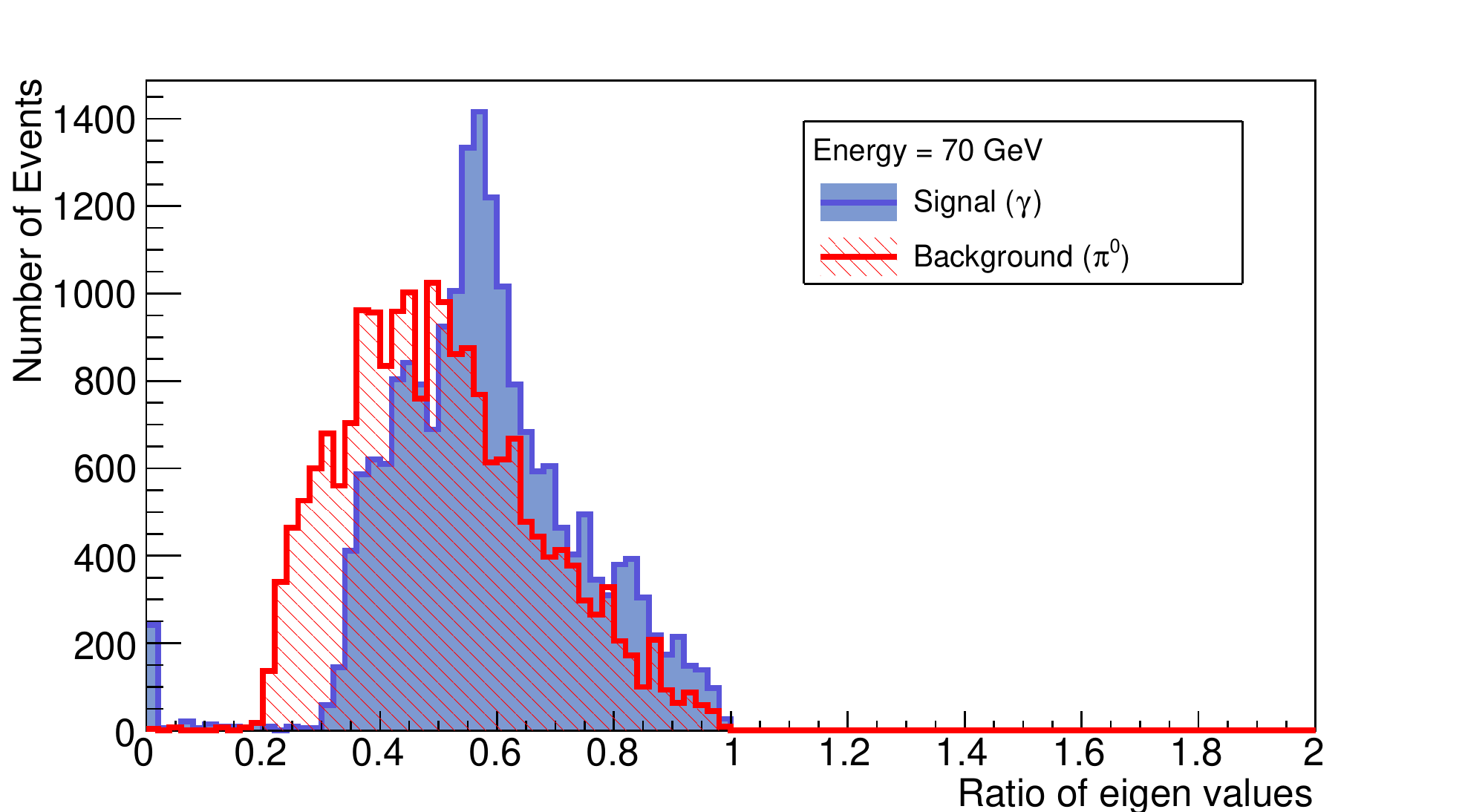}&
\includegraphics[width=0.45\textwidth]{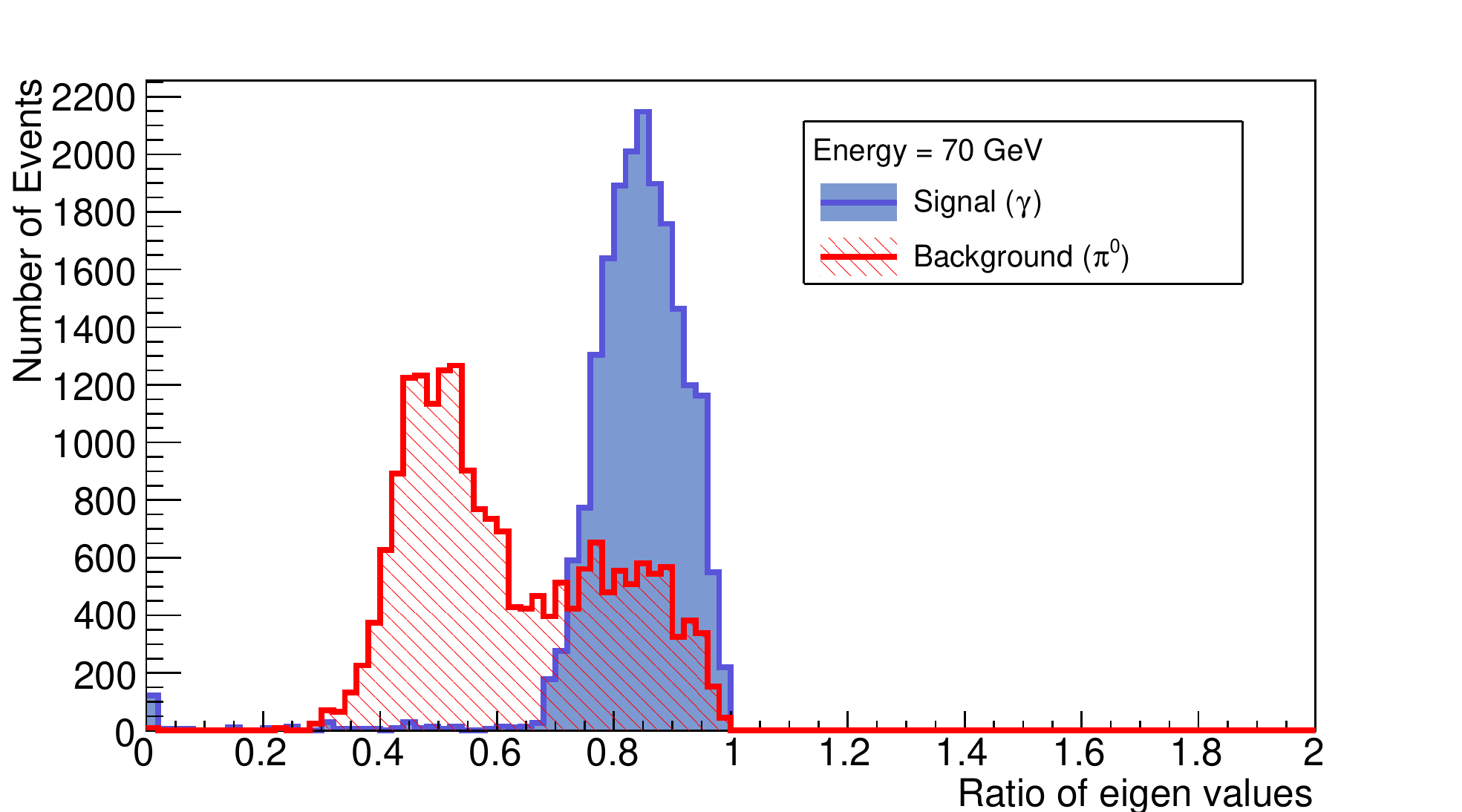}
\end{tabular}
\caption{Distribution of $\lambda_{-}$/$\lambda_{+}$ for photons and $\pi^{0}$'s 
         of 70 GeV for the case of coarse grain information (no information
         from individual fibers used). The plot on the left is done with 
         linear weights for determining the impact point while the plot on
         the right utilizes log weights.}
\label{fig:ratioplots_70}
\end{figure}

\begin{figure}[htbp]
\centering
\begin{tabular}{cc}
\includegraphics[width=0.45\textwidth]{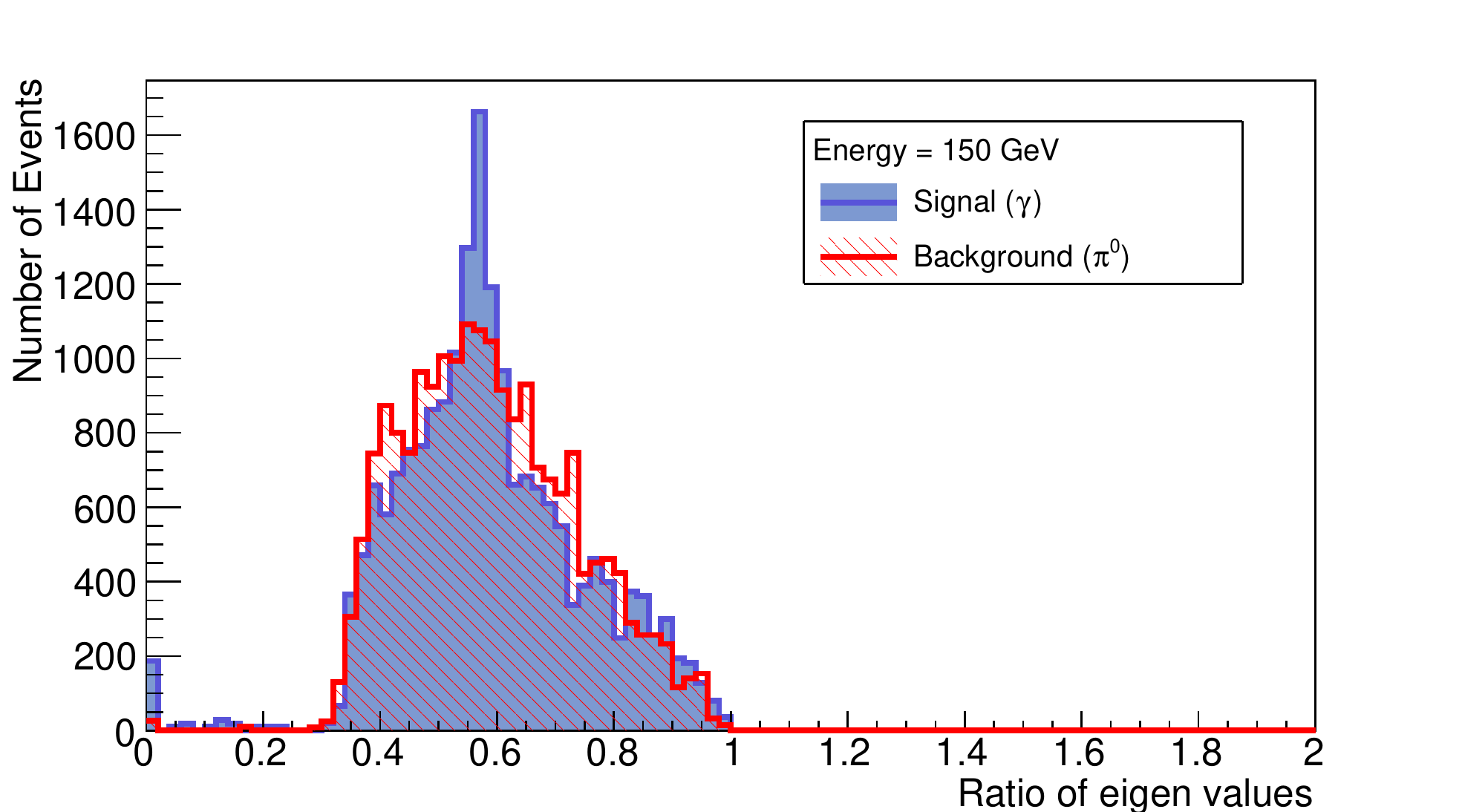}&
\includegraphics[width=0.45\textwidth]{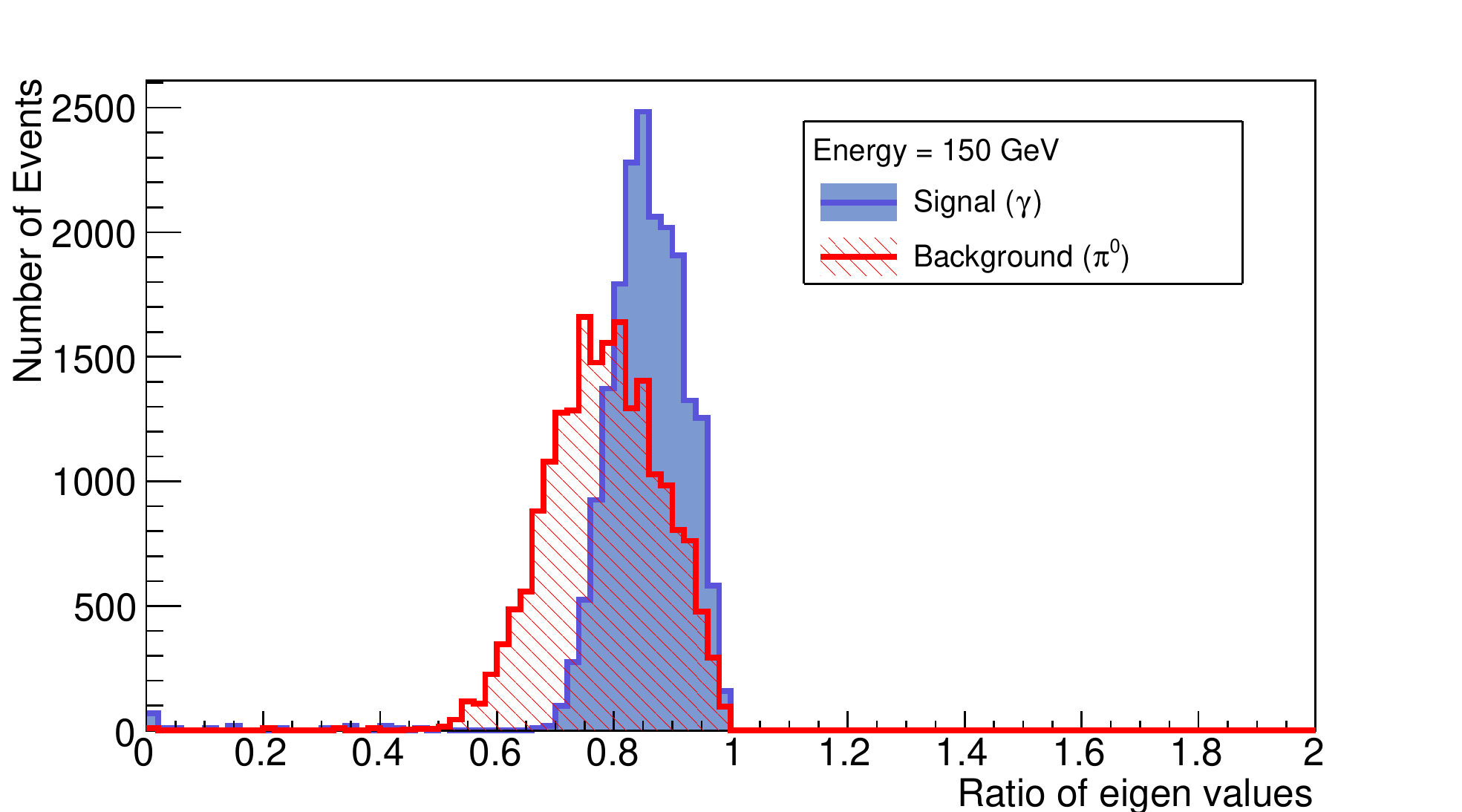}\\
\end{tabular}
\caption{Distribution of $\lambda_{-}$/$\lambda_{+}$ for photons and $\pi^{0}$'s
         of 150 GeV for the case of coarse grain information (no information
         from individual fibers used). The plot on the left is done with 
         linear weights for determining the impact point while the plot on
         the right utilizes log weights.}
\label{fig:ratioplots_150}
\end{figure}

In this case,  E$_{i}$ refers to the total energy recorded by $i^{th}$ tower 
(i=1-9 for $3\times 3$ array) and $x_{i}$ and $y_{i}$ are the X and Y coordinates 
of the center of $i^{th}$ tower. E$_{i}$ is obtained by summing the energy recorded 
from all four fibers of  the $i^{th}$ tower.
Both linear and log weights are used in the determination of the impact point.
In case of log weights, $w_0$ is set to be 4.7.
Figures \ref{fig:ratioplots_50}, \ref{fig:ratioplots_70} and 
\ref{fig:ratioplots_150} show the ratio of $\lambda_{-}$/$\lambda_{+}$ for
photons and $\pi^{0}$'s at 50 GeV, 70 GeV and 150 GeV. These plots clearly
demonstrate that log weights improve the discriminating power between photons 
and $\pi^{0}$'s over linear weights. 
%
\subsubsection{Fine grain information} \label{sec:result_momana_fiber}
\begin{figure}[htbp]
\centering
\begin{tabular}{cc}
\includegraphics[width=0.45\textwidth]{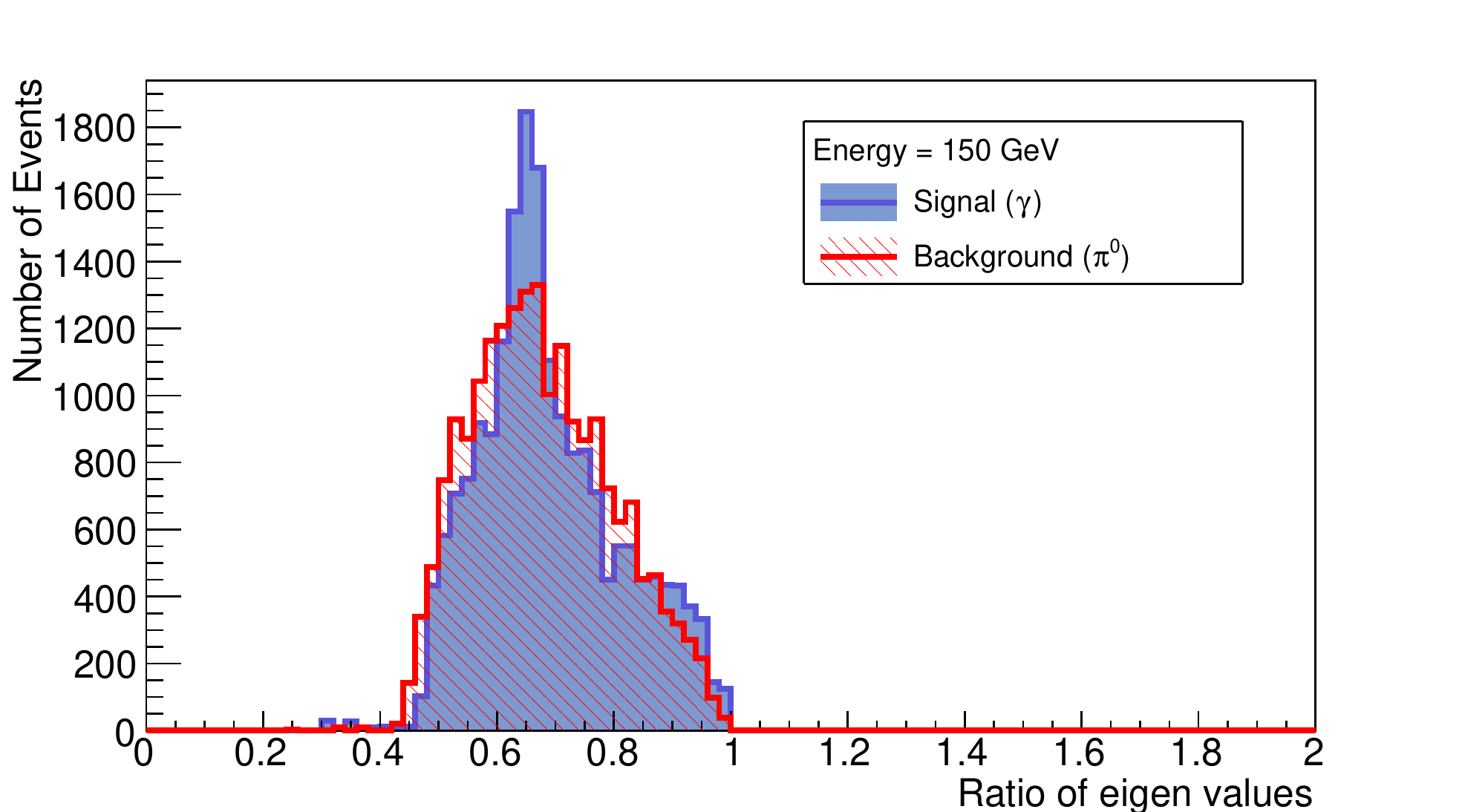}&
\includegraphics[width=0.45\textwidth]{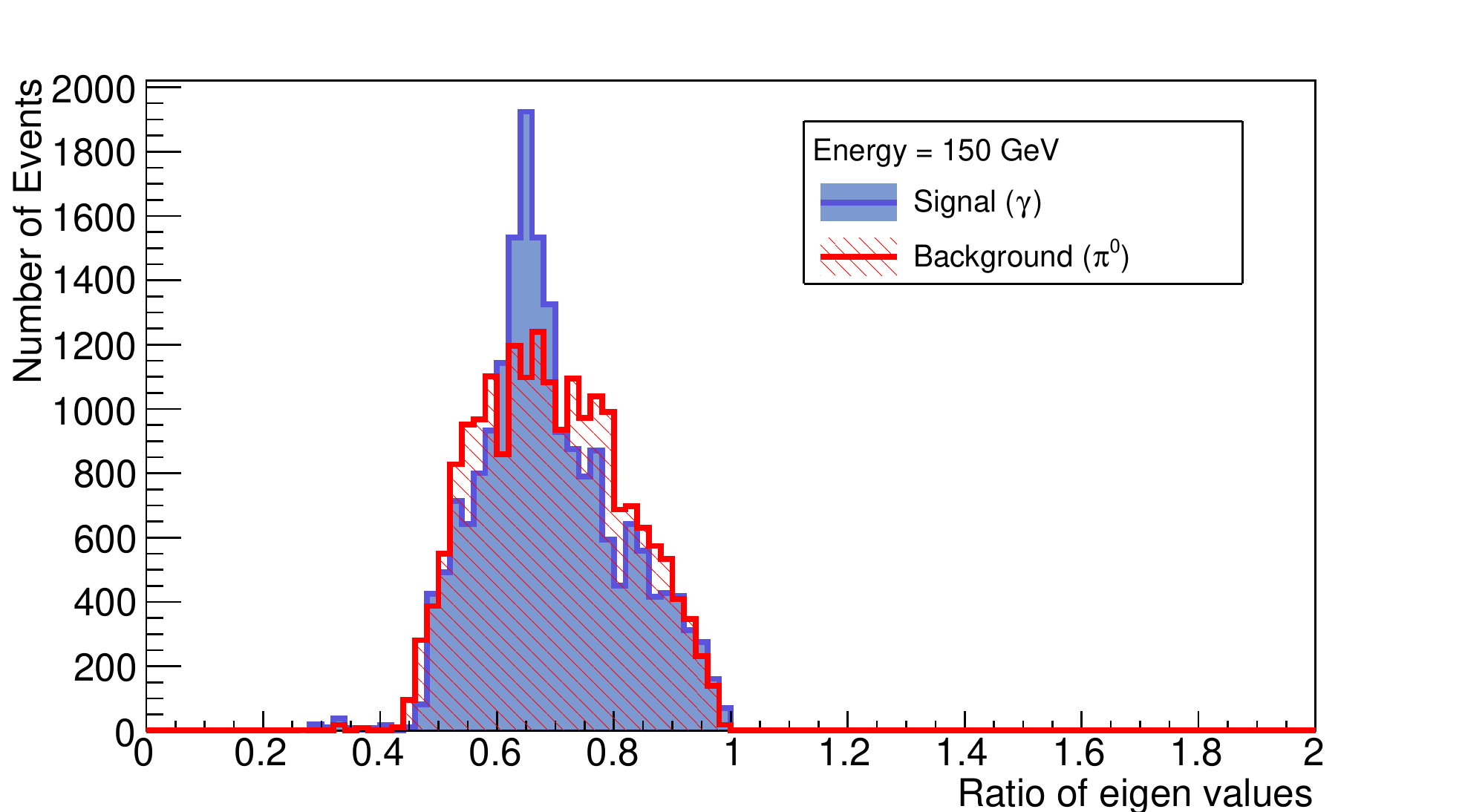}\\
\end{tabular}
\caption{Distribution of $\lambda_{-}$/$\lambda_{+}$ for photons and $\pi^{0}$'s
         of 150 GeV for the case of fine grain information (information
         from individual fibers used).  The plot on the left is done with 
         linear weights for determining the impact point while the plot on
         the right utilizes log weights.}
\label{fig:fratioplots_150}
\end{figure}

For fine grain information E$_{i}$ refers to the energy recorded by $i^{th}$ 
individual fiber (i=1-36 for $3\times 3$ array) and $x_{i}$ and $y_{i}$ are 
the X and Y coordinates of the $i^{th}$ fiber.
The impact point is determined with both linear weights and log weights.
\begin{figure}[htbp]
\centering
\begin{tabular}{cc}
\includegraphics[width=0.45\textwidth]{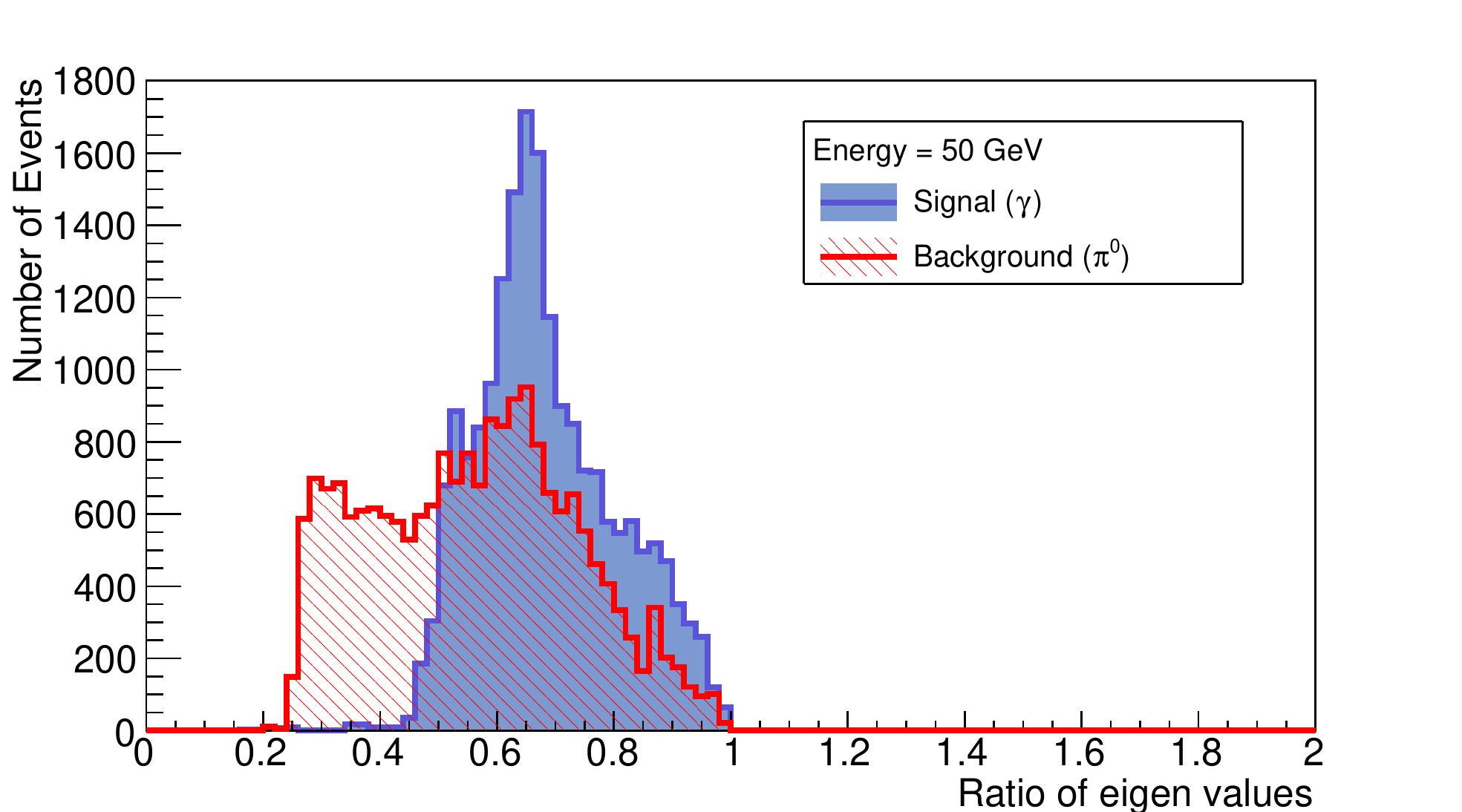}&
\includegraphics[width=0.45\textwidth]{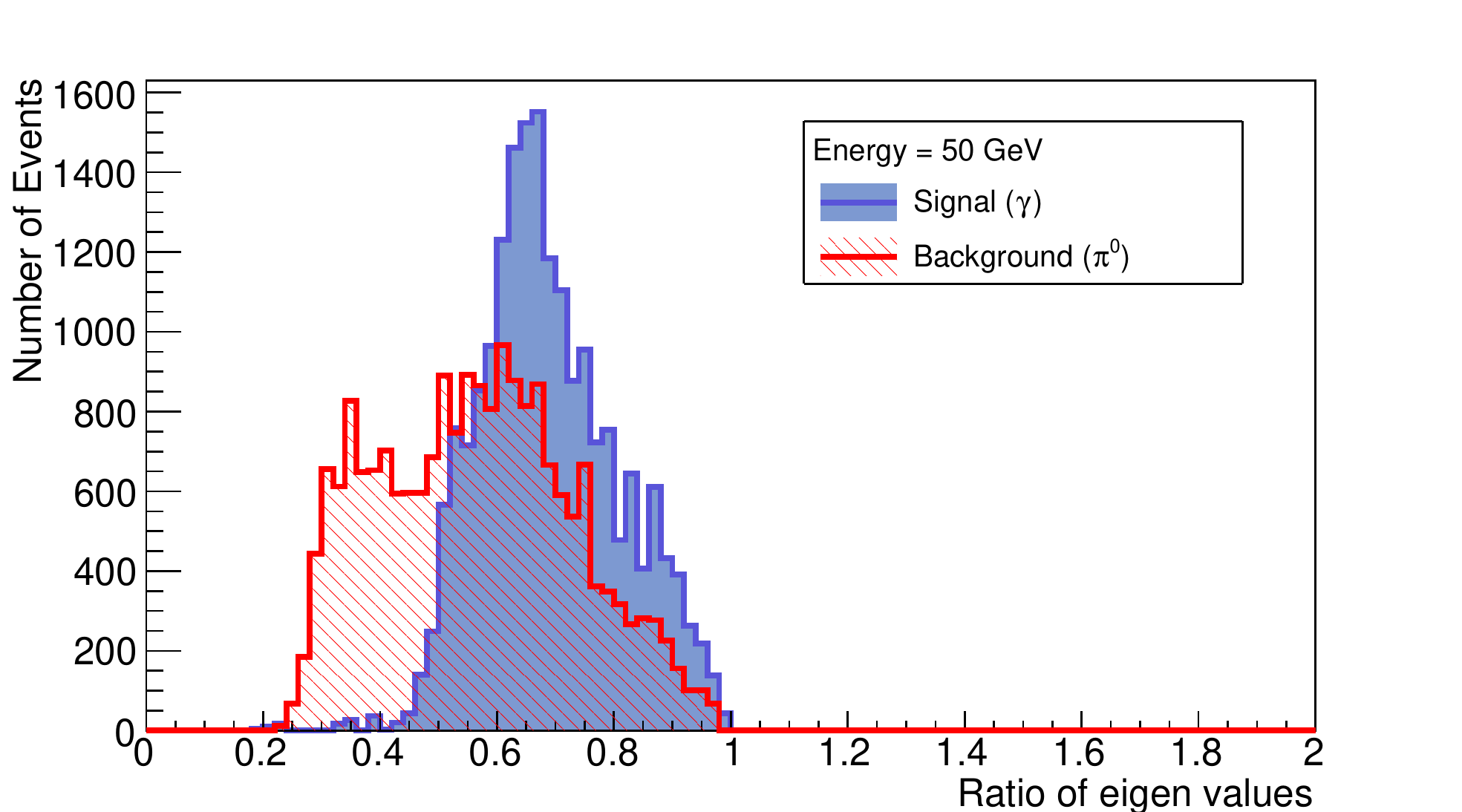}\\
\end{tabular}
\caption{Distribution of $\lambda_{-}$/$\lambda_{+}$ for photons and $\pi^{0}$'s
         of 50 GeV for the case of fine grain information (information
         from individual fibers used). The plot on the left is done with 
         linear weights for determining the impact point while the plot on
         the right utilizes log weights.}
\label{fig:fratioplots_50}
\end{figure}
\begin{figure}[htbp]
\centering
\begin{tabular}{cc}
\includegraphics[width=0.45\textwidth]{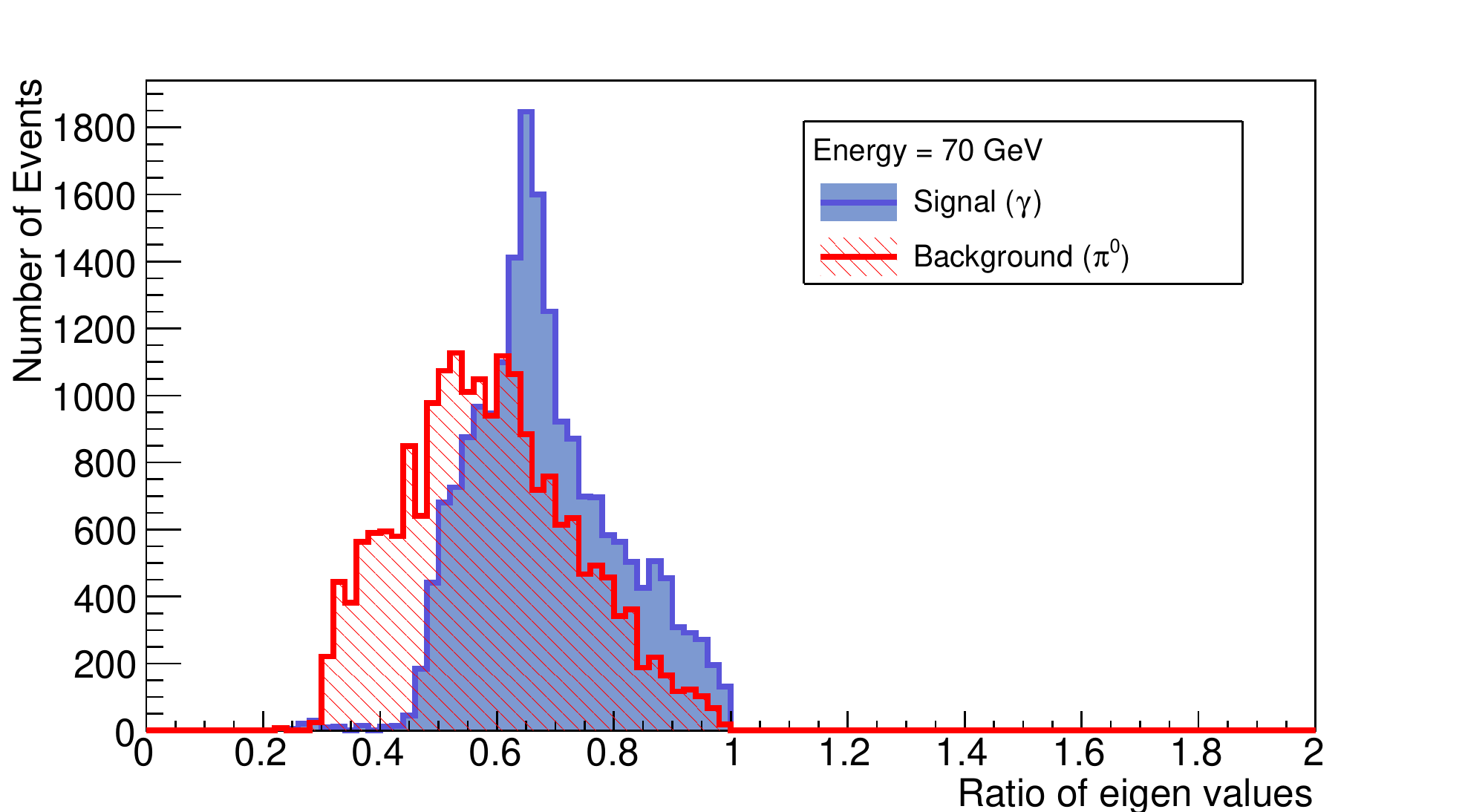}&
\includegraphics[width=0.45\textwidth]{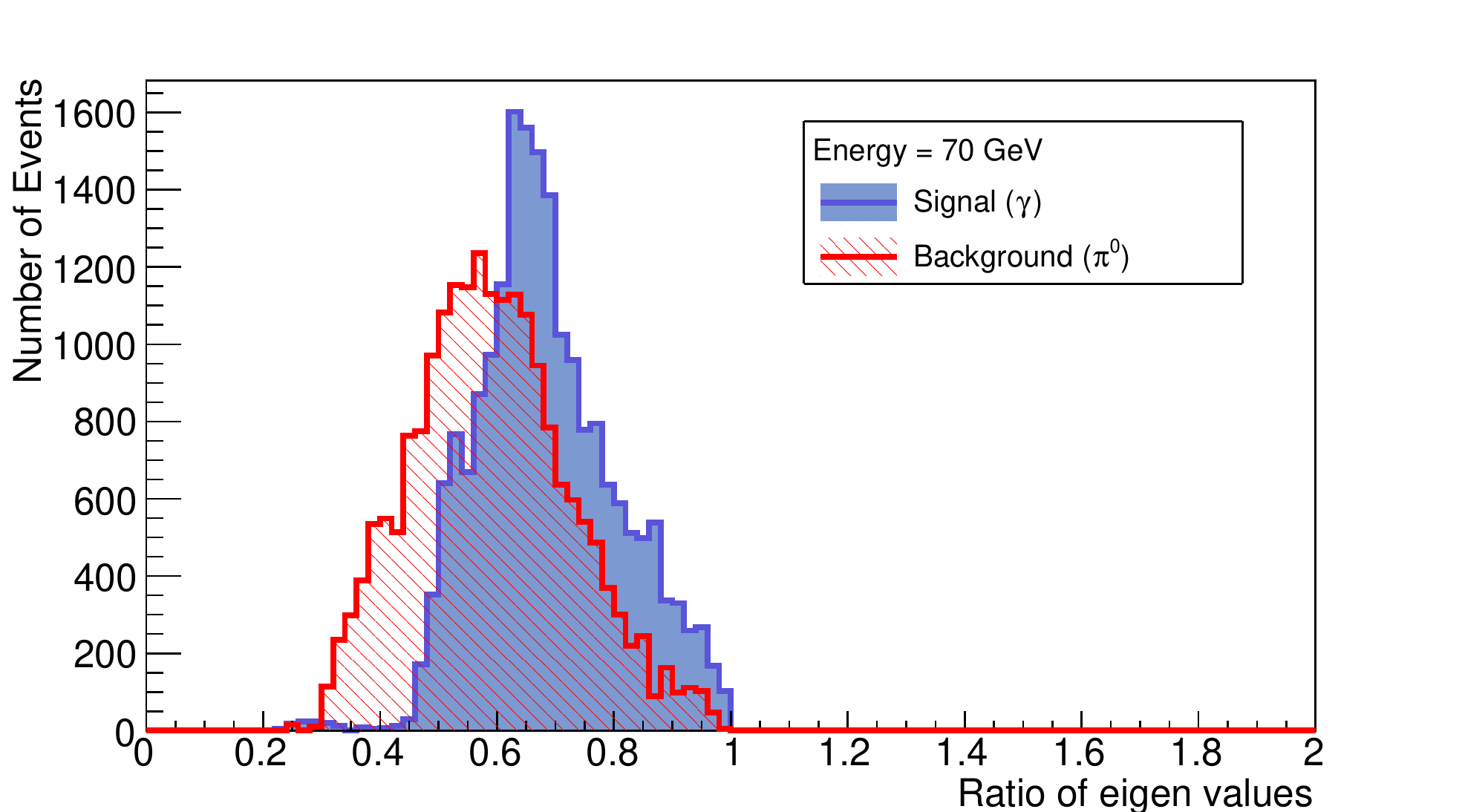}\\
\end{tabular}
\caption{Distribution of $\lambda_{-}$/$\lambda_{+}$ for photons and $\pi^{0}$'s
         of 70 GeV for the case of fine grain information (information
         from individual fibers used). The plot on the left is done with 
         linear weights for determining the impact point while the plot on
         the right utilizes log weights.}
\label{fig:fratioplots_70}
\end{figure}
A comparison between the two methods using photons and $\pi^{0}$'s of 50 GeV, 
70 GeV and 150 GeV indicates a higher discriminating power of the log weights.
By comparing the set of plots with fine grain information in Figure \ref{fig:fratioplots_50}, \ref{fig:fratioplots_70} and \ref{fig:fratioplots_150} 
with the corresponding plots for coarse grain information,  
it can be seen that not much additional sensitivity is added in the 
discriminating power by using information from individual fibers. 
Beyond 150 GeV, both the ways (coarse grain information and fine grain 
information) fail to discriminate between photons and $\pi^{0}$'s.
\subsection{Study using Multivariate Analysis (MVA)}
\label{sec:tmvaana}
As it has been described in the previous Sections, \ref{sec:showershapes}, 
\ref{sec:result_momana_crys} and \ref{sec:result_momana_fiber}, the 
discriminating power is reduced significantly for $\pi^{0}$'s of energy 
above $100$ GeV. An analysis has been carried out exploring the discriminating 
power gained by employing multivariate techniques to the problem of separating 
$\pi^0$s from photons using the spatial pattern of energy deposition 
in the ECAL. In this analysis, the classification 
problem is to separate $\pi^{0}$'s from prompt photons. The following MVA 
classifiers are examined in this analysis: 
\begin{enumerate}
\item Boosted Decision Tree (BDT)
\item Gradient Boosted Decision Trees (GBDT)
\item Artificial Neural Network (ANN)
\end{enumerate}
In this analysis, the TMVA \cite{bib:tmva} package within ROOT \cite{bib:root}  
is used. 
Each MVA is trained separately on a sample of photons and $\pi^{0}$'s. 
Energy from each individual tower in 3$\times$3 array, or energy 
from each fiber in the 3$\times$3 array is fed into the MVA. 
This analysis is done using photons and $\pi^{0}$'s at 200 GeV.
Two types of samples are produced:
\begin{description}
\item[Fixed gun sample:] These are produced with
the gun position fixed at (0 mm, 4 cm) in $(x,y)$, with z at 3.2~m.
\item[Random gun sample:]  
In this case, the gun positions are uniformly distributed both in
X and Y direction between $-7$ mm and $+7$ mm i.e. within the central tower.
\end{description}
\subsubsection{Training and testing of the MVA using fixed gun samples} \label{sec:fixedgunana}
\begin{figure}[!h]
\begin{center} \begin{tabular}{cc}
   \includegraphics[width=0.45\textwidth]{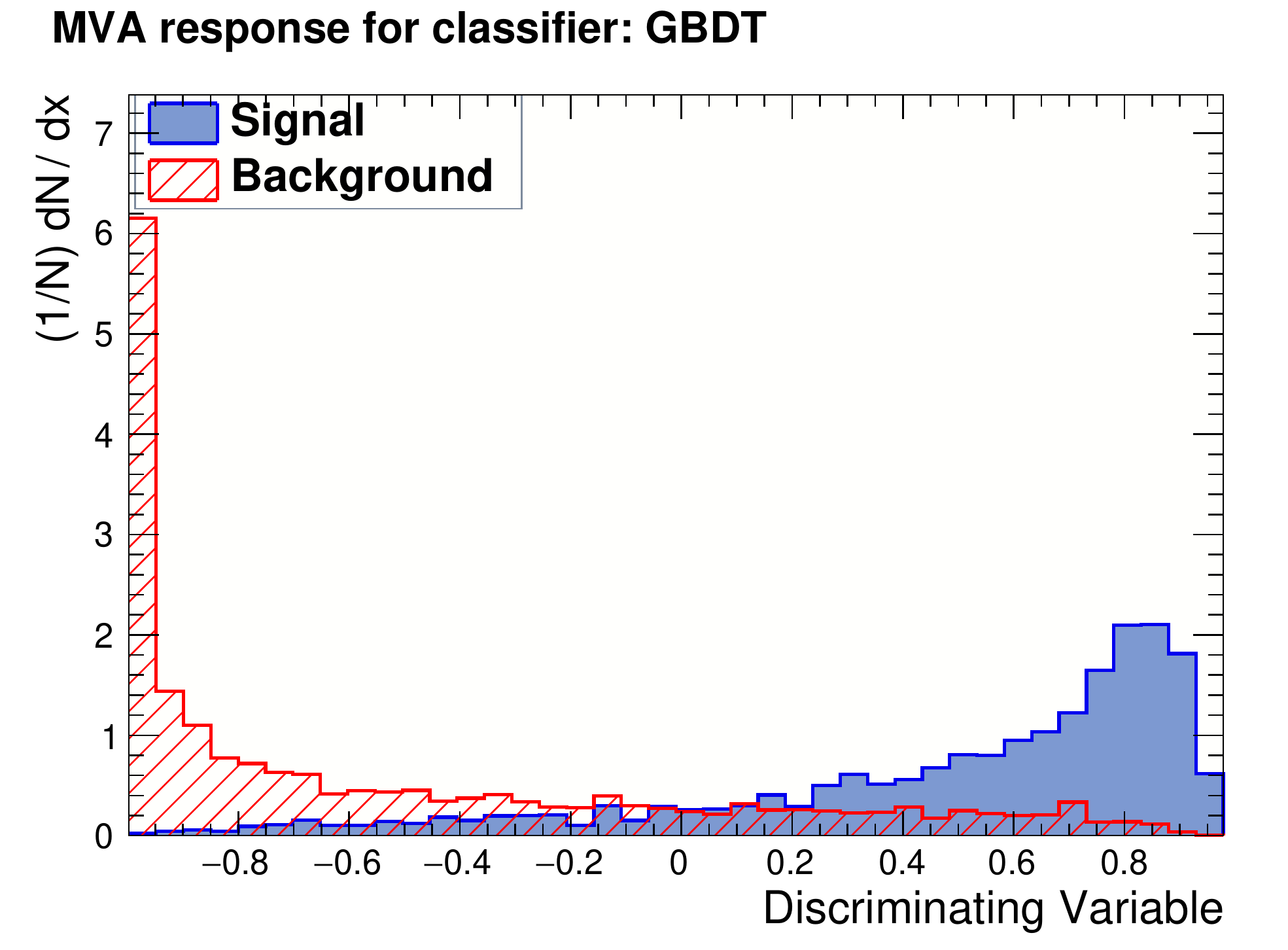}&
   \includegraphics[width=0.45\textwidth]{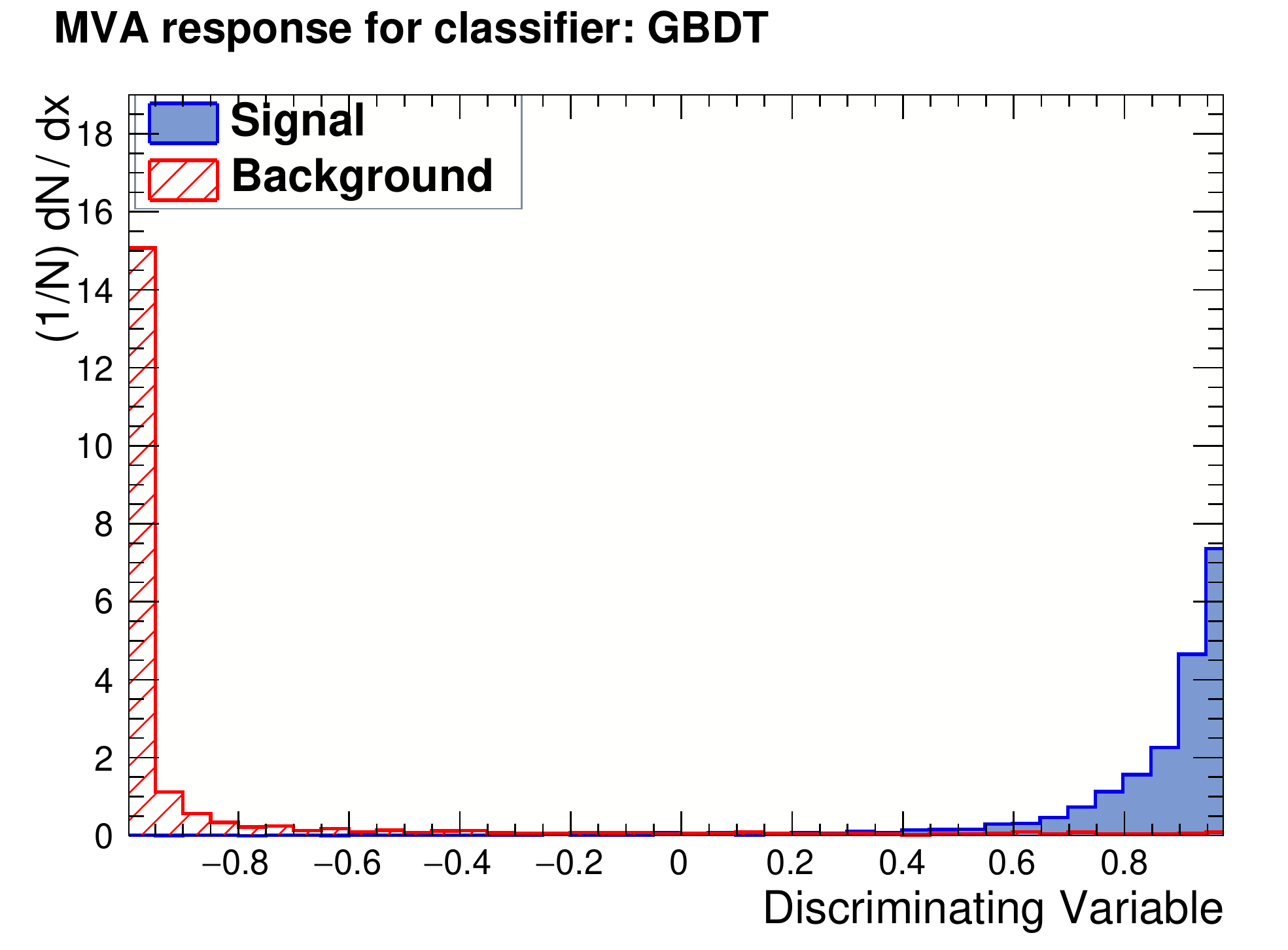}\\
   \includegraphics[width=0.45\textwidth]{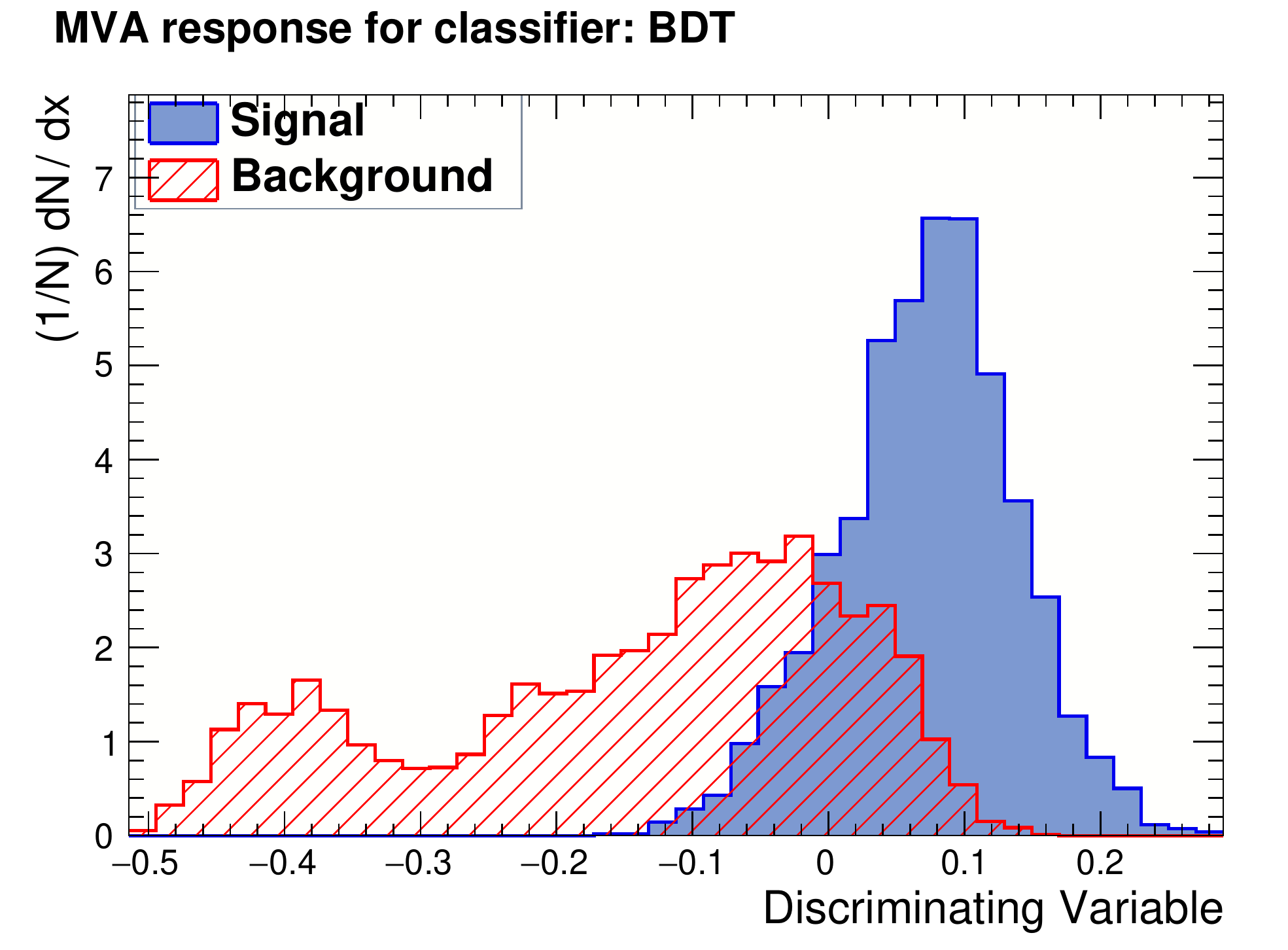}&
   \includegraphics[width=0.45\textwidth]{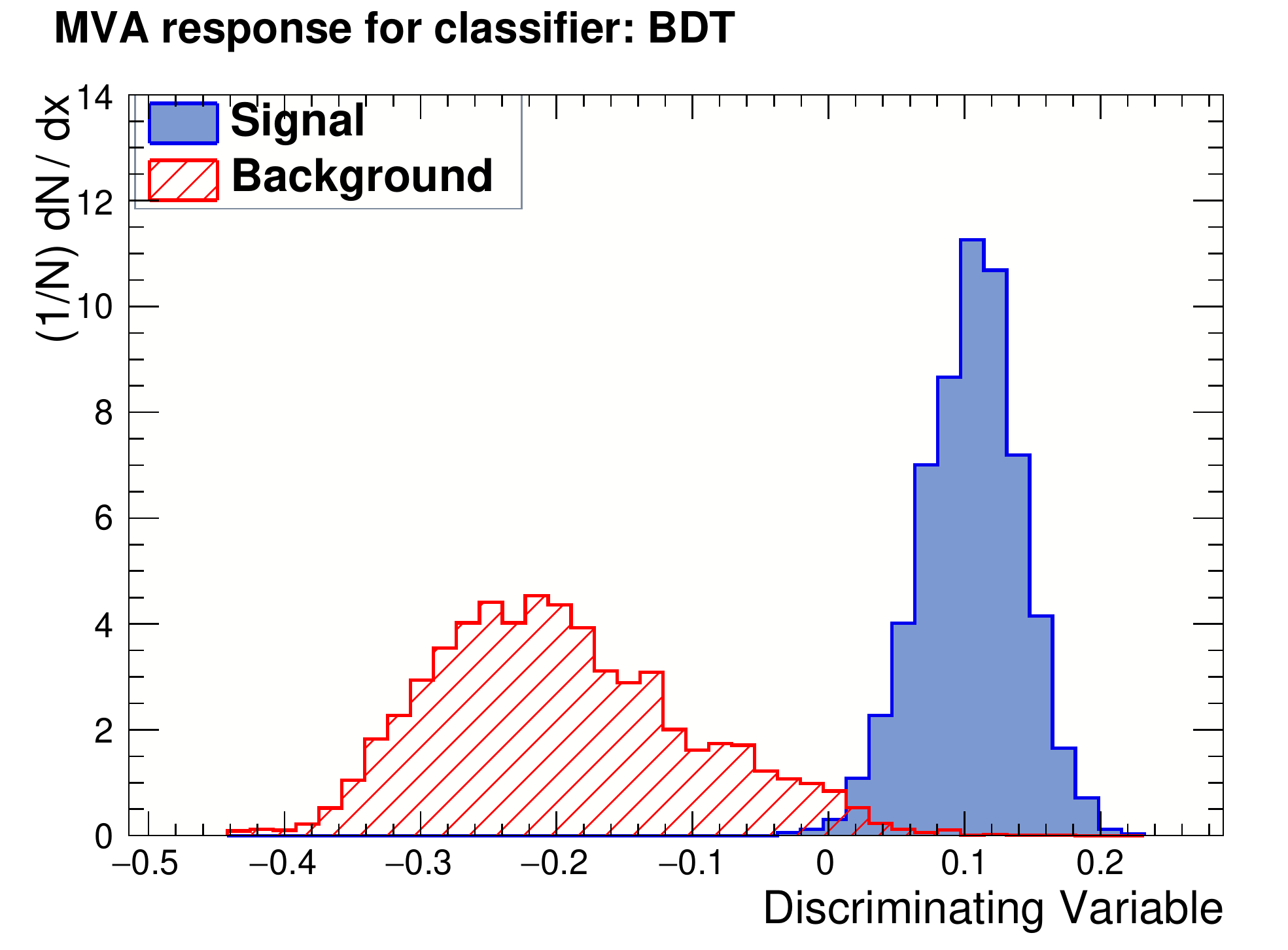}\\
   \includegraphics[width=0.45\textwidth]{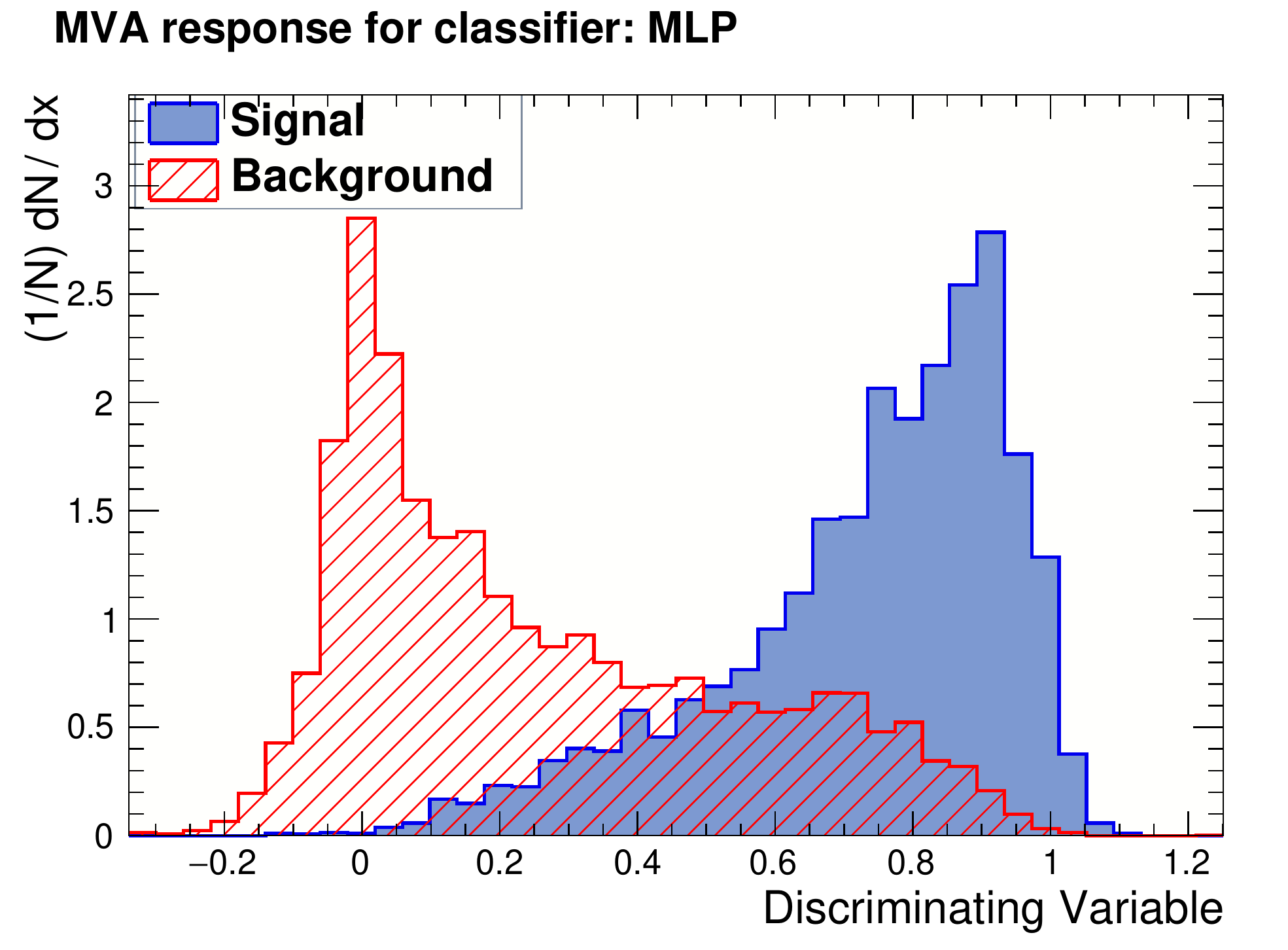}&
   \includegraphics[width=0.45\textwidth]{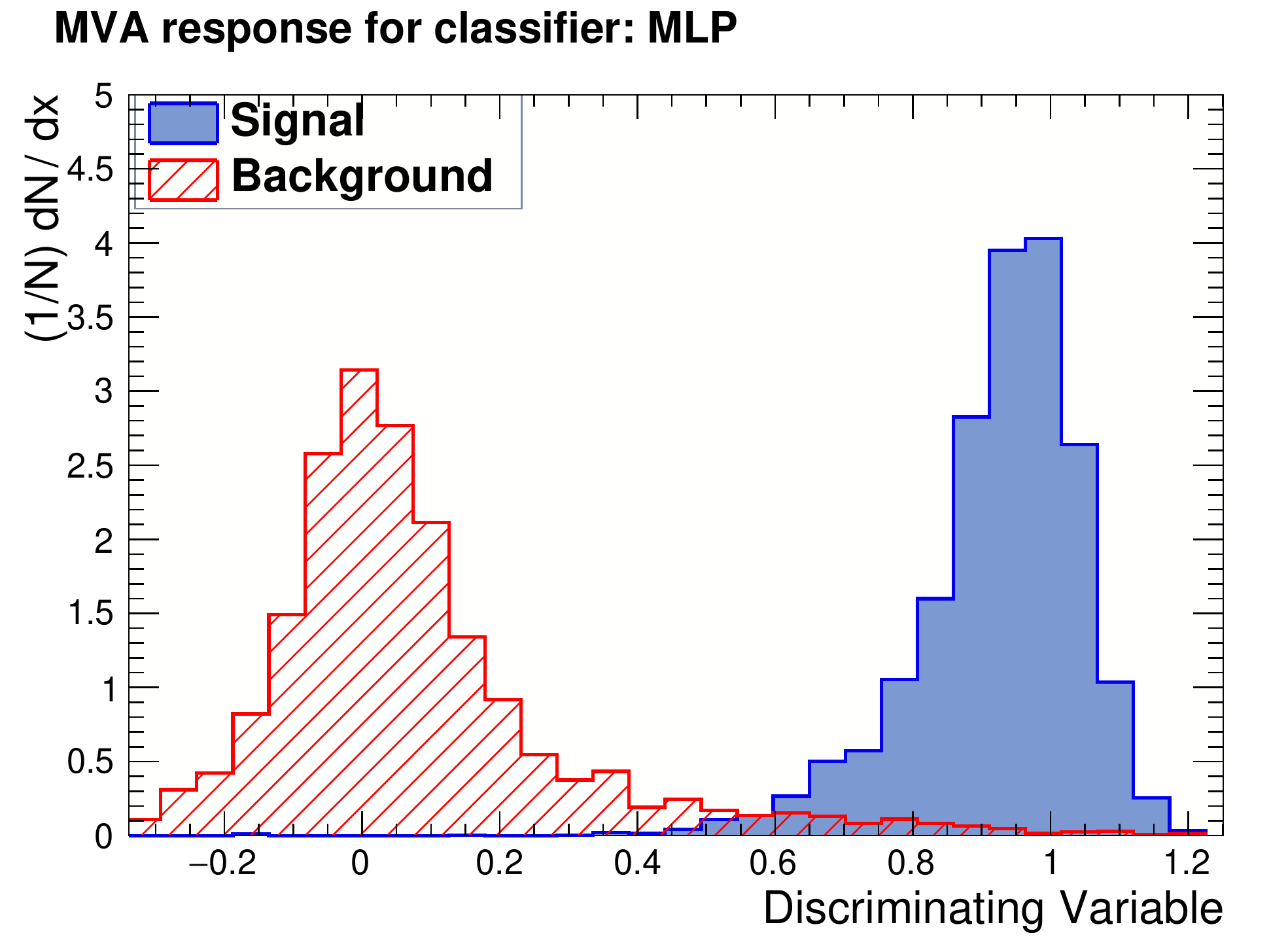}
\end{tabular} \end{center}
\caption{The output response of different MVA methods for fixed gun sample. The figures on the left
         are for coarse grain information and those on the right refer to
         fine grain information.}
\label{fig:fixed:outputMvaRes}
\end{figure}

\begin{figure}[!h]
  \begin{center} \begin{tabular}{cc}
    \includegraphics[width=0.45\textwidth]{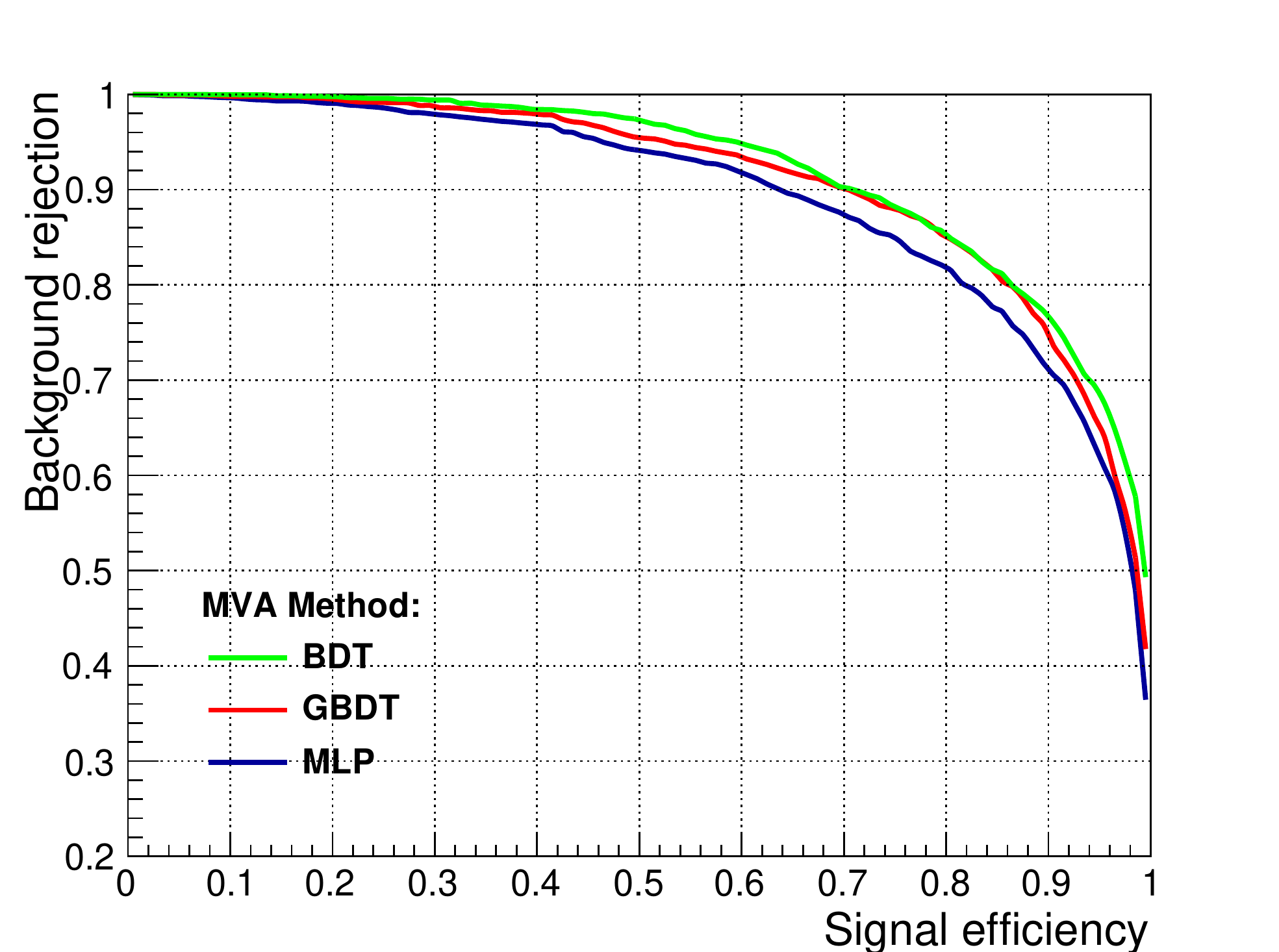} & 
    \includegraphics[width=0.45\textwidth]{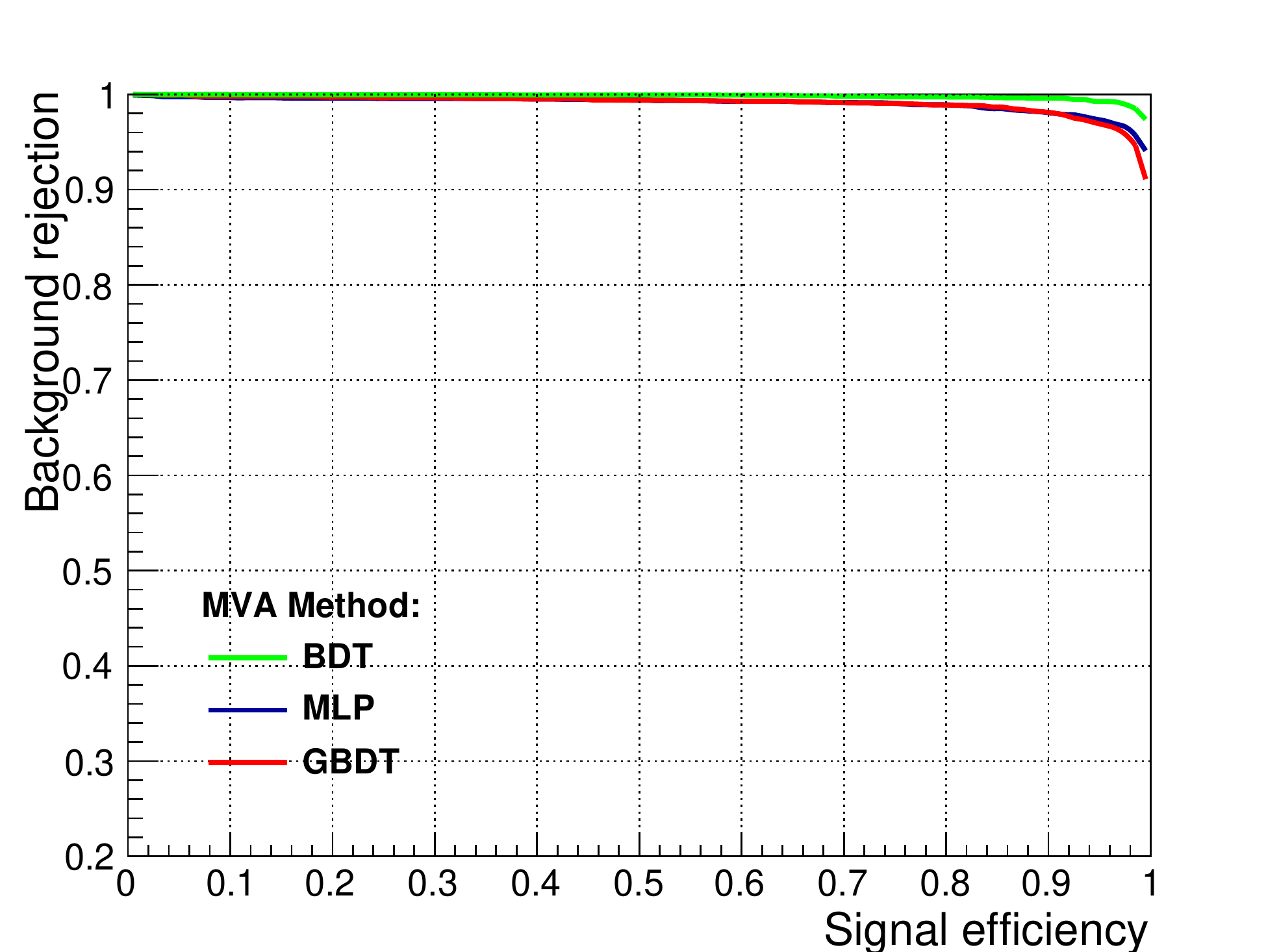}\\
\end{tabular} \end{center}
\caption{The background rejection versus the 
        signal efficiency curve for various MVA methods using fixed gun sample. Left figure for 
        coarse grain information and the figure on the right for fine grain information \cite{ref:confpaper}. }
\label{fig:fixed:outputBrejvsS}
\end{figure}

The MVA is trained using 20000 events from fixed gun samples of 200 GeV 
photons and $\pi^{0}$'s. The following two sets of training variables are used 
separately to train MVA:
\begin{description}
\item[Coarse grain information:] Input to MVA is the ratio of energy from each tower in the
3$\times$3 array to total energy in 3$\times$3 array. 
\item[Fine grain information:] Input to MVA is the ratio of energy from each individual fiber 
in 3$\times$3 array to total energy in 3$\times$3 array.
\end{description}
These energies are scaled to the total energy collected in the 3$\times$3 array.
Figure \ref{fig:fixed:outputMvaRes} shows the output response of the different MVA 
classifier for the case of both coarse grain and fine grain information.
Figure \ref{fig:fixed:outputBrejvsS} shows the background 
rejection versus signal efficiency curve for the case of coarse grain as well as fine grain information. 
It can be seen from the Figures \ref{fig:fixed:outputMvaRes} and 
\ref{fig:fixed:outputBrejvsS}, that the fine grain information is better for discriminating signal from from background.
\subsubsection{Training and testing of the MVA using random gun samples}
In a realistic scenario a particle can hit anywhere on the face of a tower. 
Keeping this in mind a training is done on a sample of 20000 photons and $\pi^0$s of energy 200 GeV 
from random gun sources.
The hit positions of the photons and $\pi^0$'s are uniformly distributed 
in X and Y direction between $-7$ mm and $+7$ mm.
Two different samples are used to train the MVA:
\begin{figure}[htbp]
\begin{center} \begin{tabular}{cc}
   \includegraphics[width=0.45\textwidth]{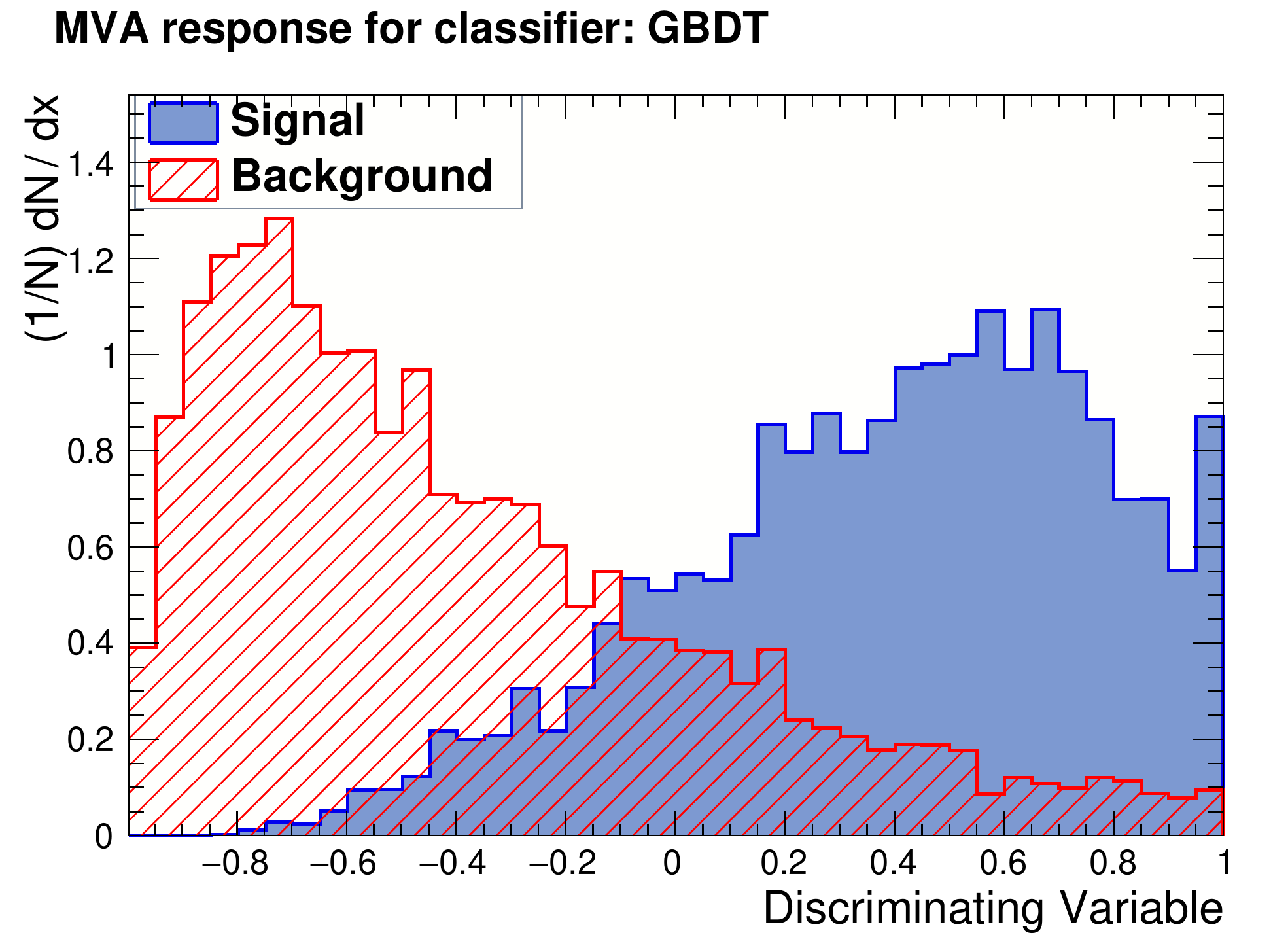}&
   \includegraphics[width=0.45\textwidth]{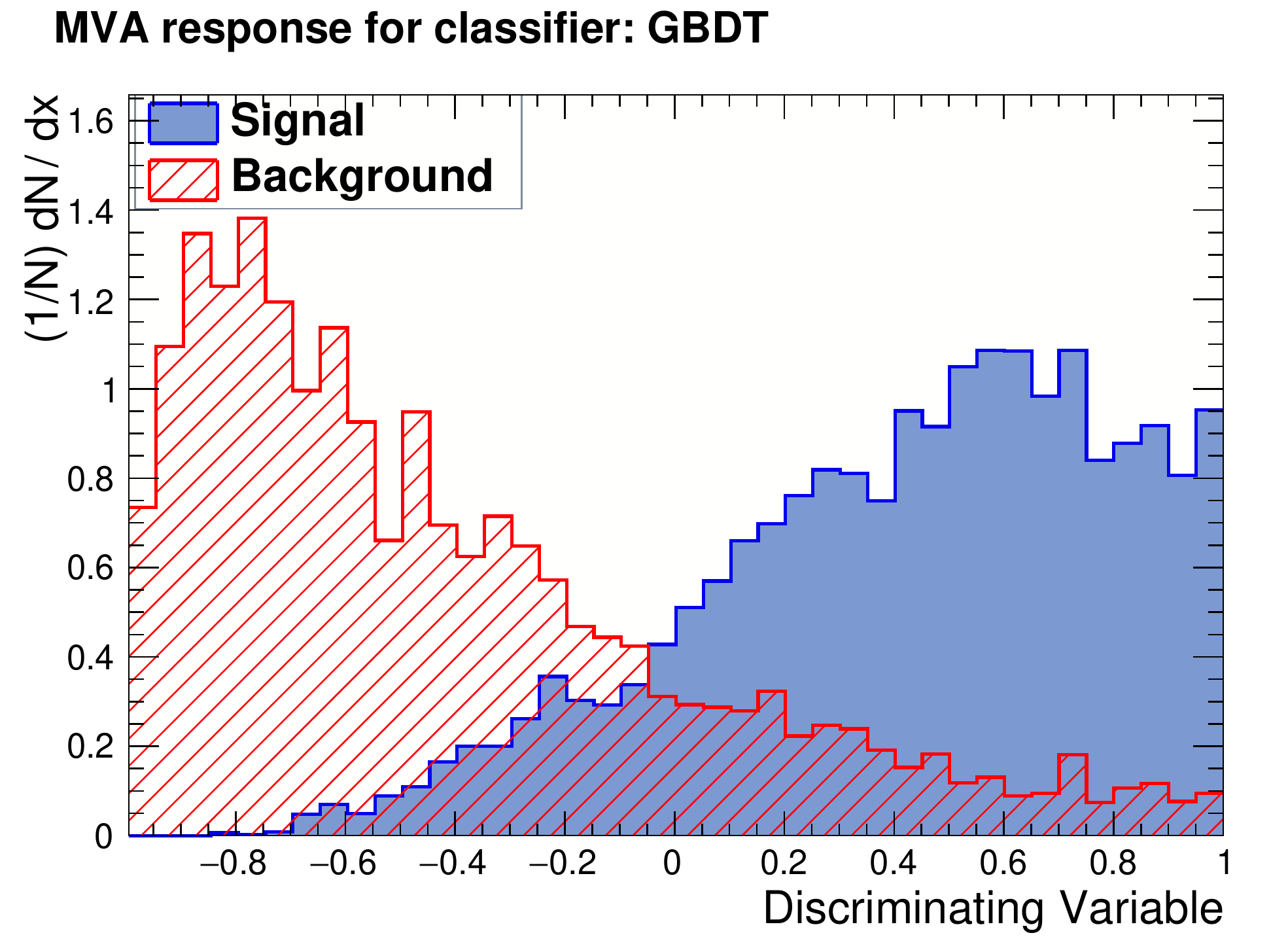}\\
   \includegraphics[width=0.45\textwidth]{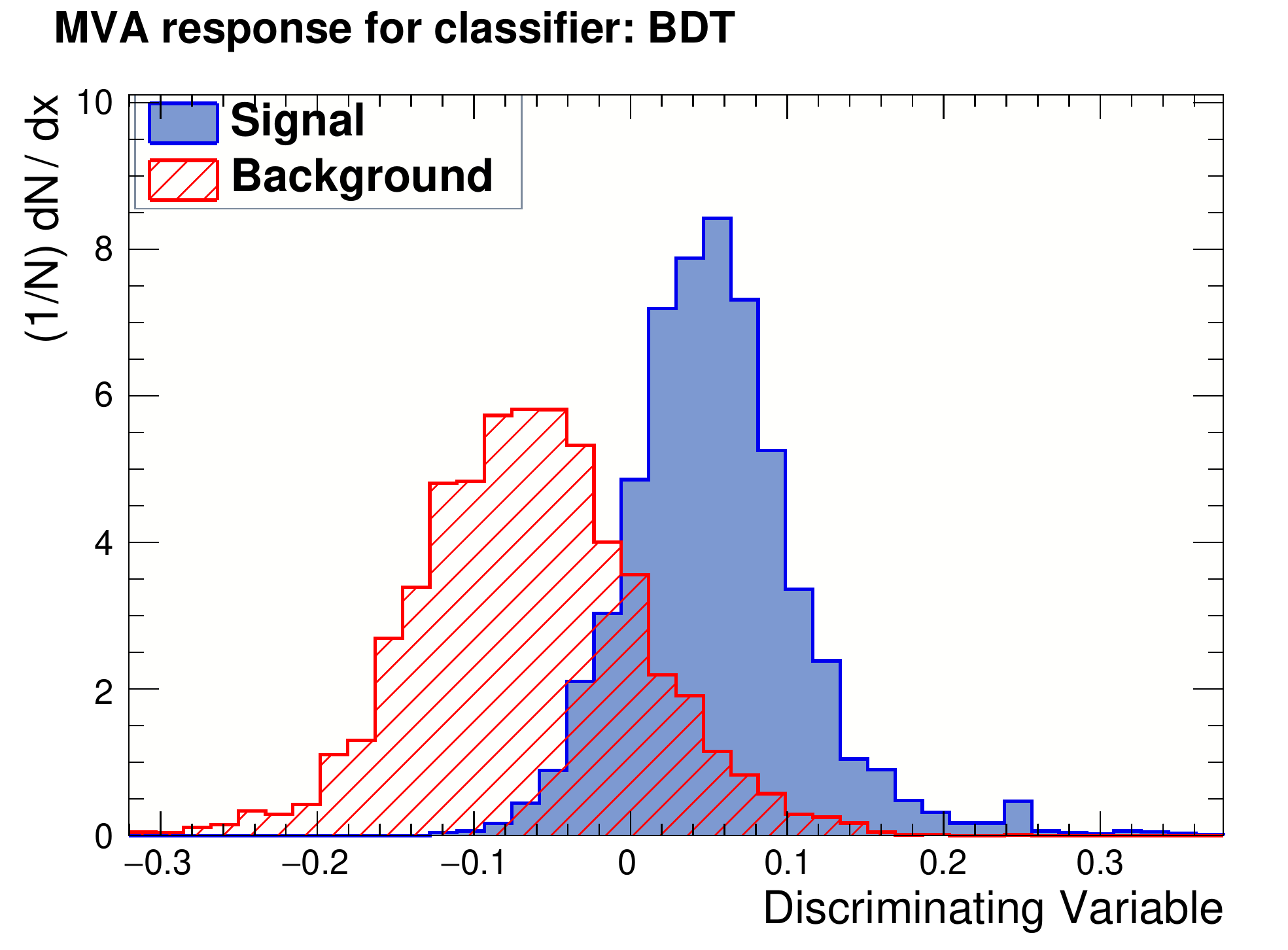}&
   \includegraphics[width=0.45\textwidth]{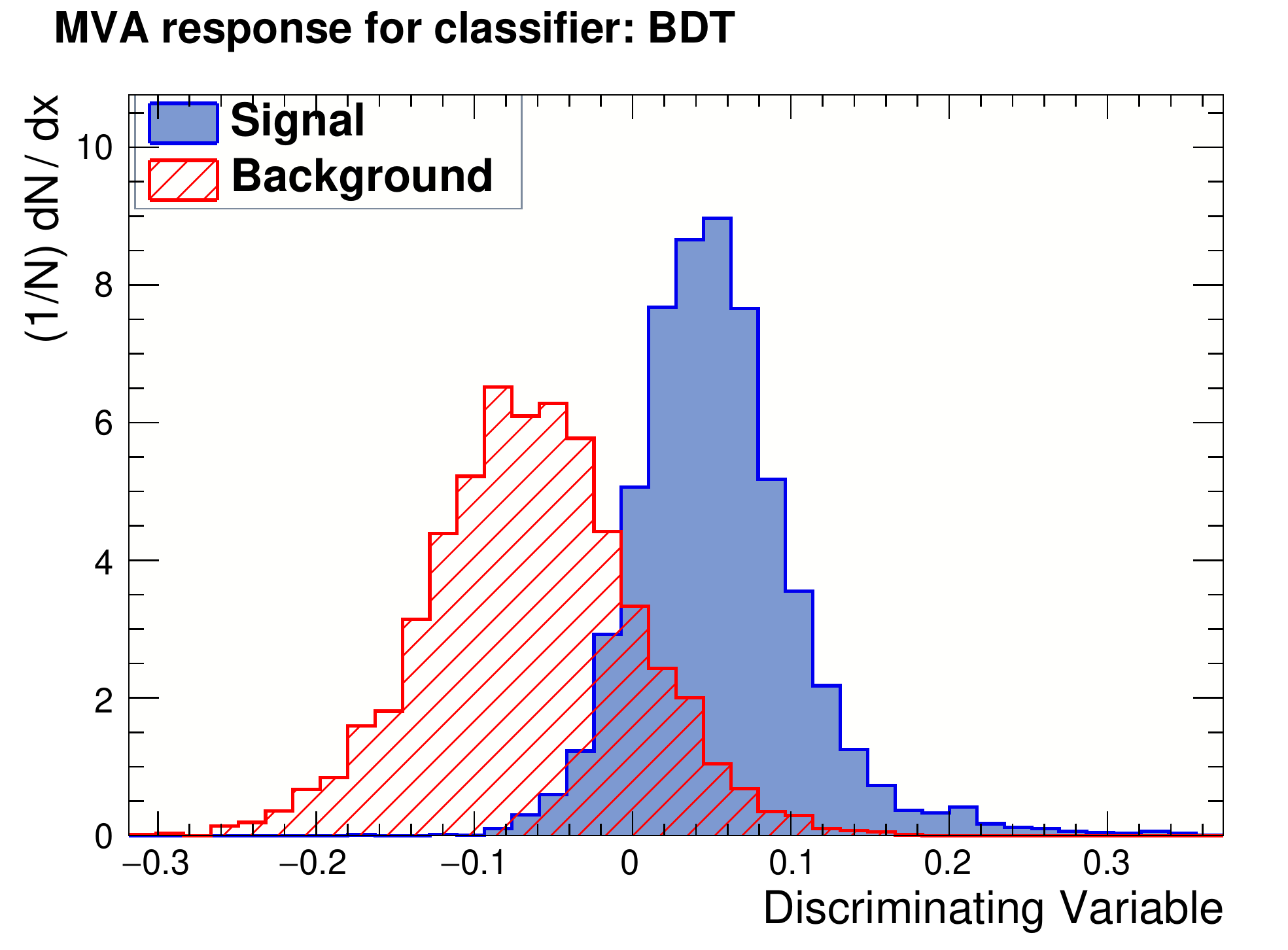}\\
   \includegraphics[width=0.45\textwidth]{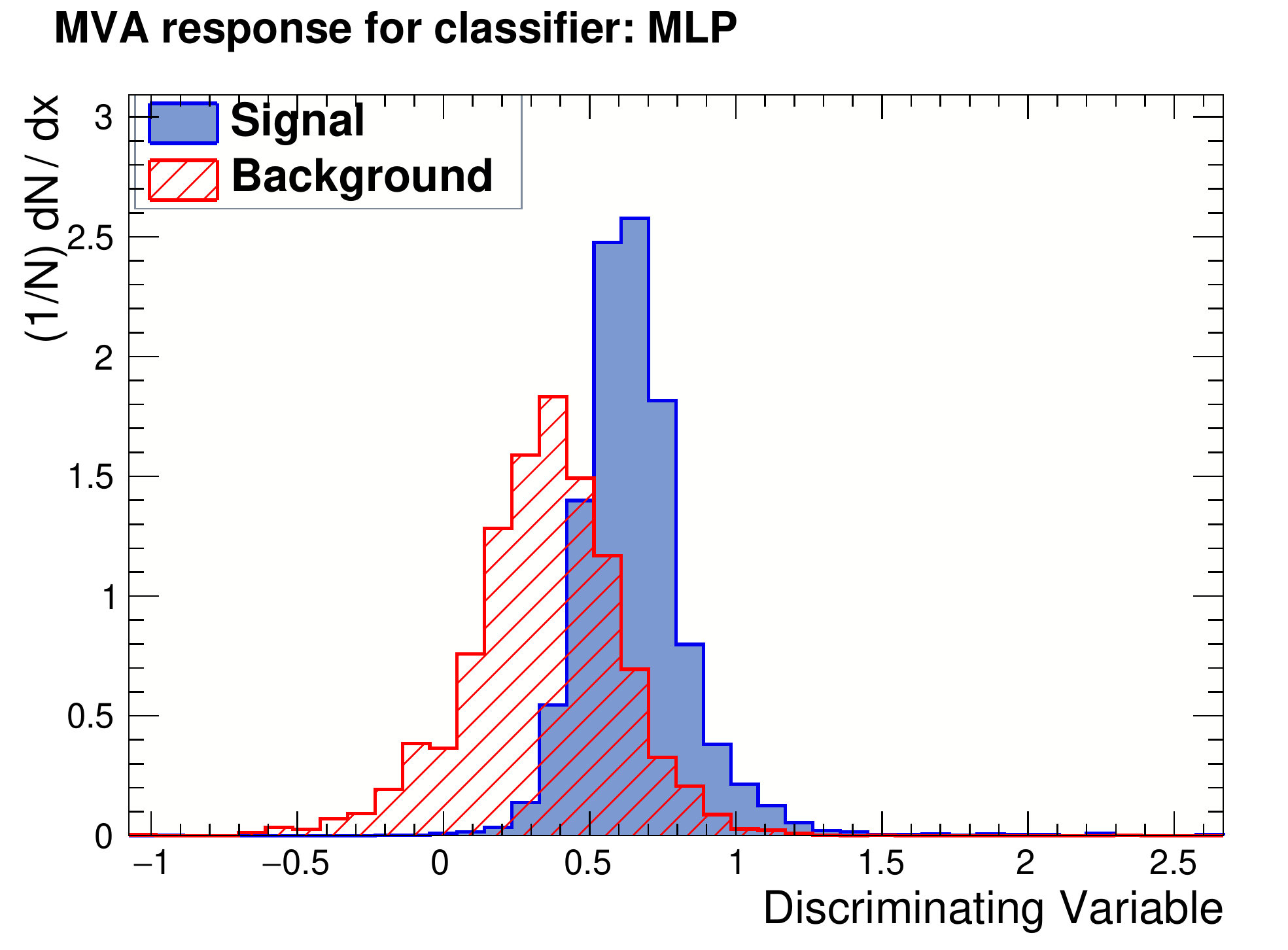}&
   \includegraphics[width=0.45\textwidth]{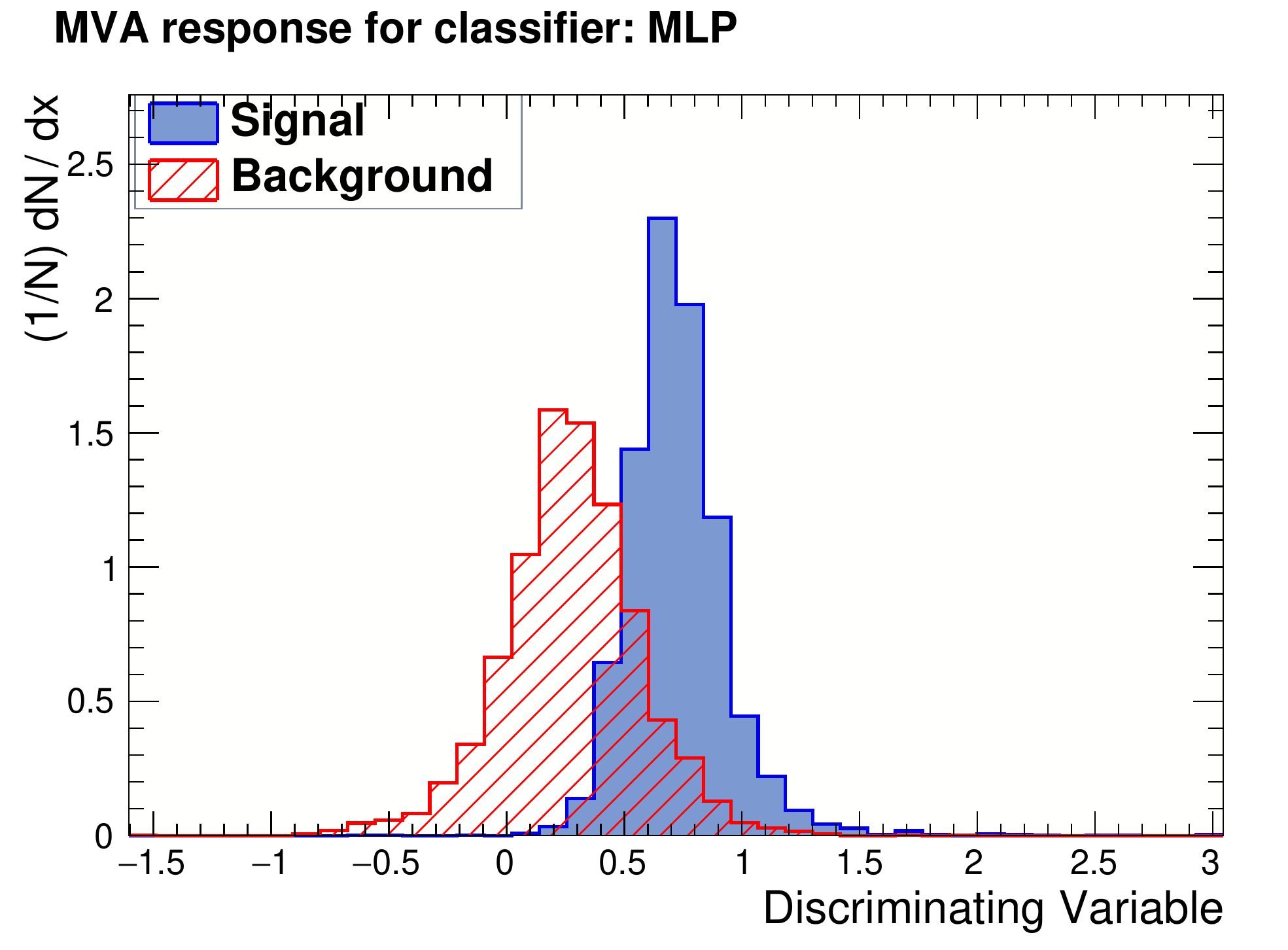}
\end{tabular} \end{center}
\caption{The output response of different MVA methods for unbinned random gun sample. 
         The figures on the left are for coarse grain information and those on 
         the right refer to fine grain information.}
\label{fig:random:outputMvaRes}
\end{figure}
\begin{figure}[htbp]
  \begin{center} \begin{tabular}{cc}
    \includegraphics[width=0.45\textwidth]{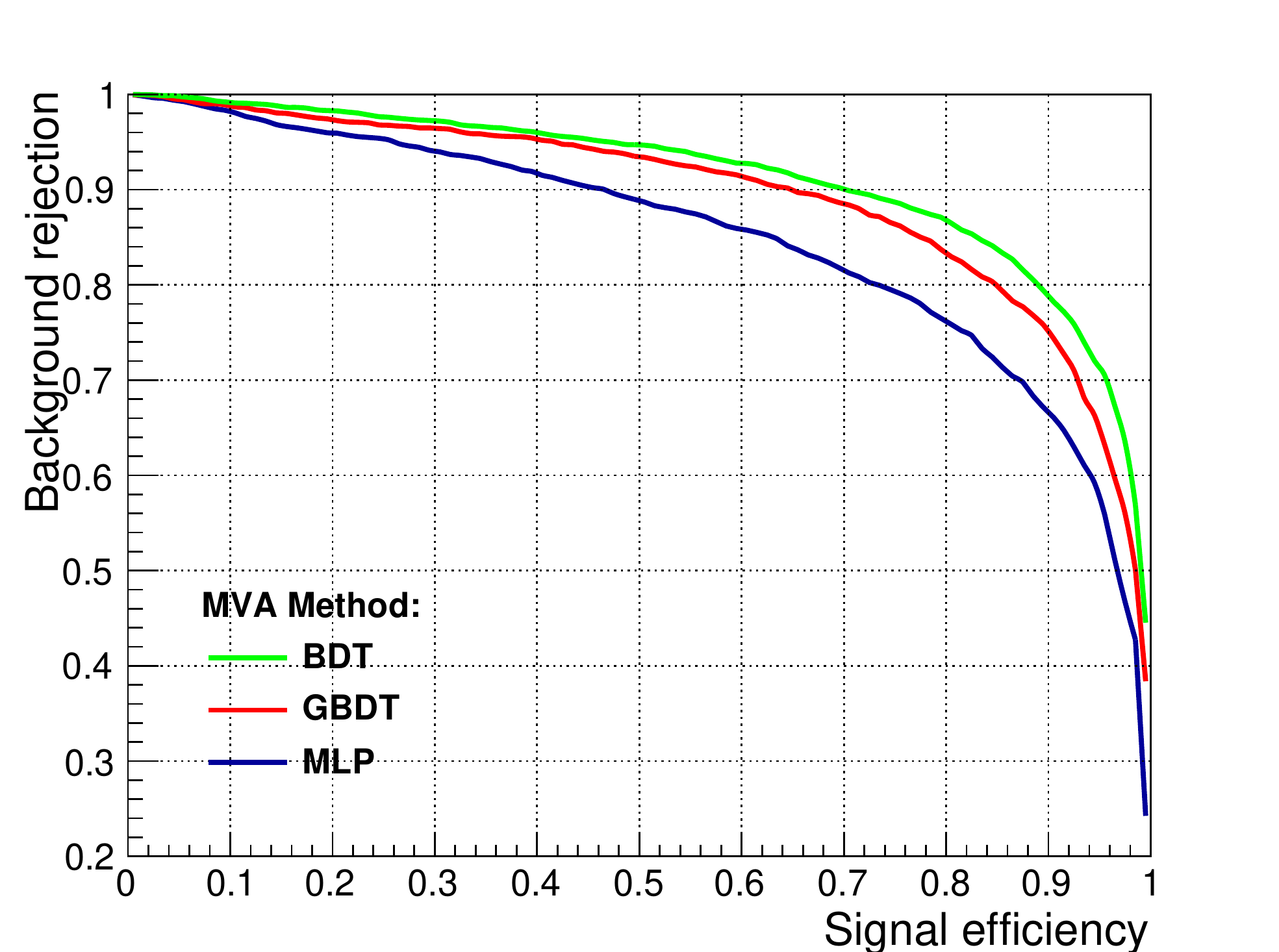}&
    \includegraphics[width=0.45\textwidth]{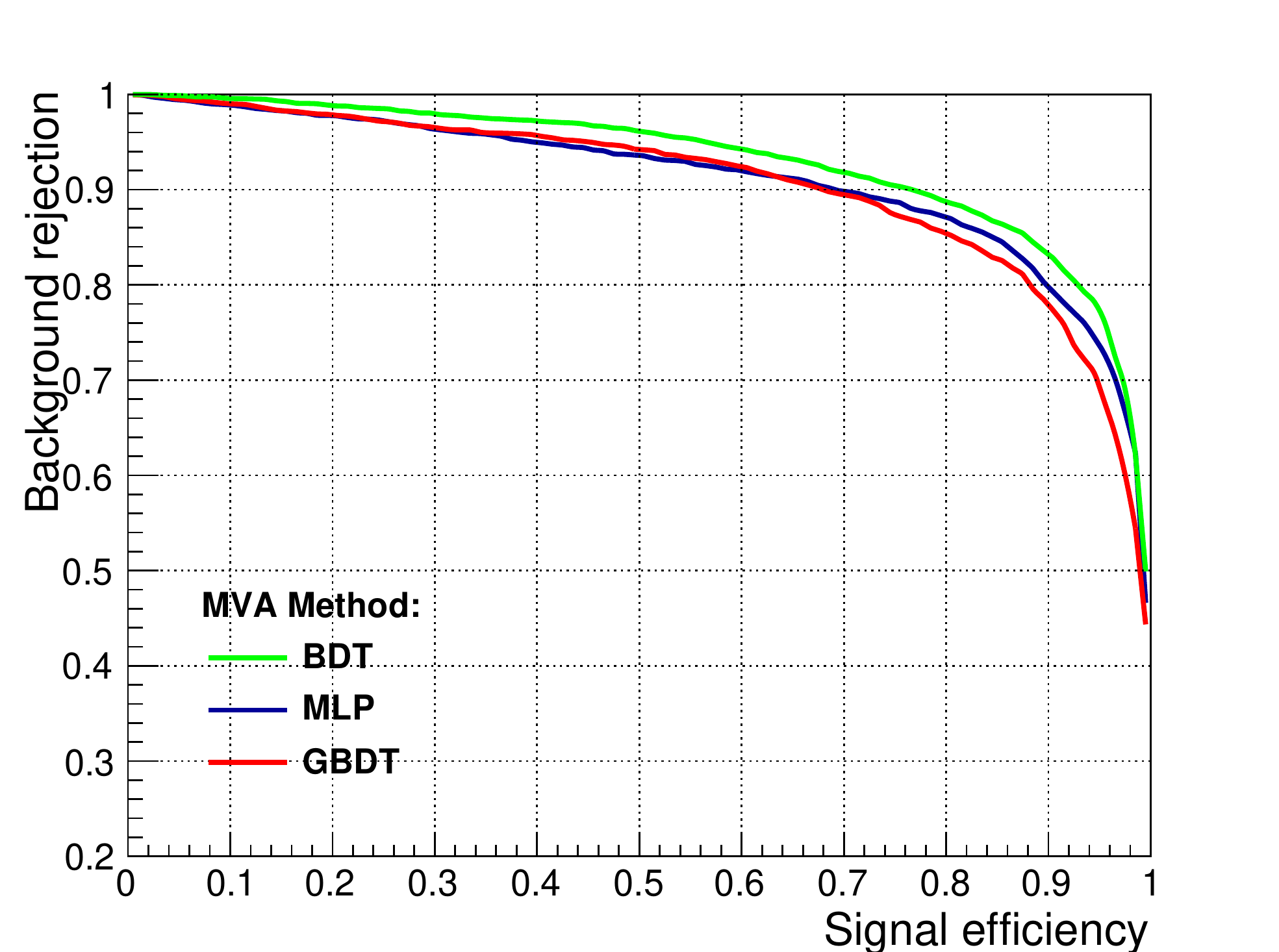}
\end{tabular} \end{center}
\caption{The background rejection versus the
        signal efficiency curve for various MVA methods using unbinned random gun sample. Left figure for
        coarse grain information and the figure on the right for fine grain information \cite{ref:confpaper}. }
\label{fig:random:outputBrejvsS}
\end{figure}
\begin{description}
\item[Unbinned random sample:] In this case the event sample is produced over entire face 
of the central tower. 
Figure \ref{fig:random:outputMvaRes} shows the output response of the different MVA
classifiers for the case of both coarse grain and fine grain information.
Figure \ref{fig:random:outputBrejvsS} shows the background
rejection versus signal efficiency curve for the case of coarse grain as well as fine grain information using random gun sample.
It can be seen from the Figures \ref{fig:random:outputMvaRes} and
\ref{fig:random:outputBrejvsS}, that 
 the fine grain information improves the discrimination for the case of random gun sample also.
\item[Binned random sample:] 
The energy deposit pattern in the matrix of Shashlik towers can vary considerably 
based on the location of the hit on the central tower. To take into account the 
dependence of the energy deposit pattern on the hit location, a hit location 
based MVA training is used, to further improve the separation power of the MVA. 
For this the central tower is divided into 7$\times$7 matrix of virtual square regions 
(or virtual cells) each of dimention 2mm$\times$2mm. 
Event samples are produced in each virtual cell independently. 
49 separate trainings are done to obtain 49 separate trees (or networks) - 
one for each virtual cell. 
At the time of testing/using the MVA, the tree to be used is chosen 
using the information of the measured hit position.
\end{description}
\begin{figure}[htbp]
  \begin{center} \begin{tabular}{cc}
    \includegraphics[width=0.45\textwidth]{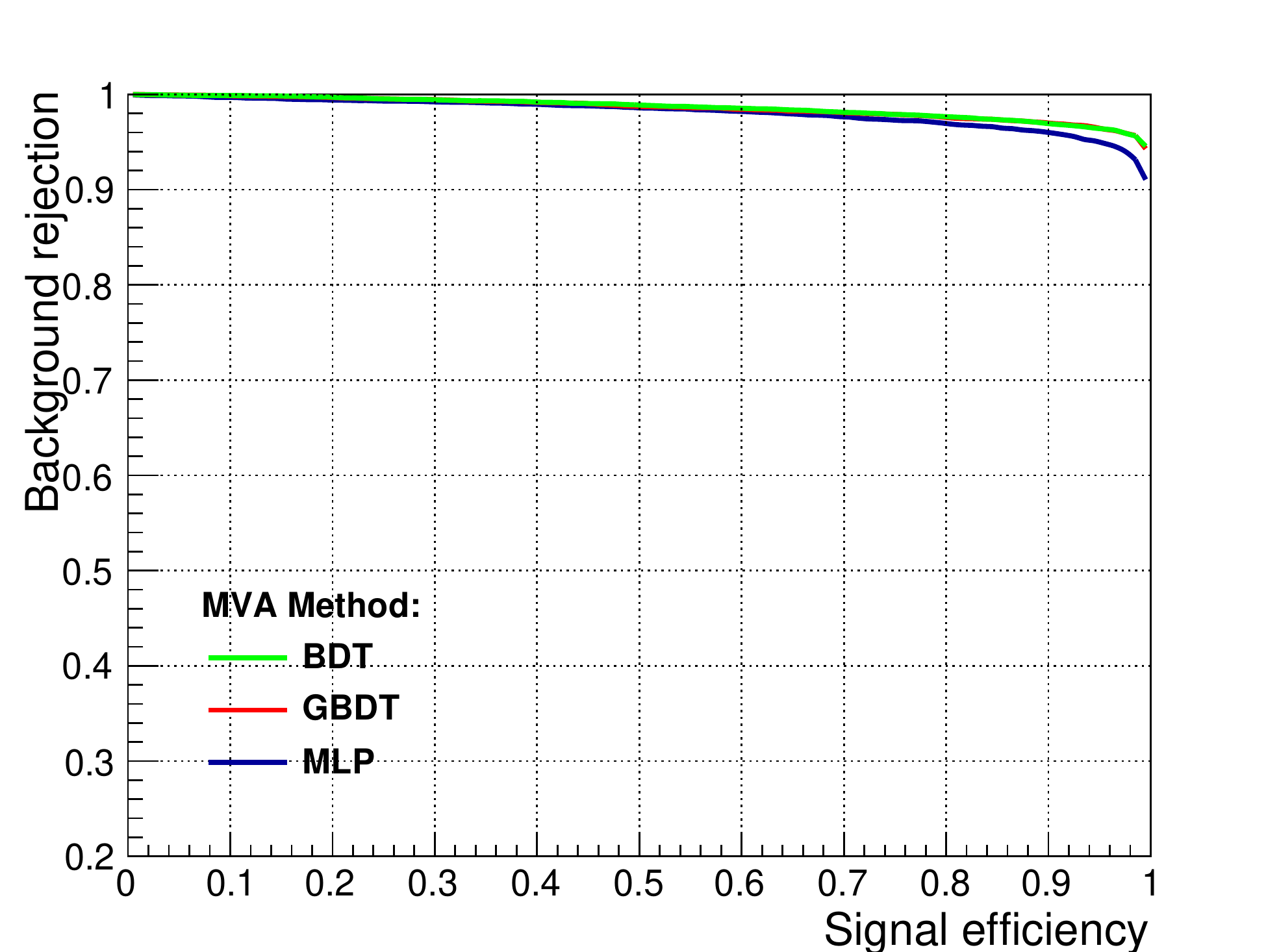} &
    \includegraphics[width=0.45\textwidth]{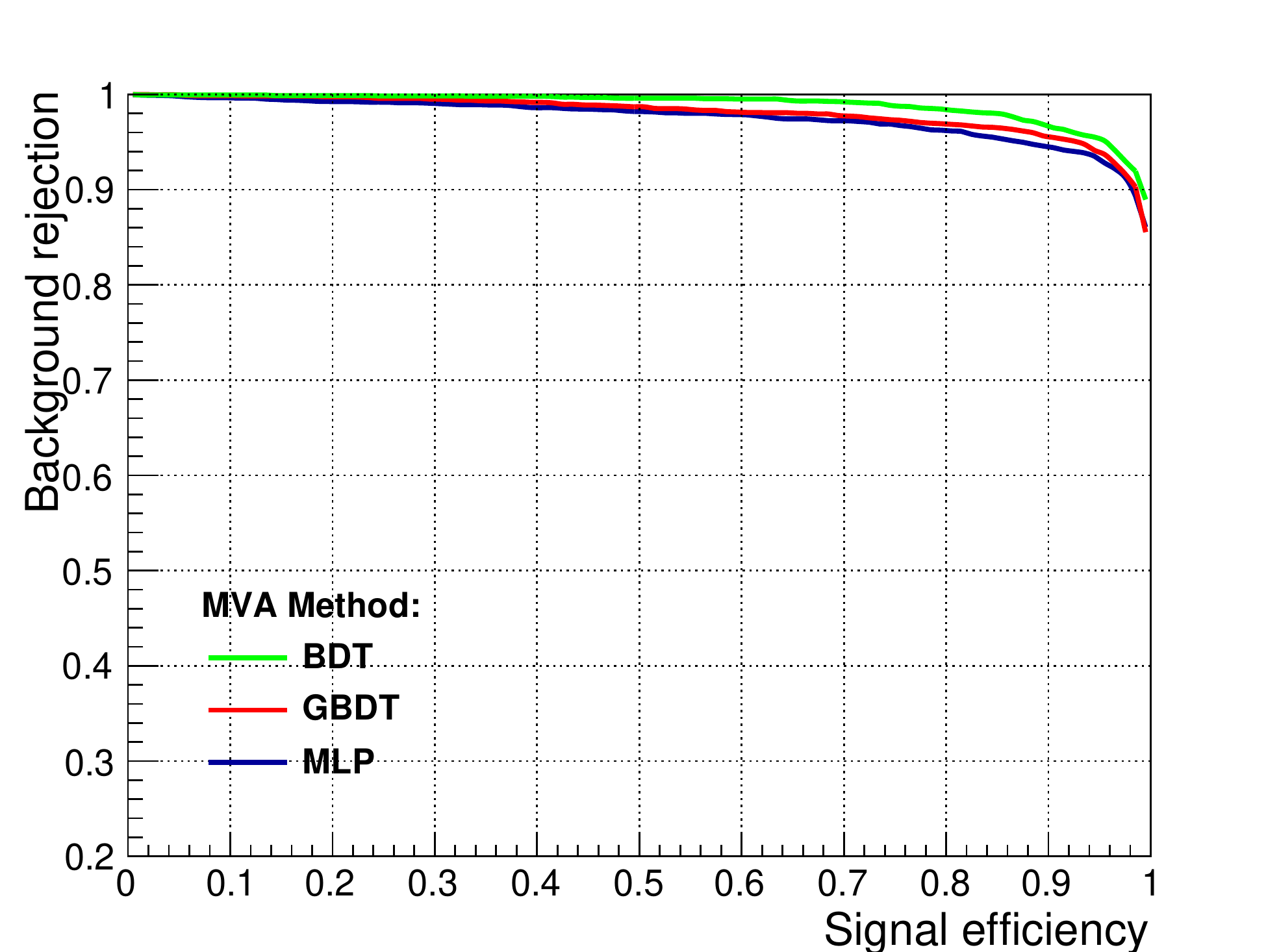}
  \end{tabular} \end{center}
  \caption{Background rejection versus signal efficiency of two virtual cells 
           for the case of binned random samples. 
           The left figure is for a virtual cell where the sample is randomized 
           over a region from $+3$ mm to $+5$ mm in both X and Y direction and the
           right figure is for a region from $-1$ mm to $+1$ mm in the
           X direction and $+3$ mm to $+5$ mm in the Y direction with respect to 
           the centre of the central tower~\cite{ref:confpaper}.}
  \label{fig:random:rejBvsS}
\end{figure}
\begin{figure}[htbp]
\begin{center} \begin{tabular}{cc}
   \includegraphics[width=0.45\textwidth]{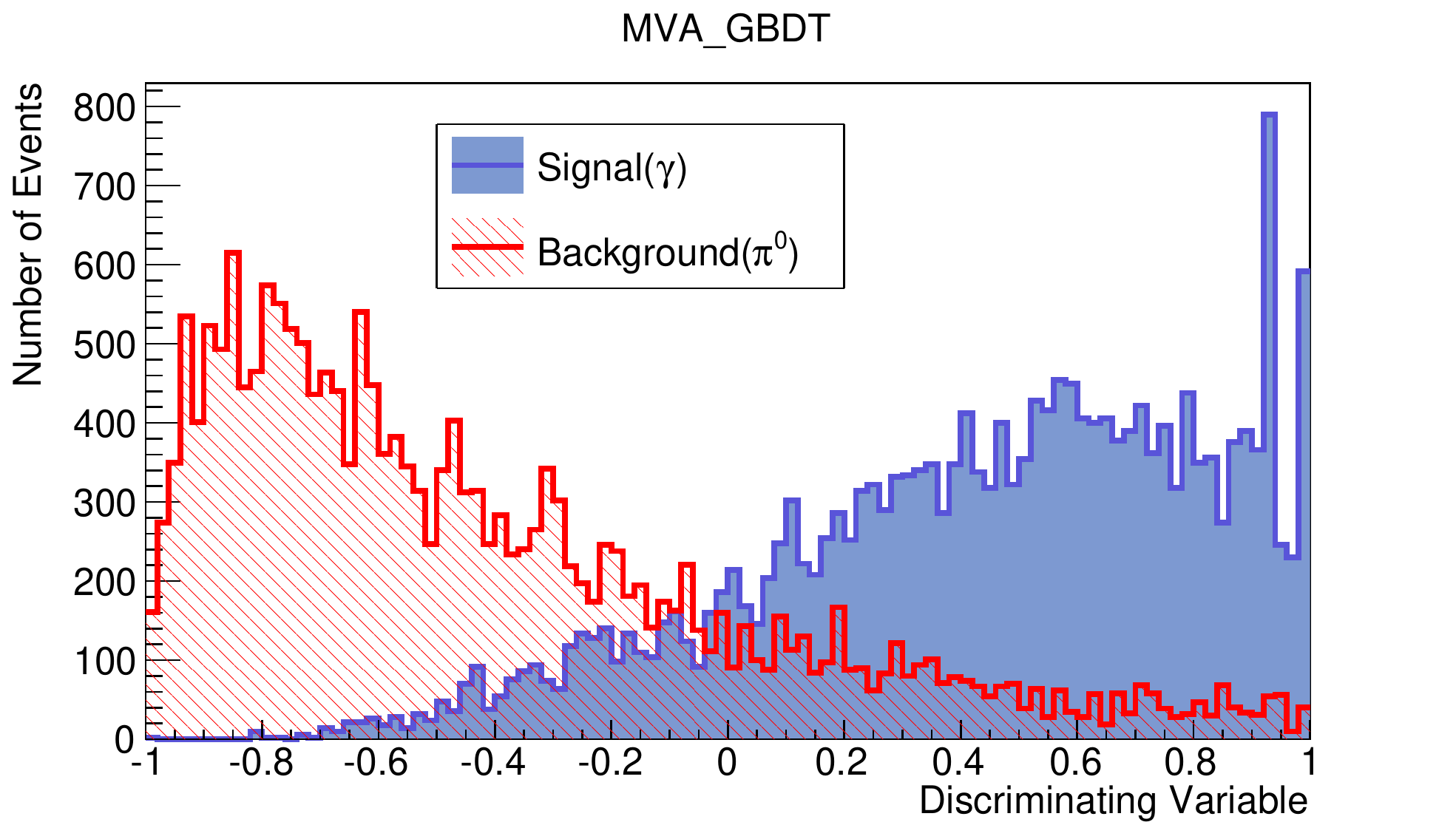}&
   \includegraphics[width=0.45\textwidth]{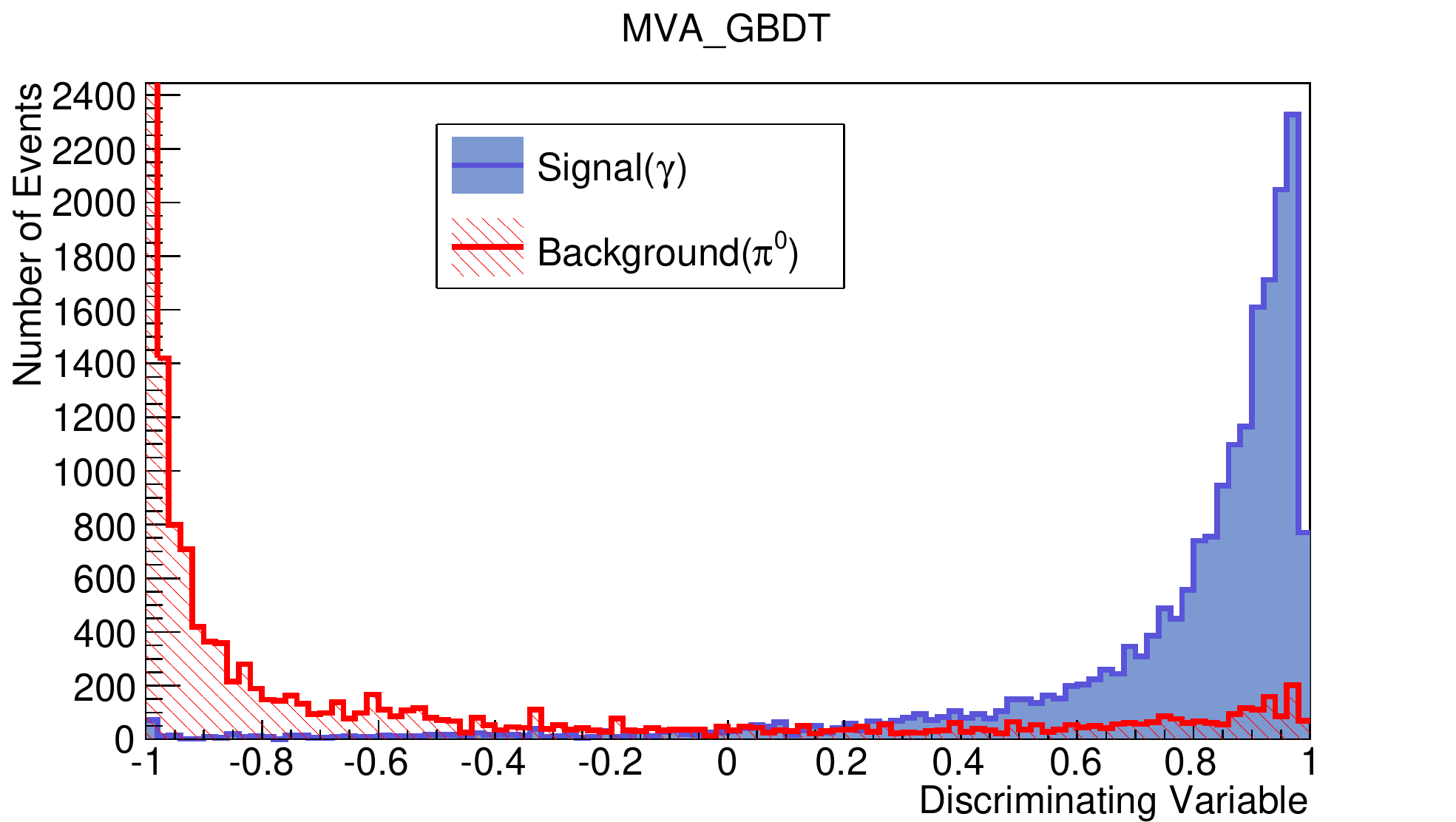}\\
   \includegraphics[width=0.45\textwidth]{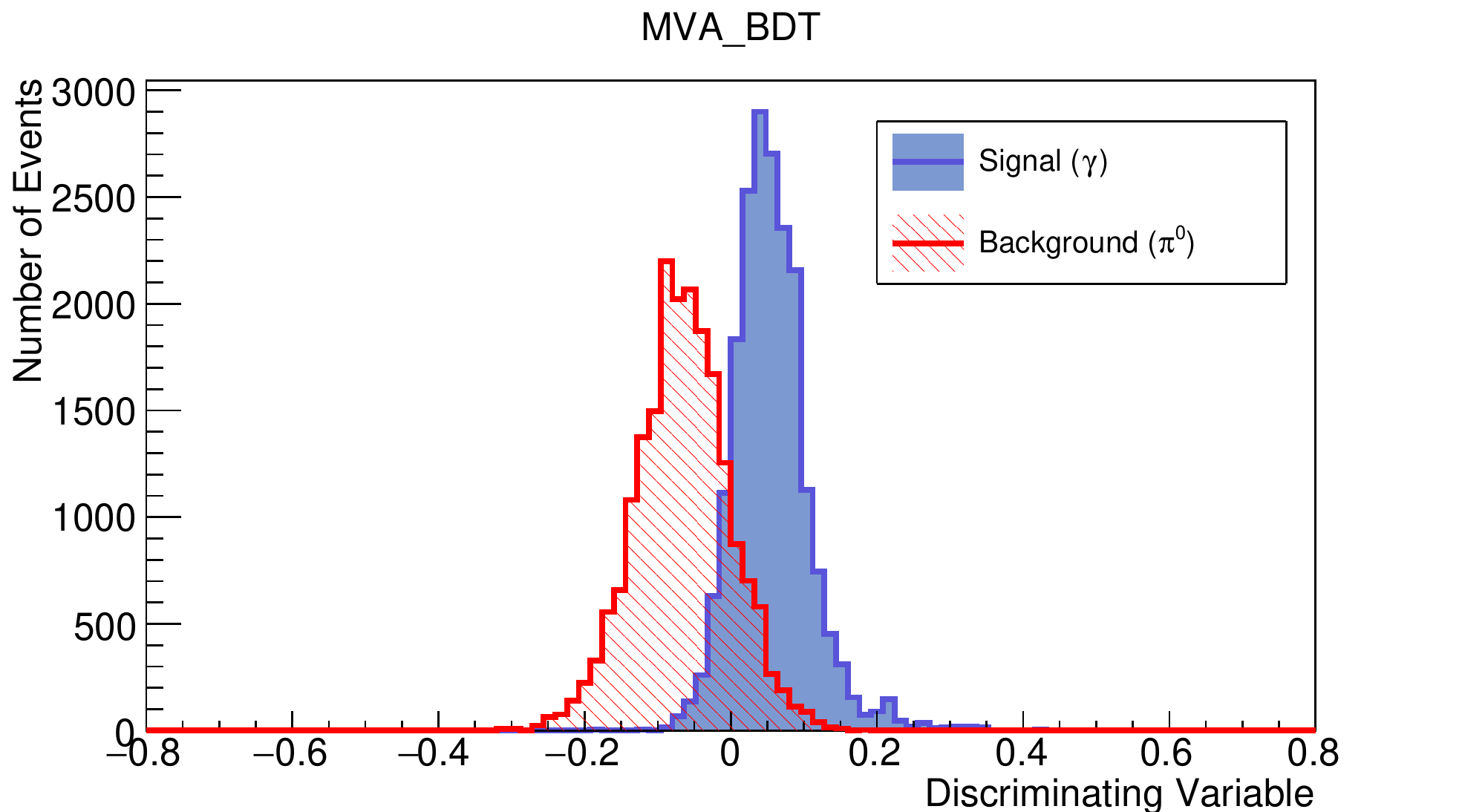}&
   \includegraphics[width=0.45\textwidth]{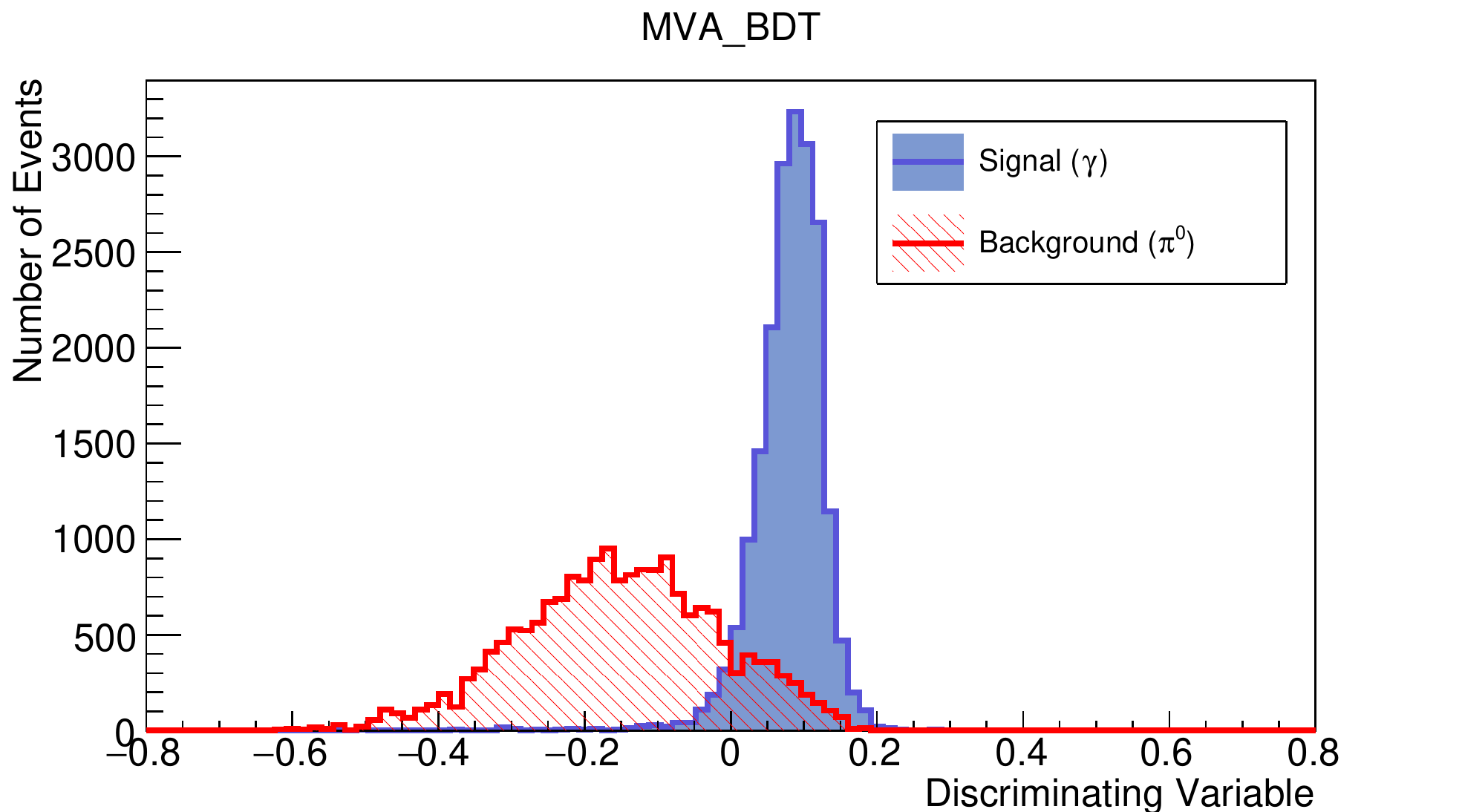}\\
   \includegraphics[width=0.45\textwidth]{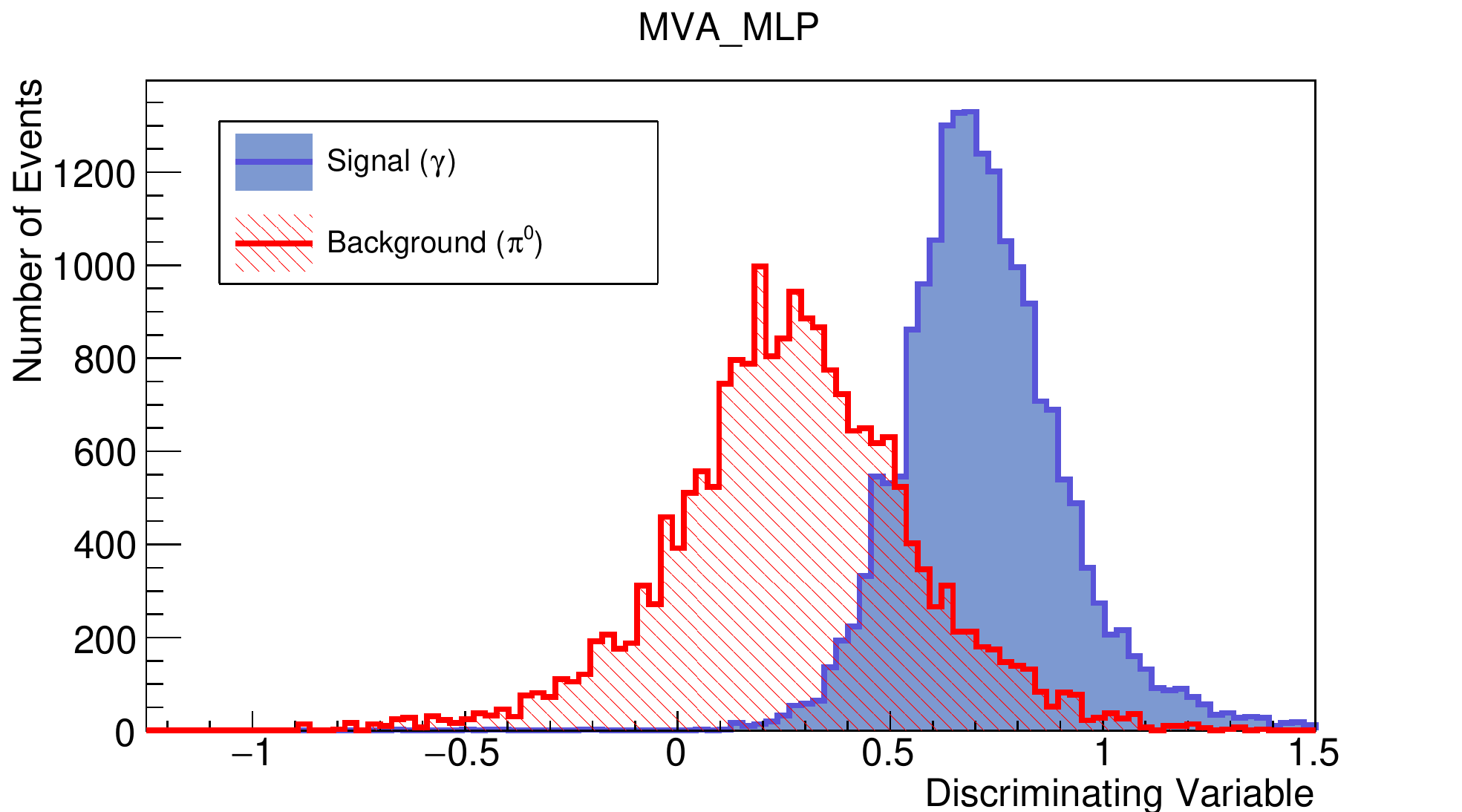}&
   \includegraphics[width=0.45\textwidth]{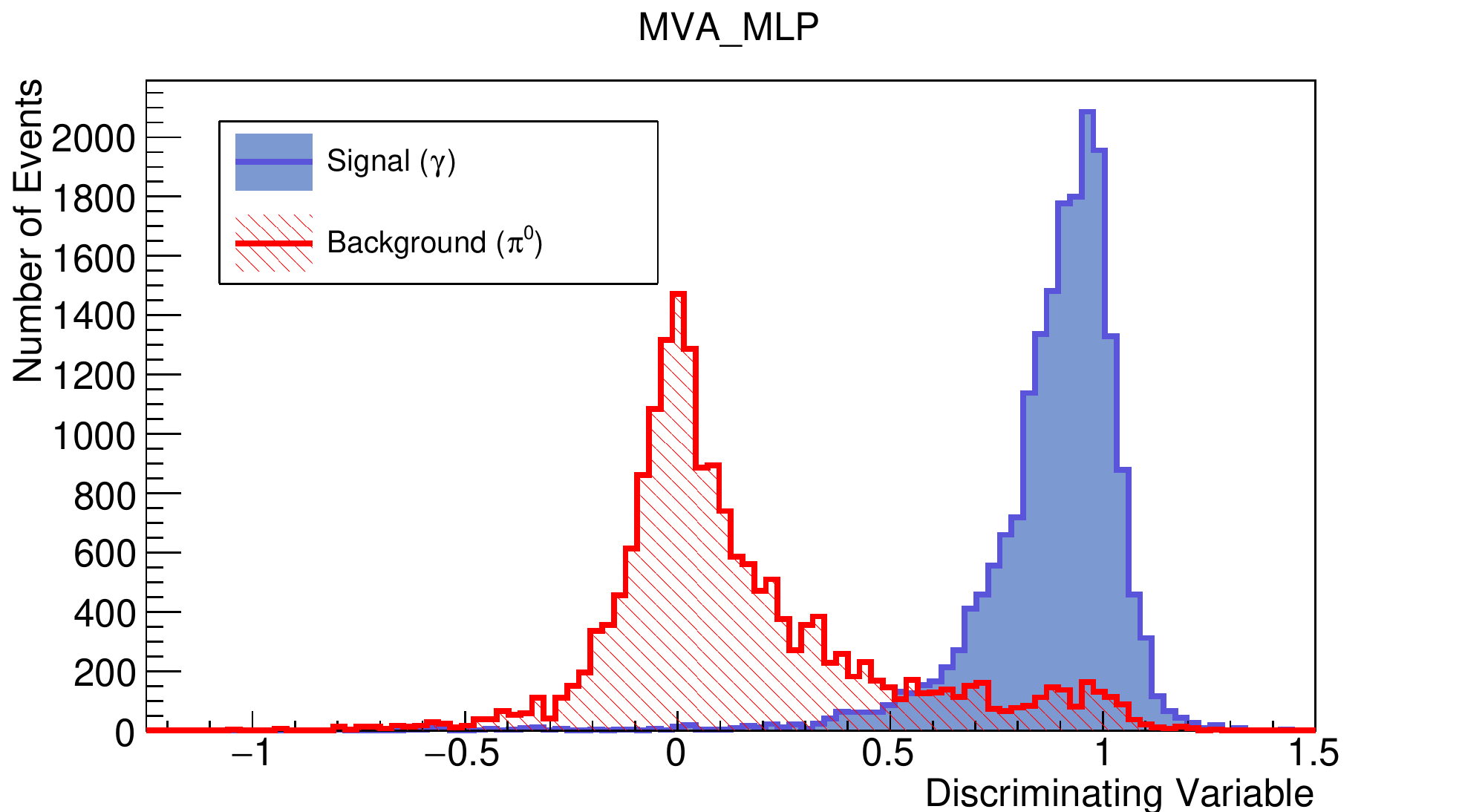}
\end{tabular} \end{center}
\caption{The output response of different MVA methods. The figures on the
         left show the response from the MVA trained with unbinned sample 
         and the ones on the right are for the MVA trained with binned sample.
        }
\label{fig:random:outrestmva}
\end{figure}
Figure \ref{fig:random:rejBvsS} shows background rejection versus signal 
efficiency plots for binned random sample.
Figure \ref{fig:random:outrestmva} shows the output response of both the MVA methods, 
namely the MVA trained with unbinned smaples and the MVAs trained with binned training samples.  
For both the cases, the same test sample is used.
\subsection{Comparison of various methods} \label{sec:comparison}
A comparison in performance is made among all the methods described in the
previous sections. This comparison is shown for 200 GeV photons 
and $\pi^{0}$'s.

\begin{figure}[htbp]
  \begin{center} \begin{tabular}{cc}
    \includegraphics[width=0.45\textwidth]{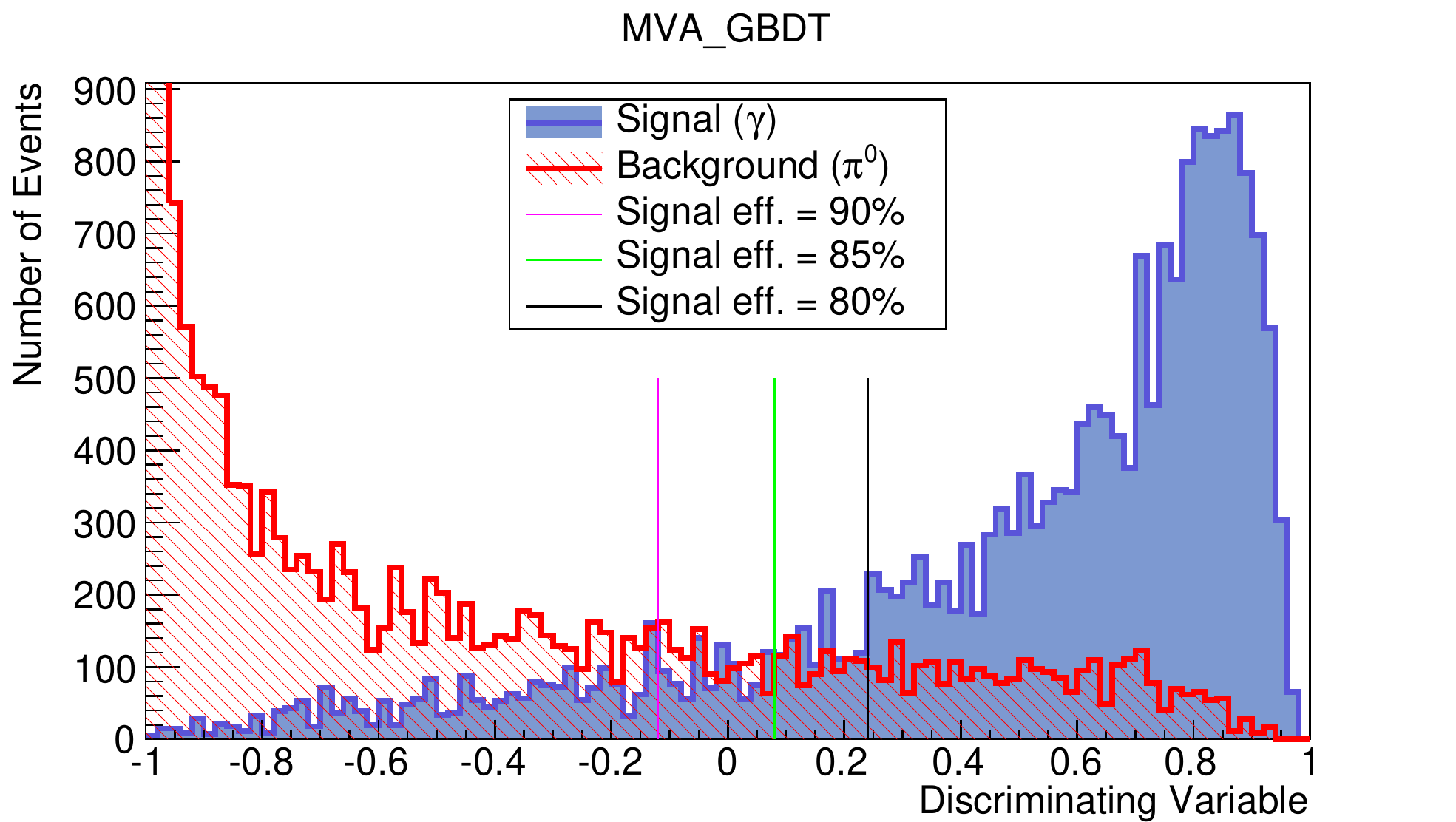}&
    \includegraphics[width=0.45\textwidth]{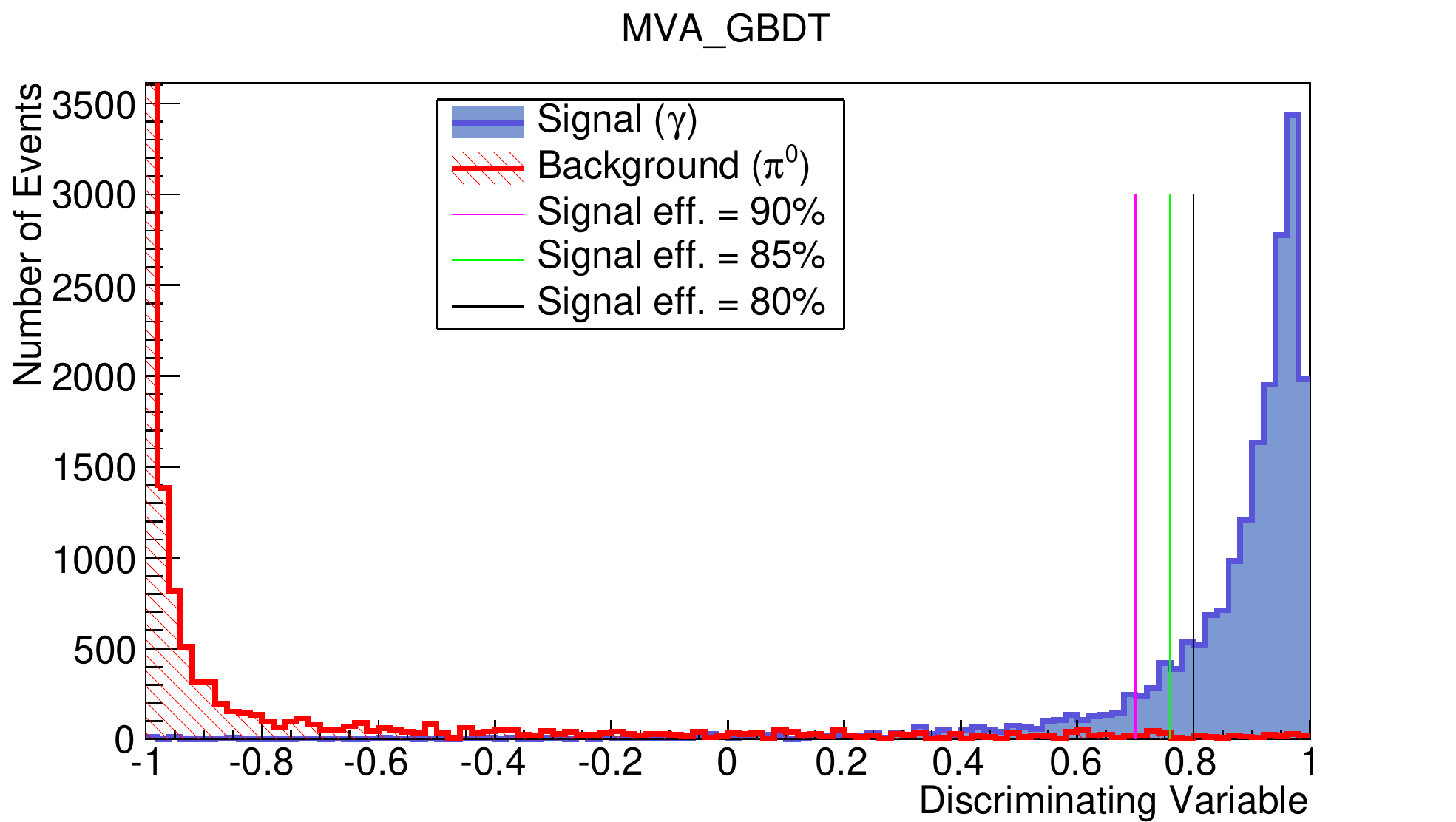}
  \end{tabular}\end{center}
  \caption{MVA response for 200 GeV photons and $\pi^{0}$'s.
           The plots on the left refers to GBDT training with coarse grain 
           information and the plots on the right makes use of GBDT algorithm trained with
           fine grain information, using fixed gun sample for (x,y):(0.0 mm, 4.0 mm). 
           The black, blue and pink lines show the point where the cut 
           is applied to achieve 80$\%$, 85$\%$ and 90$\%$ signal efficiency 
           (events falling on the right side of the line are selected).}
\label{fig:comp_showershapes2}
\end{figure}
\begin{figure}[htbp]
  \begin{center} \begin{tabular}{cc}
    \includegraphics[width=0.45\textwidth]{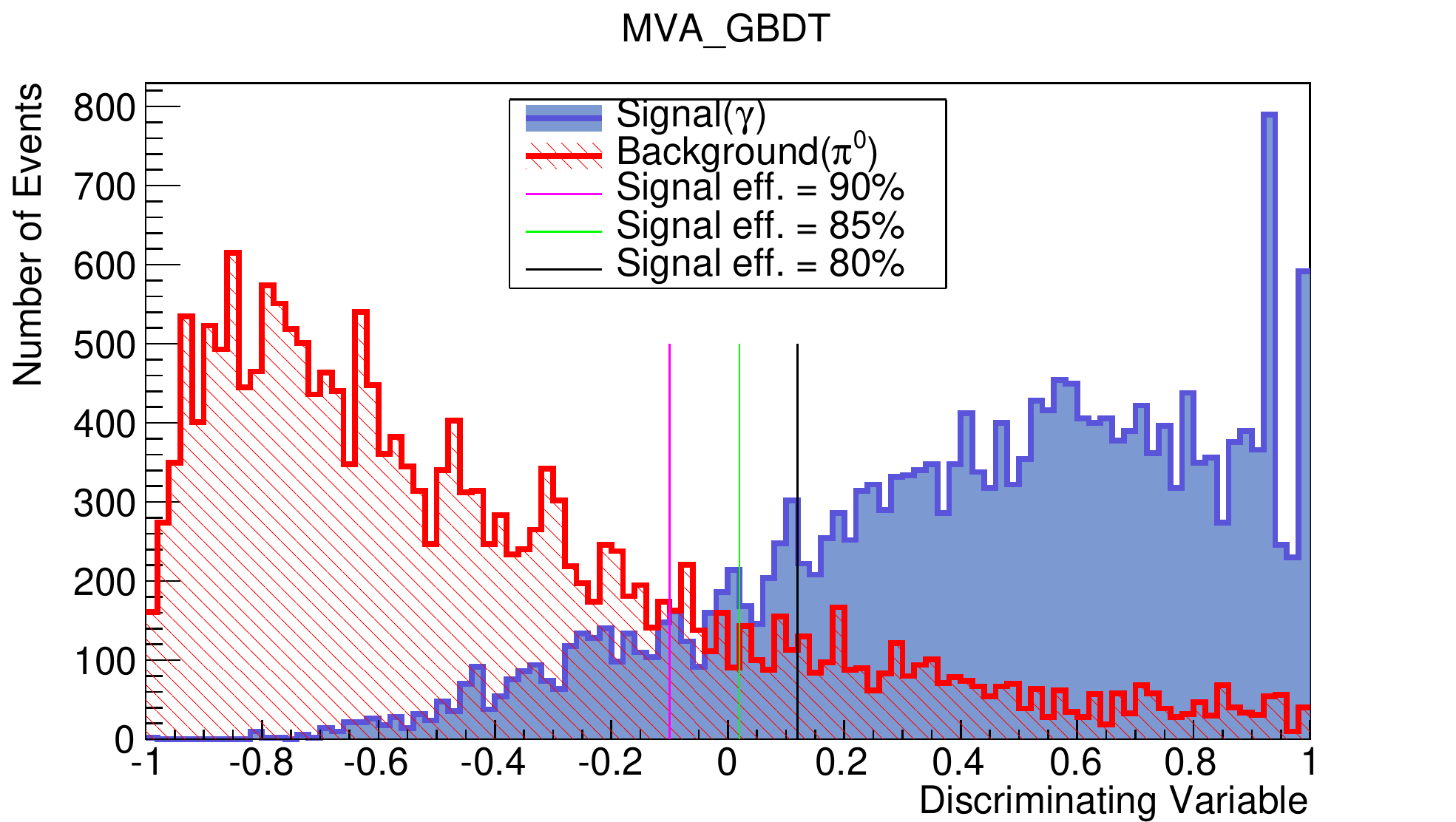}&
    \includegraphics[width=0.45\textwidth]{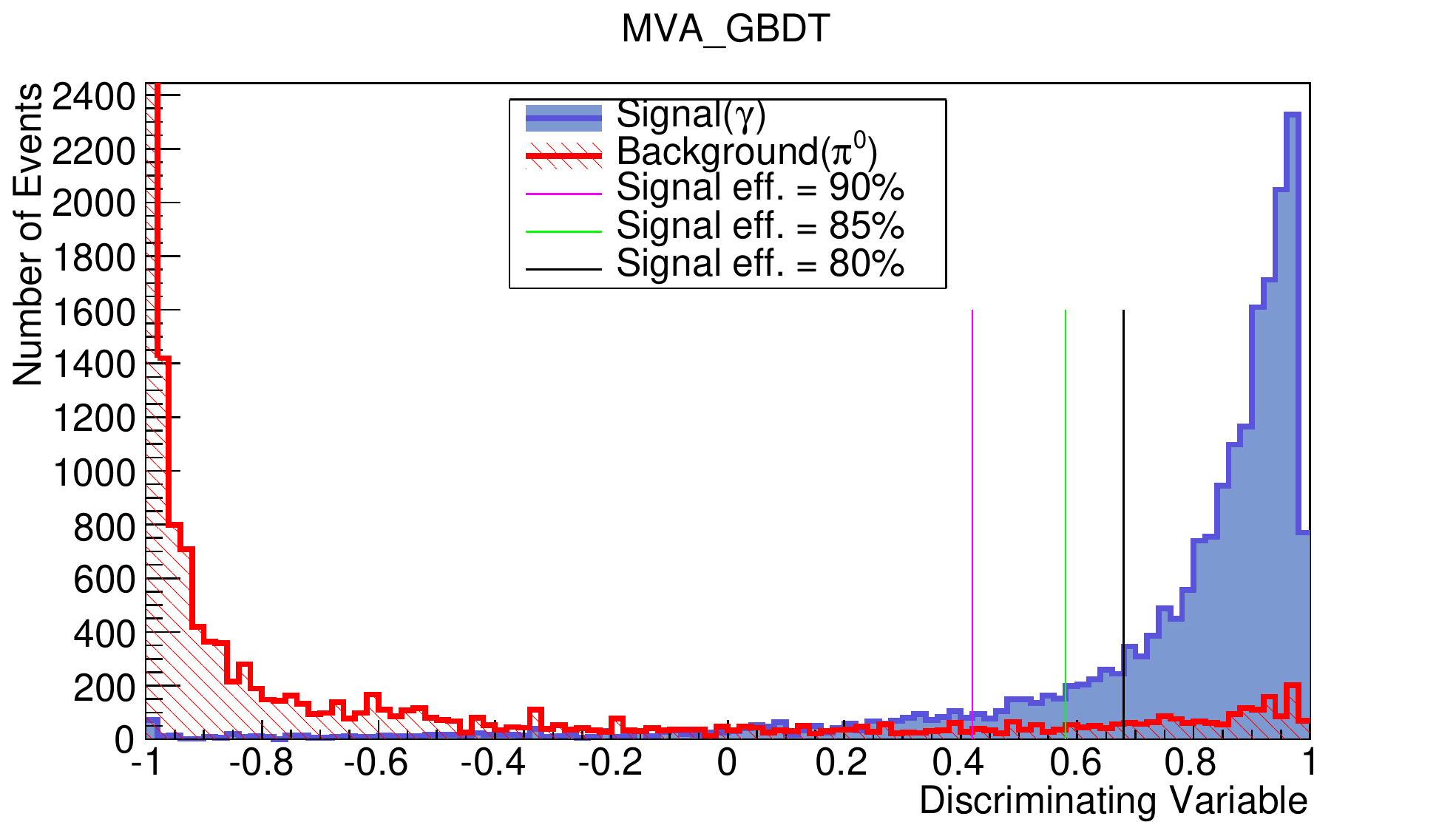}
  \end{tabular}\end{center}
  \caption{Comparison of GBDT MVA method for 200 GeV photons and $\pi^{0}$'s. 
           The plots in the left refers to GBDT training with unbinned sample and 
           the plots on the right makes use of GBDT algorithm trained with binned sample.
           The black, green and pink lines are the point where
           the cut is applied to achieve 80$\%$, 85$\%$ and 90$\%$ signal 
           efficiency respectively \cite{ref:confpaper}.}
  \label{fig:comp_bdtg}
\end{figure}
\begin{figure}[htbp]
  \begin{center} \begin{tabular}{cc}
    \includegraphics[width=0.45\textwidth]{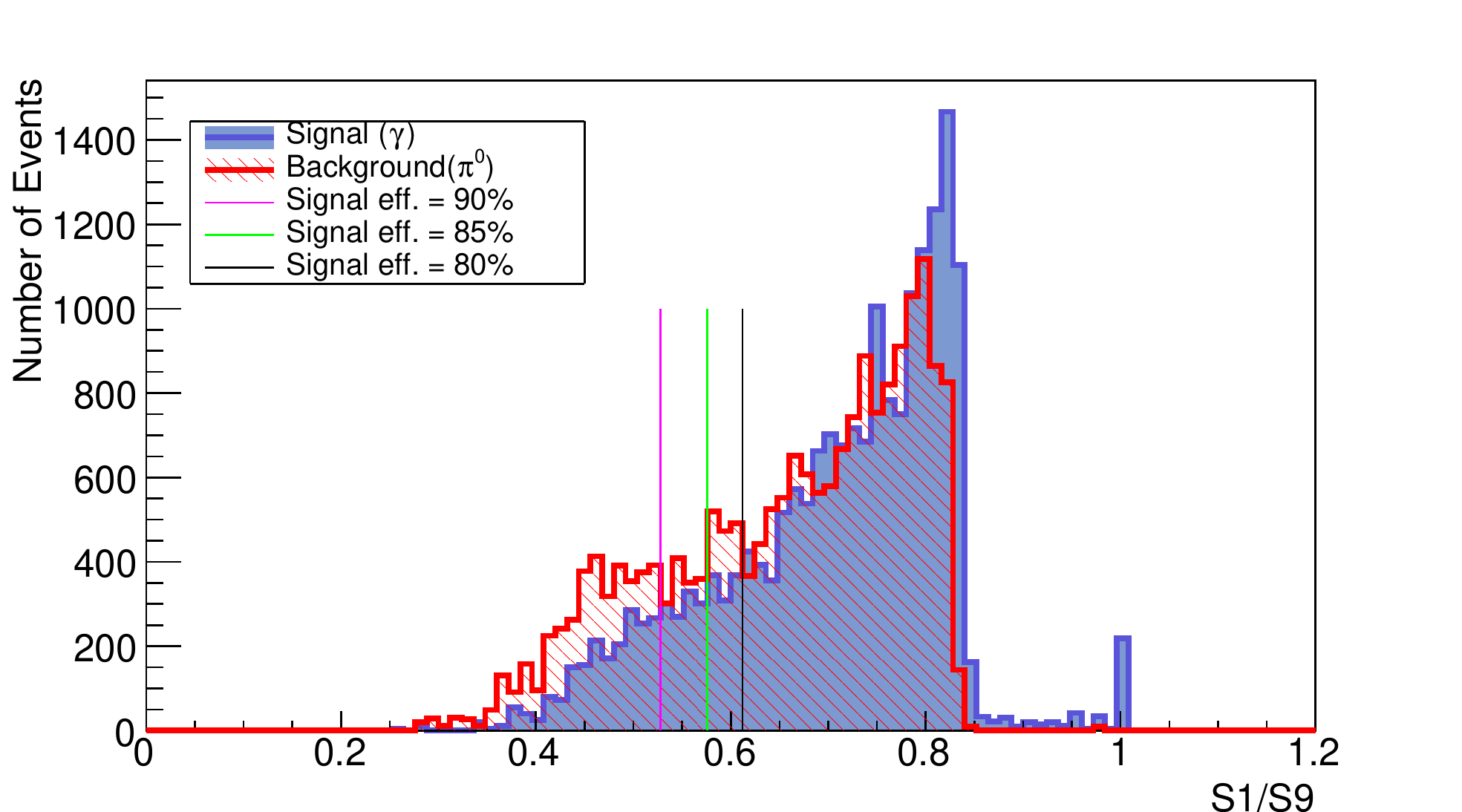}&
    \includegraphics[width=0.45\textwidth]{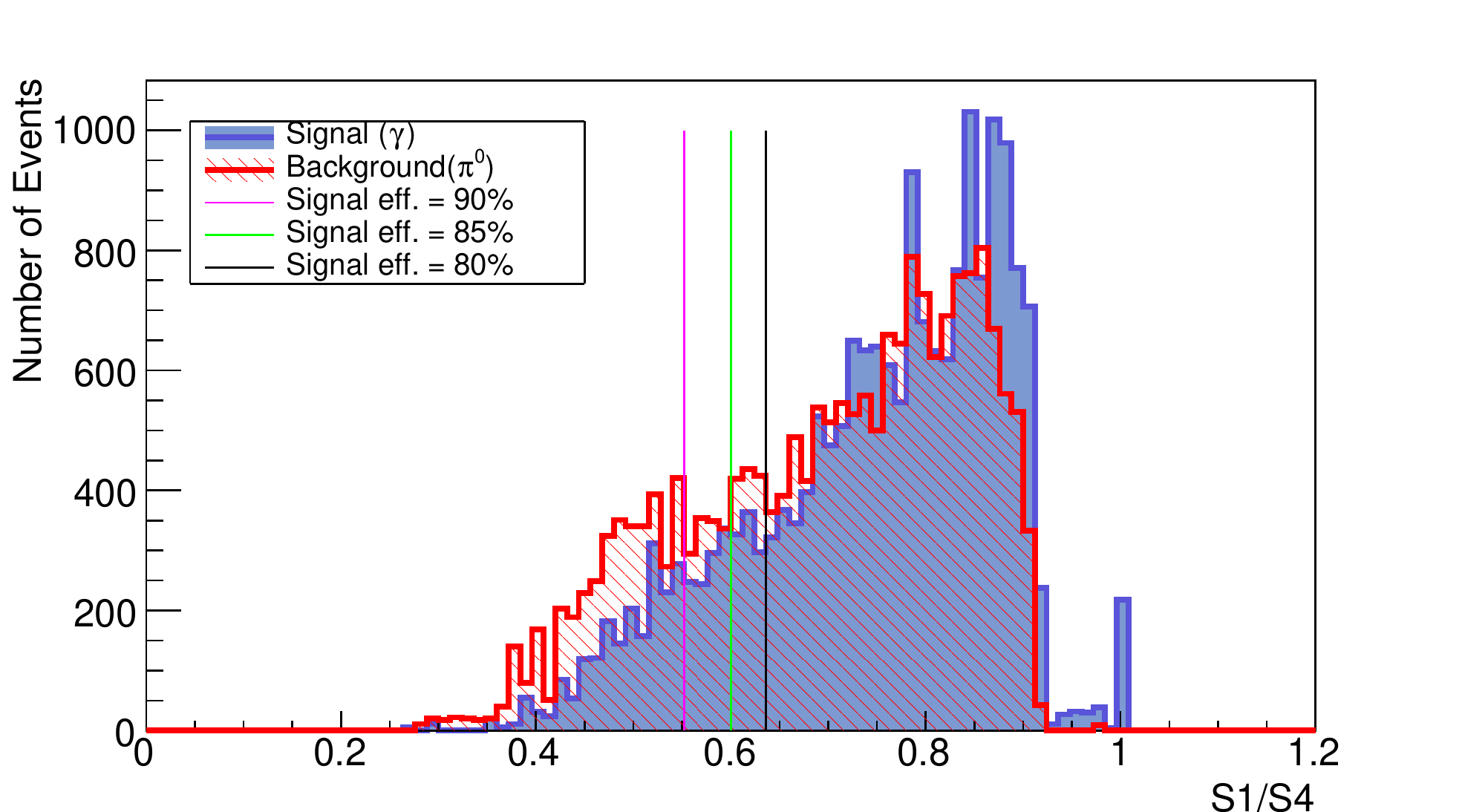}\\
    \includegraphics[width=0.45\textwidth]{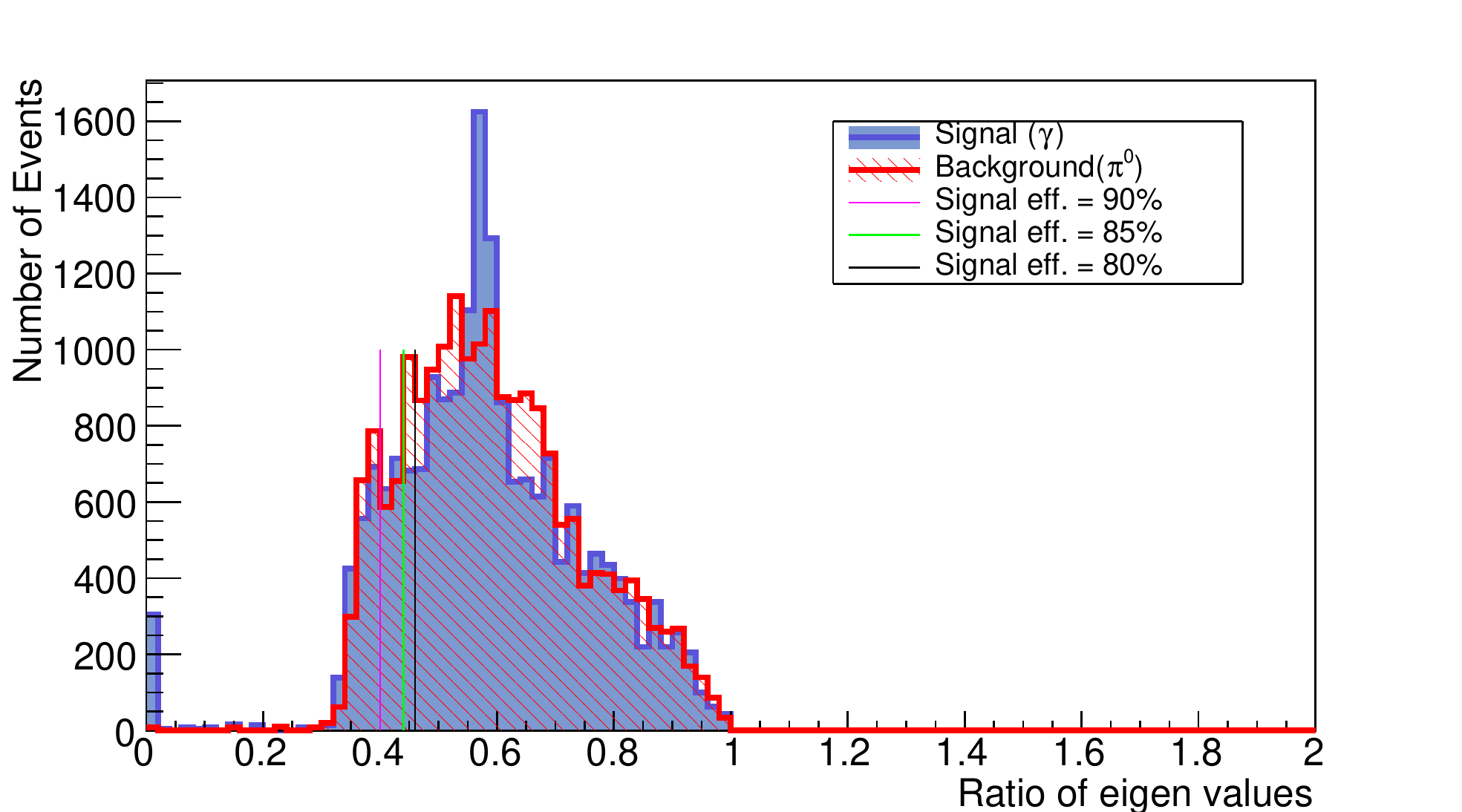}&
    \includegraphics[width=0.45\textwidth]{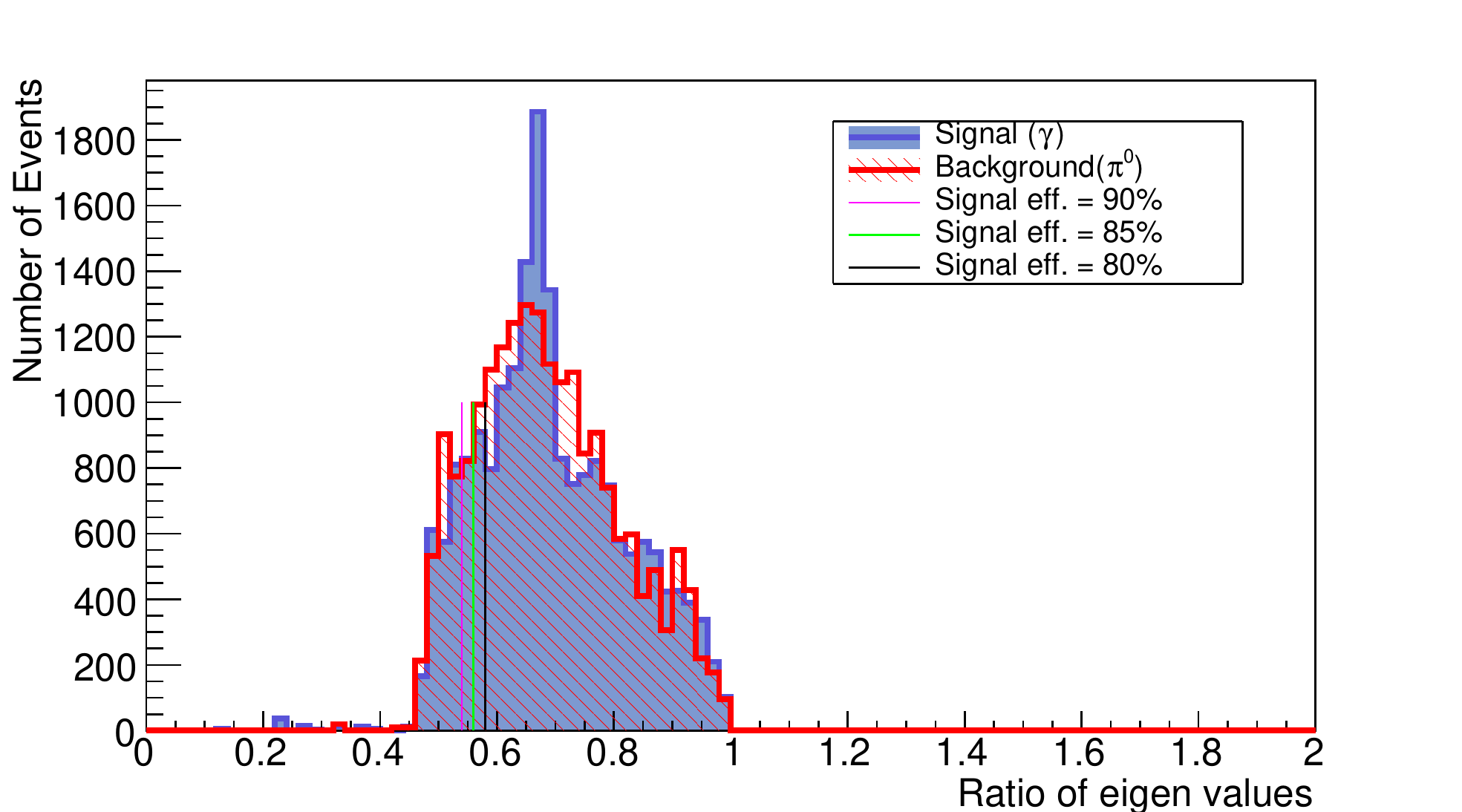}\\
    \includegraphics[width=0.45\textwidth]{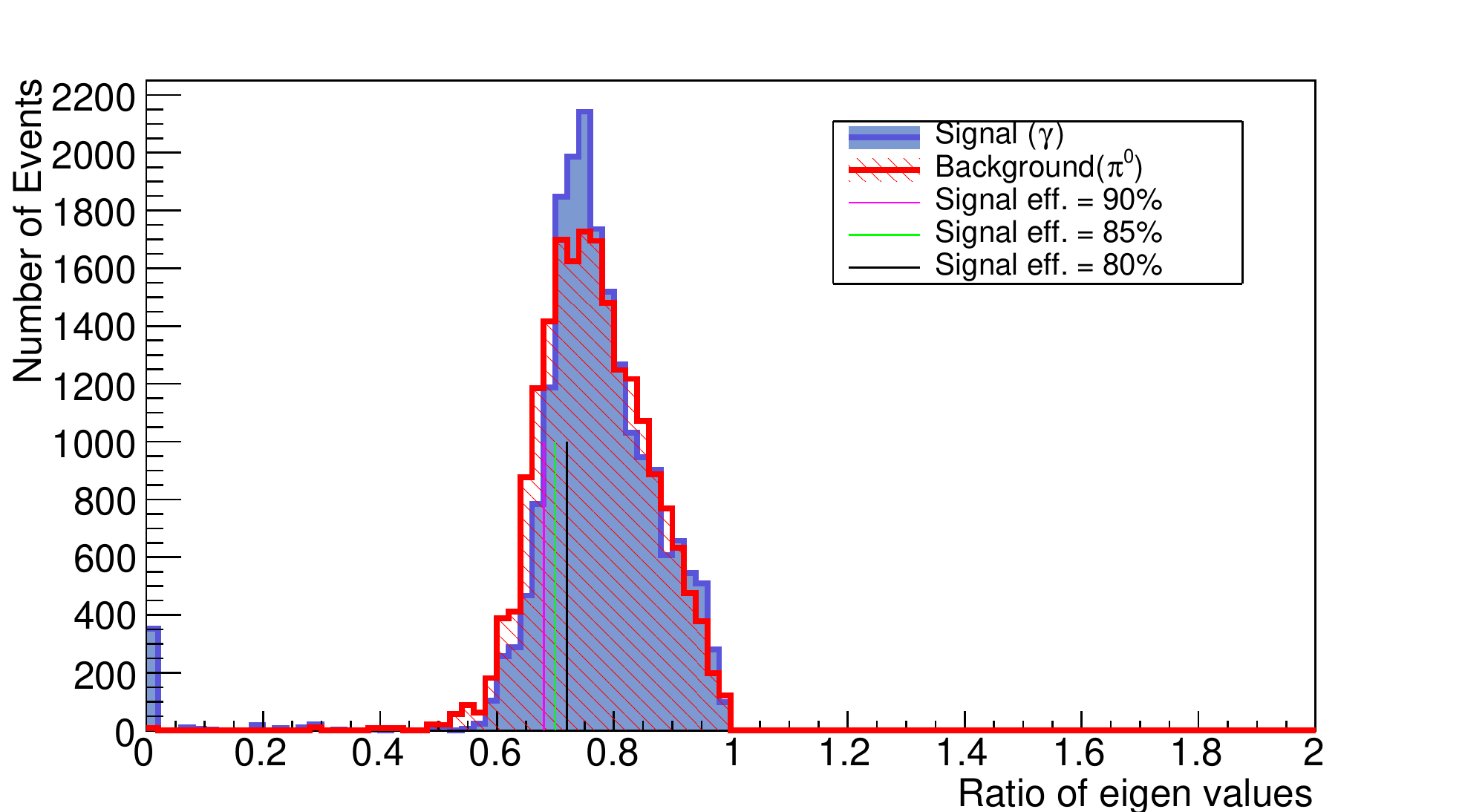}&
    \includegraphics[width=0.45\textwidth]{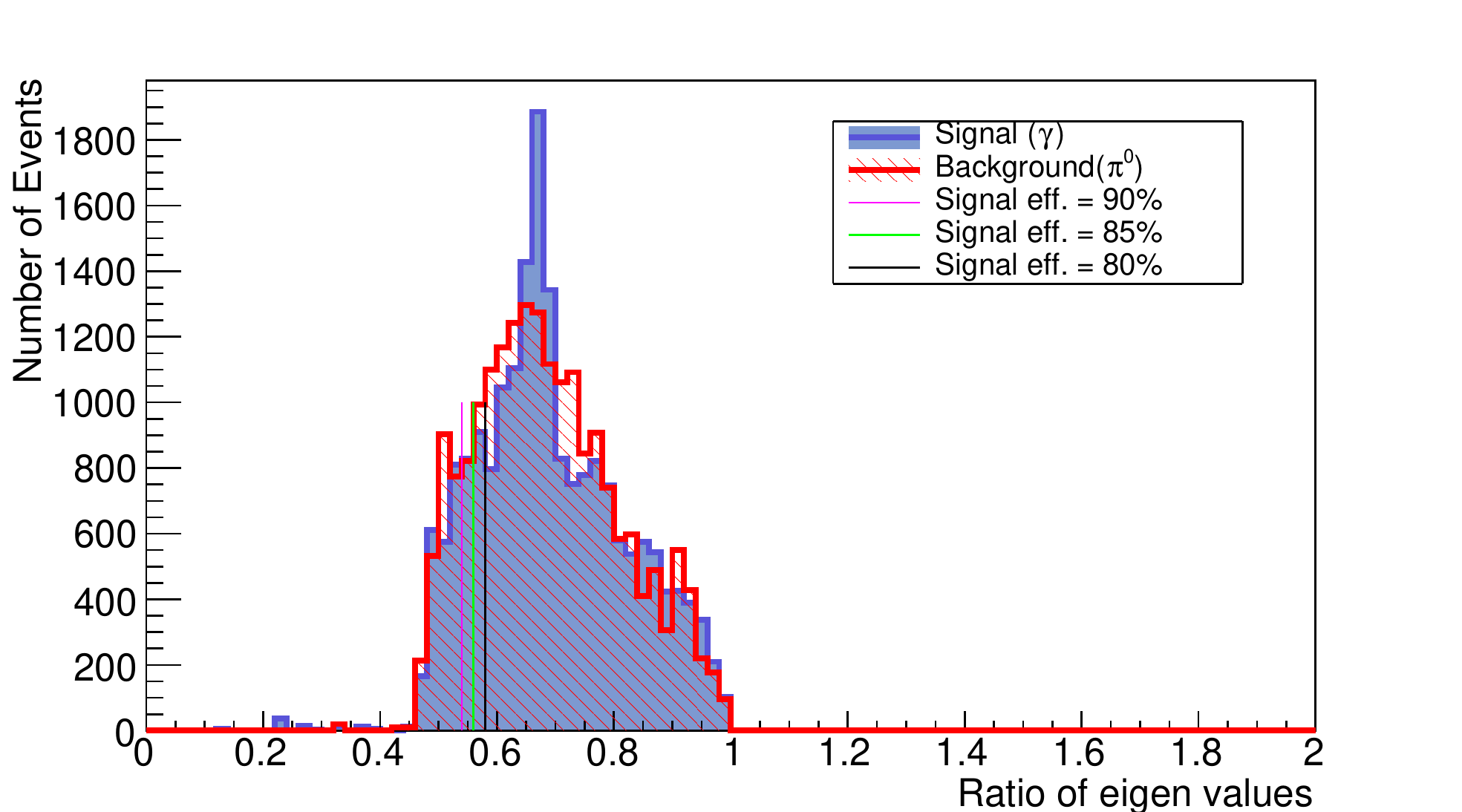}
  \end{tabular} \end{center}
\caption{Distributions for 200 GeV photons and $\pi^{0}$'s of (a) $S1/S9$ on the
         top left; (b) $S1/S4$ on the top right; (c) ratio of eigenvalues using
         coarse grain information and linear weights on the middle left; (d) 
         ratio of eigenvalues using fine grain information and linear weights 
         on the middle right; (e) ratio of eigenvalues using coarse grain 
         information and log weights on the bottom left; (f) ratio of 
         eigenvalues using fine grain information and log weights on the
         bottom right. The black line shows the point where the cut is applied
         to achieve 80$\%$ signal efficiency, blue line for 85$\%$ signal 
         efficiency and the pink line for 90$\%$ signal efficiency.}
\label{fig:comp_showershapes1}
\end{figure}

\begin{table}[htbp]
\begin{center}
\begin{tabular}{|p{5cm}|c|c|c|} \hline
Variable & \multicolumn{3}{c|}{Background rejection ($\%$)} \\ \cline{2-4}
 & $\epsilon_{signal}$=80$\%$ & $\epsilon_{signal}$=85$\%$ & 
   $\epsilon_{signal}$=90$\%$ \\ \hline
S1/S9 & 33.2 & 27.7 & 21.2\\ \hline
S1/S4 & 32.7 & 26.8 & 20.1\\ \hline
Linear weights: Ratio of eigenvalues (coarse grain) & 21.1 & 15.3 & 9.6\\ \hline
Linear weights: Ratio of eigenvalues (fine grain)   & 23.0 & 18.9 & 12.2\\ \hline
Logarithmic weights: Ratio of eigenvalues (coarse grain) & 41.1 & 36.9 & 32.3\\ \hline
Logarithmic weights: Ratio of eigenvalues (fine grain)   & 23.1 & 18.4 & 13.2\\ \hline
MVA(GBDT): coarse grain (fixed gun sample) & 86.3 & 82.0 & 76.5\\ \hline
MVA(GBDT): fine grain  (fixed gun sample) & 99.1 & 99.0 & 98.5\\ \hline
MVA(GBDT): fine grain  (unbinned random gun sample)& 86.3 & 83.3 & 78.9\\ \hline
MVA(GBDT): fine grain  (binned random gun sample)& 92.3 & 91.0 & 89.3\\ \hline
\end{tabular}
\end{center}
\caption{Table showing the background rejection for signal efficiencies of 
         80\%, 85$\%$ and 90\% for various methods. This is shown for energy 
         point of 200 GeV \cite{ref:confpaper}.}
\label{tab:comparison}
\end{table}

If a method shows good separation power for this high energy point, then it 
is good for lower energy points as well. 

BDT gives the best response for the case of both coarse grain and fine grain information 
as can be seen from the figures \ref{fig:fixed:outputBrejvsS},\ref{fig:random:outputBrejvsS} and \ref{fig:random:rejBvsS}. The response of BDTG and BDT are similar. Here the comparison is made using the BDTG. 

Table \ref{tab:comparison} shows the background rejection for all the various 
methods for a signal efficiencies of 80$\%$, 85$\%$ and 90$\%$.
\clearpage
\section{Summary}  \label{sec:summary}
A simulation study of energy and position resolution of a Shashlik detector
is presented. The energy resolution is dominated by 
the sampling fluctuation which contributes to the stochastic term. The
constant term is found to be better than 1\% while the stochastic term
is found to be 10.3\%/$\sqrt{E}$ for light yield value of 4000 p.e./MeV. 
The energy resolution is found to be
similar for lead/LYSO and tungsten/LYSO configurations and the optimum
number of layers is found to be 28 which corresponds to $\sim$25 radiation
lengths deep detector. 
For 125 GeV Higgs boson decaying to a 
pair of photon, this detector will achieve a mass resolution of 0.71 GeV 
when both the photons are detected in the Shashlik detector. 

The position resolution using information of the Shashlik detector alone
is 2.0~mm for photons of 100 GeV. The resolution improves with energy of
the photon and a better resolution is obtained when the center of gravity
method uses logarithmic weighting (to 0.34~mm) or a correction is made for
the S-shape (to 0.22~mm). 

A study of the $\pi^{0}-\gamma$ separation presented in this paper
shows that the fine grain information of the shower profile collected
by individual fibers is useful for separation between $\pi^{0}$ and
$\gamma$ at high energies.
With the MVA technique a background rejection efficiency of $90\%$ with signal efficiency $90\%$ was
achieved, which is approximately three times better than the best background rejection
that could be achieved by cut-based methods.
We proposed a method of virtual slicing of the hit tower and impact point
based training of the network, which gives an additional improvement of 8-10\%.
We conclude that the $\pi^{0}-\gamma$ separation power of the Shashlik calorimeter
can be improved significantly by emplyoying an MVA based method with fine
grain information as input and impact point based training.
In this study we have considered a Shashlik detector of a specific dimension and material. However
the methodology described in this paper for the resolution studies as well as the techniques 
employed for distinguishing between the spatial patterns of energy deposits by a 
photon and a  $\pi^{0}$, can be easily adapted to any sampling calorimeter.

\acknowledgments
We would like to thank Alexander Ledovskoy for useful discussions at various stages of the study.

\end{document}